\def\be{\begin{equation}}
\def\ee{\end{equation}}
\def\bq{\begin{eqnarray}}
\def\eq{\end{eqnarray}}
\def\bi{\begin{itemize}}
\def\ei{\end{itemize}}
\def\ben{\begin{enumerate}}
\def\een{\end{enumerate}}
\def\mnras{MNRAS}
\def\apj{ApJ}
\def\prd{PRD}
\def\apjs{ApJ Supplement Series}
\def\apjl{ApJ Letters}
\def\physrep{Phys Reps}
\def\aj{A J}
\def\aap{A\&A}
\def\nat{Nature}
\def\araa{Annual review of astronomy and astrophysics}
\documentclass[12pt,a4paper]{report}
\usepackage{duthesis}		
\usepackage{epsfig}
\usepackage{natbib}
\usepackage{graphicx}
\usepackage{graphics}
\usepackage{latexsym}
\usepackage{setspace}		
\usepackage{appendix}

\usepackage{xspace}		
\renewcommand{\baselinestretch}{1.5}	 
\def\lcdm{$\Lambda{\rm CDM}$}
\def\g{\gamma}

\def\th{\Theta}
\def\x{{\bf x}}
\begin{document}
\title{Nature of Clustering of Large Scale Structures}
\author{JASWANT KUMAR YADAV \\
 DEPARTMENT OF PHYSICS AND ASTROPHYSICS \\
 UNIVERSITY OF DELHI \\
 DELHI $-$ 110 007 \\
 INDIA
}
%
%
\supervisor{Dr. T.~R.~Seshadri}		
%
\coguidefalse				
%
\hod{Prof. D.~S.~Kulshreshtha}

\vspace{6cm}

\submitdate{\it December,~2008}
\copyrightfalse				
\figurespagetrue			
\tablespagetrue
\beforepreface				


\chapter*{}
\vspace{6cm}
\renewcommand{\baselinestretch}{3.5}
\hspace{10cm} {\bf \bf To my parents....} 
\renewcommand{\baselinestretch}{1.5}

\prefacesection{Acknowledgements}
I would like to thank my supervisor Dr. Terizhandhur Rajagopalan  Seshadri for suggesting this topic of research for my thesis. His able and continuous guidance over the years has been invaluable. During the whole period of my thesis he has given me a lot of freedom, which has contributed immensely to my growth as a researcher.  He has always been there like a brother, discussing research as well as any personal difficulty if I had. Thanks Sesh, you have been an excellent motivator, very friendly and above all a wonderful supervisor. 

I am also deeply grateful to my collaborators and friends in this field. I have benefitted immensely from my collaborations with Somnath Bharadwaj and J.S. Bagla, whose insights into physics and computers have taught me a lot. I thank them for enriching my vision of science. It has been interesting and stimulating to work with Biswajit Pandey. I have also had many interesting discussions on scientific and philosophical matters with Sanil and Keshwarjit. 

I sincerely thank  Kulshreshtha sir and all the other teachers for their continuous encouragement and guidance since my M. Sc. days.  
Over the last five years I have received a great deal of help about administrative matters from Mrs. Dawar. Thanks are also due to staff at the documentation center, Library and at finance section. I thank CSIR, Govt. of India, for providing me research fellowship during a part of my thesis. Computational work for a part of this thesis was carried out at the cluster computing facility in the Harish-Chandra Research Institute (http://cluster.hri.res.in).  

On the personal front - I am grateful to all the friends and non-friends at Room number 184 and IRC in the department. It is not practical to name everyone but I would like to thank everyone by thanking the  ``student's director'' Mr. Ranjit Kumar at 184 and Miss Shalini at IRC. I have had a great time with all of them sharing different softwares and other scientific information.  There are also many people outside Delhi University who have been a lot of help to me. I collectively thank all the friends at HRI, CTS-IITKGP and at IUCAA for all the help and support. 

Ma and Pita Ji, I dedicate this thesis to you. You both are my ever lasting source of inspiration and energy. My respected brother Ashok, my Bhabhi Ji and my cute, little nephew Golu are too near to be thanked ! But for your confidence and the faith you put in me, it would have been impossible for me to survive these long years of academic
pursuit ! It is difficult to say it in person, but let me take this opportunity to say that this is the tribute I found most suitable for all your love, support and understanding ! I wish and pray, I can offer you much more in days to come. I also thank my cousin Satbir Singh for all the encouragement.

I must acknowledge my wife and best friend, Suman, without whose love, support, understanding and encouragement, I would not have finished this thesis. Thank you, Suman, for being what you are. 
\vspace{3cm}

Date:\hspace{9cm}  Jaswant Kumar Yadav

Place: Delhi University, Delhi

\prefacesection{List of publications}
This thesis is based on following publications.
\vspace{0.5cm}

\begin{enumerate}
  \item [1.] {\it ``Testing homogeneity on large scales in Sloan Digital Sky Survey Data Release One''}, \\
{{\bf Jaswant Yadav}, S. Bharadwaj, B. Pandey \& T. R. Seshadri}\\
{\bf \mnras, 2005, 364, 601}
  \item [2.] {\it ``Fractal Dimensions of a Weakly Clustered Distribution and the Scale of Homogeneity'' } \\
{J. S. Bagla, {\bf Jaswant Yadav} \& T.R. Seshadri} \\
{\bf \mnras, 2008, 390, 829 }
\end{enumerate}

\vspace{0.3cm}

{\large \bf {Presentations in International Conferences}}
\begin{enumerate}
\item [1.] {\it ``Fractal Dimension as a probe of Homogeneity''} \\
	{{\bf Jaswant Yadav}, J.S. Bagla \& T. R. Seshadri}\\
	{To appear in proceedings of conference on ''Frontiers in Numerical Gravitational Astrophysics'' held at {\it ERICE, ITALY } from {\it June 27 -July 5, 2008}}
  \item [2.] {\it ``Effects of finite number and clustering on fractal Dimension''}\\
{\bf Jaswant Yadav} \\
{International Conference on Gravitation and Cosmology, Inter University Centre for Astronomy and Astrophysics, Pune, India, December 2007}
\item [3.]  {\it ``Fractal Dimension of a weakly clustered Distribution''}\\
{\bf Jaswant Yadav} \\
{International Workshop on the probes of large scale structures, Inter University Centre for Astronomy and Astrophysics, Pune, India, July 2008}
\end{enumerate}

\prefacesection{Abstract}
In the standard cosmological model the Universe is assumed to have begun approximately $13$ billion years ago when it began expanding from an inconceivably hot and dense state. Since then, the Universe has continued the process of expansion and cooling, eventually reaching the cold sparse state that we observe today. Galaxy surveys carried out in the $20^{th}$ century have revealed that the distribution of galaxies in the Universe is far from random at least on the scales of the survey. This distribution is highly structured over a range of scales. Surveys being currently
undertaken and being planned for the future will provide a wealth of information about these structures. The ultimate goal of this exercise is not only to describe galaxy clustering  but also to explain how this clustering arose as a consequence of evolutionary processes acting on the initial conditions that we see in the Cosmic Microwave Background Anisotropy data.

In order to achieve this goal, we would like to describe cosmic structures quantitatively. We need to build a  mathematically quantifiable description of structures or distribution of points. Identifying the region where the scaling laws apply to these distributions and the nature of these scaling laws is an important part of understanding as to which physical mechanisms have been responsible for the organizations of the clusters, superclusters of galaxies and voids between them. Finding the region where these scaling laws are broken is equally important since it indicates the transition to different underlying physics of structure formation.

The present thesis focuses on characterizing the distribution of points and galaxies using {\it \bf multifractal analysis}. In this attempt the main emphasis is on calculating the Minkowski-Bouligand fractal dimension ($D_q$) of the distribution of points over different scales and hence finding the scale of homogeneity of the distribution. Effect, of finite size of the sample and clustering in the distribution, on the $D_q$ has been studied in detail. The assumption that the large scale distribution of matter in the Universe is homogeneous has been verified with  multifractal analysis of the data from Sloan Digital Sky Survey.

The thesis starts with a broad introduction to standard model of cosmology with special emphasis on the formation and distribution of structures in the Universe. A review of different analytical formalisms and important observations has been presented. A set of notations of different physical and statistical quantities of interest has been provided.

A detailed review of literature, regarding  various statistical techniques for the characterizing  the distribution of matter over large scales, has been presented. The standard analysis of two point correlation function has been discussed. The need to look for a statistical technique which does not presuppose the homogeneity of the distribution on the scale of the sample region has been motivated. In this direction fractal dimension as an alternative to N point correlation functions has been discussed. Various definitions of fractal dimension which are useful
to quantify distribution of points in various density environments have been  presented. A correct prescription to describe the galaxy distribution in the Universe has been presented in the form of Minkowski Bouligand dimension.

A detailed derivation of Minkowski Bouligand dimension for both homogeneous   as well as weakly clustered distribution has been presented. The benchmark dimension to quantify the finite size homogeneous distribution of points has been obtained. An analytical expression for the contribution of weak clustering to the deviation of fractal dimension from the euclidian dimension has been derived. Baryon acoustic oscillations prior to matter radiation decoupling give rise to a bump in the correlation function at a scale of $\sim 100 \,{\it h^{-1}} {\rm Mpc}$. The effect of this bump in correlation function on the behavior  of fractal dimension of clustered distribution has also been discussed. The multifractal technique has been applied to the  unbiased (e.g. the $L_*$ type of galaxies) as well as biased (e.g. Large Redshift Galaxies) tracers of underlying matter distribution in the concordant model of cosmology.

In the end the application of multifractal analysis to the distribution of galaxies in the Sloan Digital Sky survey has been presented. This exercise has been undertaken to obtain the scale of homogeneity of the Universe. The galaxy distribution  from the SDSS has been projected on the equatorial plane and a 2-dimensional multi-fractal analysis has been carried out by counting the number of galaxies inside circles of different radii $r$ in the range $5 \, h^{-1} {\rm Mpc}$ to $150 \, h^{-1} {\rm Mpc}$. The comparison of the galaxy distribution with different realizations of  point distributions from an N-Body simulation has been presented. It has been concluded that the galaxy distribution in the volume limited subsamples of sloan digital sky survey is homogeneous on large scales well within the survey region.
\afterpreface			
%
\chapter{Introduction}\label{chap1}
Cosmology is the Scientific study of the cosmos a whole. An essential part of cosmology
is to test theoretical models with observations. During the last decades
we have witnessed an unprecedented advance in both theory and observations of the Universe.
For the first time we have the tools to answer some of the most fundamental questions
in cosmology.

\begin{figure}
\begin{center}
\rotatebox{0}{\scalebox{0.5}{\includegraphics{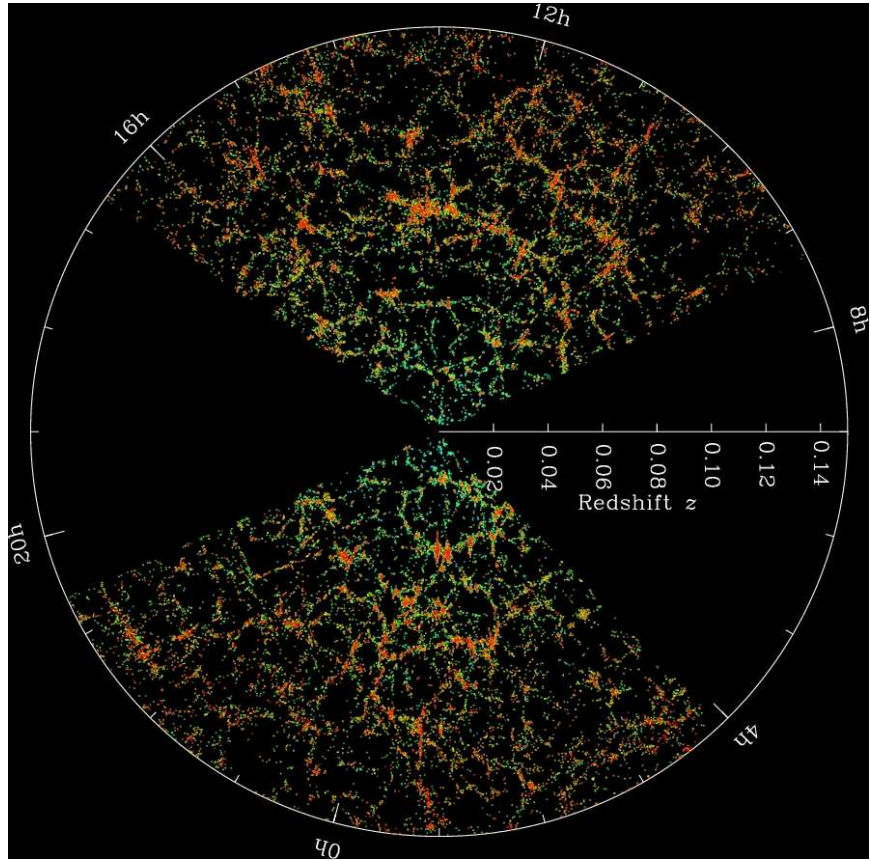}}}
\caption[\sf Large scale structures in SDSS main galaxy redshift sample]{\sf Slices through the SDSS $3-$dimensional map of the distribution of galaxies. Earth is at the center,  and each point represents a galaxy,  typically containing about 100 billion stars. Galaxies are colored according to the ages of their stars,  with the redder,  more strongly clustered points showing galaxies that are made of older stars. The outer circle is at a distance of two billion light years. The region between the wedges was not mapped by the SDSS because dust in our own Galaxy obscures the view of the distant universe in these directions. Both slices contain all galaxies within $-1.25$ and $1.25$ degrees declination. Figure Courtesy : Michael Blanton and the Sloan Digital Sky Survey team}
\label{sdsspie}
\end{center}
\end{figure}
The current paradigm of cosmology states that the the Universe originated some 13.7
billion years ago as an extremely energetic event out of which all matter,  energy and
indeed spacetime emerged into existence. This is known as the Hot Big Bang Theory
\citep{1950QB51.H8........}. This extremely dense and hot Universe expanded and cooled
down. The evolution of the Universe is dictated by gravity which is the weakest
force in nature. The Big Bang Theory provides an answer to the evolution of the Universe and its global
properties. However,  understanding of the theory of structure formation is still incomplete within the framework of Big Bang Cosmology.

The large galaxy surveys indicate
that the galaxy and matter distribution on scales even up to a few dozen Megaparsecs  is
far from homogeneous (see figure \ref{sdsspie}). Starting with systematic redshift surveys
like the CfA survey \citep{1982ApJ...257..423H,  1986ApJ...302L...1D} and the Las
Campanas Redshift Survey \citep{1996ApJ...470..172S} up to the major 2dFGRS and SDSS
mapping campaigns \citep{2003astro.ph..6581C},  we have learnt that galaxies are large associations
of different objects from a few up to hundreds of Megaparsec \citep{1983ARA&A..21..373O}.

The most outstanding concentrations of galaxies are the clusters of galaxies \citep{1989HiA.....8..435B}. They are the most massive,  and most recently fully collapsed and virilized
objects in the Universe. The richest clusters contain many tens ($\sim 50$ to $1000$) of galaxies within a
relatively small region  of only a few Megaparsecs in length scale. A typical example of a rich cluster
is the Coma cluster $A1656$.  Clusters of galaxies contain dense and
compact concentrations of dark matter,  representing overdensities $\Delta \approx 1000$. Galaxies
and stars only form a minor constituent of clusters,  they are trapped and embedded in the
deep gravitational wells of dark matter. These are best identified as a bright source of
X-ray emission,  emerging from the diffuse extremely hot $(107 - 108 K)$ intracluster gas
trapped inside them \citep{2001A&A...369..826B}.
A richly structured network of elongated filaments bridges the space in between massive
clusters. They form highly coherent canals along which matter is accreted on to the
clusters located at the nodes of the network. The canonical example of a filament is the
Pisces-Perseus supercluster,  a system of clusters and filaments extending over more than
$100 h^{-1} Mpc$. It includes the massive Perseus cluster which is one of the most prominent clusters
in the nearby Universe.

Filaments appear to frame tenuous
planar agglomerations known as walls. Because of their low surface density walls are
usually difficult to identify. Walls and filaments define the boundaries of vast near-empty
regions of space,  the voids,  with dimensions ranging up to $30 - 50 h^{-1} Mpc$
\citep[see e.g.][]{1981ApJ...248L..57K,  1987ApJ...314..493K, 1986ApJ...302L...1D, 2003astro.ph..6581C}. Voids play a dominant role in the spatial organization of matter on Megaparsec scales. While they
occupy most of space,  their narrow spacing define a framework of interconnected clusters,  filaments
and walls that pervades the whole of the visible Universe. This pattern has become
known as the Cosmic Web \citep{1996Natur.380..603B, 2005MNRAS.364.1105S}.
\section{\bf  Motivation of our investigation}
A large amount of theoretical work has been directed towards understanding the formation
and properties of the elements of the Cosmic Web as a result of the gravitational
growth of initially tiny random density and velocity fluctuations \citep{1980lssu.book.....P, 2005MNRAS.364.1105S}. These studies describe the formation and properties of the structural elements
of the Cosmic Web based on the primordial density field. Some of them can be used
to obtain a general or statistical description of its individual components \citep{1970A&A.....5...84Z, 1986ApJ...304...15B, 1996ApJS..103....1B, 2006ApJ...645..783S} while others go one
step further and elucidate the complex relation between them \citep{1996Natur.380..603B}; see also
\citep{2002ASSL..276..119V,  2006astro.ph..7539V}.
The different morphologies of the Cosmic Web define unique cosmic environments
in terms of local density,  dynamics and gravitational influence. This is reflected in their
internal structure and particular dynamics. The influence of the Cosmic Web is also seen
on galactic scales. The same processes that give rise to the Megaparsec scale matter
distribution also affect the properties of the galaxies.

A detailed statistical analysis of the Cosmic Web is very much relevant for our understanding of the formation of structure in the Universe. It is also important for defining the diverse
cosmic environments in which galaxies form and evolve.  There are some techniques for  identifying and quantifying the morphological elements in the Cosmic Web,  however,  many of them have several limitations.

A proper characterization of the Cosmic Web is crucial in order to identify,  differentiate,
select and isolate the different morphological and dynamical environments. The
availability of such a method would open up unprecedented possibilities for a much better,
focused and well-defined study of the cosmic web. It will provide a physically better
definition of cosmic environment than hitherto available and pave the way for a crisp
and considerably improved assessment and understanding of the influence on the formation
of galaxies.

In the rest of this chapter we review the basic theoretical background that will be used
in this thesis. Section \ref{HBG} describes the Hot big Bang model of the Universe. We describe a theoretical framework for growth of structures
from primordial fluctuation in section \ref{TGI}. The linear as well as non linear regimes of structure formation are described in section \ref{lin}
to \ref{nlin}. The ideas of galaxy distribution and observations are explained in section \ref{gdis} and \ref{gsur}. We conclude this chapter by
defining the goals and an outline of this thesis in section \ref{got}.

For a more complete discussion we refer the reader to the textbooks by \citet{1980lssu.book.....P,  1993ppc..book.....P}; \citet{1993sfu..book.....P,  2002tagc.book.....P}; \citet{1999coph.book.....P}; \citet{2002itc..book.....N}; \citet{2002coec.book.....C}; \citet{2000cils.book.....L}; \citet{2002sgd..book.....M} and \citet{2005spcs.book.....L}. For a good and up-to-date overview of the current knowledge on the
Big Bang universe see \citet{2008arXiv0802.2005R}.

\section{\bf Hot Big Bang}\label{HBG}
The theoretical framework on which most theories of our Universe are based is the Cosmological Principle. It states that the Universe is homogeneous and isotropic. General theory of relativity,  proposed by Albert Einstein,  explains and describes gravity. General relativity is a metric theory that describes gravity as the manifestation of the curvature of spacetime.  This theory implies that the Universe should either be expanding or contracting.  This is true for universes with flat,  hyperbolic and spherical curvature. Usually these curvatures are denoted by means of the scaled curvature coefficient $k$.  It has the values $k = 0$ for a flat space,  $k = +1$ for a spherical space and $k = -1$ for a negatively curved hyperbolic space. The spacetime metric of these universes can be described by the Robertson-Walker
metric,
\be
ds^2=c^2dt^2-a^2(t)\left(dr^2+{R_c}^2{S_k}^2\left(r/R_c\right)\left({d\theta}^2+{sin}^2\theta {d\phi}^2 \right)\right)
\ee
where $R_c$ is the radius of curvature and $S_k(r)$ is the function given by
\be
S_k(x) = \left\{\begin{array}{cl}
            \sin(x)  	&k = +1 \\
            x       	&k = ~0 \\
	       \sinh(x) 	&k = -1
    \end{array}\right.
\ee
The variable $t$ is the proper \textit{cosmic time},  synchronized on the basis of Weyl's postulate\footnote{Weyl's postulate states that the world lines of galaxies form
a bundle of non-intersecting geodesics orthogonal to a series of spacelike hypersurfaces.
This series of hypersurfaces allows for a common cosmic time and
the spacelike hypersurfaces are the surfaces of simultaneity with respect to
this cosmic time}.
The dimensionless scale factor $a(t)$ describes the expansion (or contraction) of the Universe
and may be normalized with respect to the present-day value,  i.e. $a(t_0) = 1$. $c$ is the velocity of light and $r, \theta, \phi$ are the usual spherical coordinates. \citet{1922ZPhy...10..377F} solved Einstein's field equations for general homogeneous and isotropic Universe models and derived the time dependence of the expansion factor. The resulting equations are known as the Friedman-Robertson-Walker-Lemaitre $(FRW)$ equations.
They form the basis of almost all of modern cosmology,
\be
\frac{\ddot{a}}{a}=-\frac{4 \pi G}{3}\left(\rho+\frac{3p}{c^2}\right)+\frac{\Lambda}{3}
\label{frw2}
\ee
and
\be
\left(\frac{\dot{a}}{a}\right)^2 = \frac{8 \pi G \rho}{3} - \frac{kc^2}{a^2{R_0}^2}+\frac{\Lambda}{3}
\label{frw1}
\ee
In the Friedman-Robertson-Walker-Lemaitre equations $G$ is Newton's gravitational constant,
$\rho$ is the energy  density of the universe,  $p$ is the pressure of the various
cosmic components, $\Lambda$ is the cosmological constant and $R_0$ is the present-day value of the
curvature radius.

The evolution of the energy density $(\rho)$ of the Universe can be inferred from the energy equation obtained by combining the $FRW$ equation \ref{frw2} and \ref{frw1}. This is given by
\be
\dot{\rho} + 3\left(\rho+\frac{p}{c^2}\right)\frac{\dot a}{a}= 0
\label{ener}
\ee
The macroscopic nature of the medium is expressed by the equation of state,  $p = p(\rho)$,
which for most cosmologically relevant components may be expressed as
\be
p = w \rho c^2.
\label{eos}
\ee
Here $w$ is called the equation of state parameter. Equation \ref{ener} and \ref{eos} can be combined to give the evolution of energy density with
the expansion of the Universe:
\be
\rho(t) \propto a(t)^{-3(1+w)}.
\label{evol}
\ee

\subsection{Cosmic Expansion}

The expansion rate of the Universe is expressed in terms of the Hubble parameter,
\be
H(t) =\frac{\dot{a}}{a}.
\ee
The present-day value of $H(t)$,  sometimes called the Hubble ``constant'',  is often parameterized
in terms of a dimensionless factor $\, {h}$ $( = H_0 / 100 \, km^{-1}~s ~Mpc)$,  where $H_0$ is the Hubble constant expressed in units of \, $km~s^{-1}~Mpc^{-1}$. The expansion of the Universe does not only express itself in
continuously growing distances between any two objects,  it also leads to the increase of the wavelengths of photons. This resulting cosmological redshift $z$ of a
presently observed object is given by the relation
\be
1 + z = \frac{a(t_0)}{a(t)} =  \frac{1}{a(t)}
\ee
where $a(t)$ is the expansion factor of the Universe at the time the observed light was
emitted.

\subsection{Cosmic Constituents}

The evolution of the Universe is fully dictated by its energy density $\rho$ and its curvature $k$. The energy density of the Universe is conveniently expressed in terms of the density needed to produce
a geometrically flat Universe,  the \textit{ critical density}:

\be
\rho_c(t)=\frac{3 {H}^2}{8 \pi G}.
\ee
The value of critical density at present epoch is thus $\rho_{c, 0} = \rho_c(t=0)=1.9 \times 10^{-29}[{\, h}^2 g/cm^3]$. The contribution of any
component towards the energy density of the Universe may be expressed in terms of  the ratio of its energy density to the critical density. This
ratio is denoted by $\Omega(t)$,  the {\it density parameter},  and is expressed as:
\be
\Omega(t) = \frac{\rho(t)}{\rho_c(t)} = \frac{8 \pi G \rho}{3 H^2}.
\label{denp}
\ee
The value of $\Omega(t)$ at $t = t_0$ (denoted by $\Omega$) is given by
\be
\Omega =\frac{8 \pi G \rho_0}{3 {H_0}^2}.
\ee
The Universe contains a variety of components. While the contributions of e.g. magnetic
fields and gravitational waves may be held negligible,  the most important ingredients of
the Universe are radiation,  baryonic matter,  nonbaryonic dark matter and dark energy.
The equation of state parameter $w$ for radiation and  matter (baryonic as well as non baryonic) is $1/3$ and $0$ respectively,  whereas for dark energy its
value is less than $-1/3$. If the dark energy is in the form of a cosmological constant,  then $w=-1$. Thus equation \ref{evol}
  suggest that radiation $(\rho_r \propto a^{-4})$,  matter $(\rho_m \propto a^{-3})$ and dark energy ($\rho_\Lambda = constant$)
have evolved differently with the expansion of the Universe.

\begin{figure}
\begin{center}
\rotatebox{0}{\scalebox{1.0}{\includegraphics{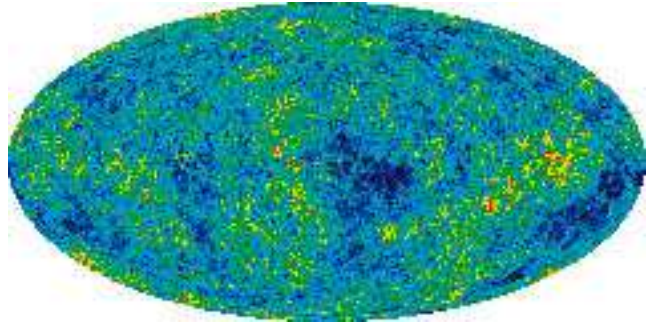}}}
\caption[\sf Sky projection of Cosmic Microwave Background]{\sf Sky projection of the Cosmic Microwave Background (CMB). The shades of gray correspond to temperature fluctuations. Courtesy of the WMAP team.}
\label{cmbr}
\end{center}
\end{figure}

As the radiation cools off as a result of the expansion of the Universe,  its spectrum peaks at microwave wavelengths and is observed today in the form of the {\bf Cosmic Microwave Background} (CMB) with a temperature of $T_0 = 2.725 K^\circ$. Since the temperature of radiation scales in inverse
proportion to the scale factor $(T \propto a^{-1}(t))$,  it must have been very high in the early Universe. The almost perfect blackbody spectrum of CMB
defines the strongest evidence for the existence of a very hot and dense phase in the early
Universe,  i.e. for the Hot Big Bang (see figure \ref{cmbr}). Technically speaking we should also
include cosmic neutrinos in the radiation bill,  even though they do not interact with any other
cosmic species beyond $z \sim 10^{11}$ and are approximately $4$ times less abundant than photons. At very early
times radiation was dynamically dominant component of the Universe. Its current density is about $10^{-34} g~cm^{-3}$ and constitutes only a fraction $10^{-5}$ of the total density.

Baryonic matter $\Omega_b$ is the normal matter we ourselves,  planets and stars are made of.
It is mainly in the form of protons and neutrons (and also electrons). However,  it only represents a minor cosmological component and accounts
for a mere $4.4\%$ of the energy content of the Universe. Nonbaryonic Dark Matter $\Omega_{dm}$ is a very important component for the formation
of structures in the Universe. It accounts for $\approx 23\%$ of the energy content of the Universe.
The combined contribution of matter (baryonic and non baryonic dark matter) to the energy density is usually expressed as ${\Omega_m}$. One of the
most pressing problems in astrophysics is the identity of this dark matter. While its presence
is unmistakably felt through its gravitational attraction,  it has as yet escaped direct
observation or detection in the laboratory. The commonly accepted view is that it is some
unknown weakly interacting particle,  presumably some of the particles predicted
by supersymmetric theories. Dark matter is pressureless and insensitive to the electromagnetic influence of radiation.

Fluctuations in the dark matter could have started growing as soon as matter began to dominate
the dynamics of the Universe at around the epoch of \textit {matter-radiation equality}
$\left( \rho_r=\rho_m \right)$. This occurs at a scale factor of $a(t) \approx 10^{-4}$. The growth of these fluctuations in the dark matter created the gravitational potential
wells. After the baryonic matter and  radiation decoupled at the epoch of recombination,  the baryonic matter started falling into these gravitation potential wells. This process is believed to have led to the formation of galaxies and stars. Without dark matter it would have been impossible to form the rich structure we observe in today's Universe.

Finally,  we now have conclusive evidence to suggest that Universe at the present epoch is undergoing an accelerated expansion (i.e $\ddot{a} > 0 $).
This could be due to the presence of an elusive medium called \textit {Dark Energy}. Dark Energy $(\Omega_\Lambda)$ is the most dominant component
of our Universe at the present epoch. It accounts for $\approx 73\%$ of cosmic energy density. The nature of Dark Energy is even more mysterious than dark
matter. All that can be said about dark energy is that it has a negative pressure. This is apparent from equation \ref{frw2} which suggests that for
$(\ddot{a} > 0)$ we need $p < -\rho/3$. Most observational studies agree with the Dark Energy being equivalent to a cosmological constant although
other options are still viable.

The influence of dark energy on the structure formation process is mainly related to its impact on the expansion
rate and timescales in the Universe. As soon as the expansion rate of the Universe becomes too high,  structure formation comes to a halt. On the
other hand,  it has stretched the time available in the past to form and evolve structure. It is once again stressed that the evolution of energy
density of radiation,  matter and dark energy is governed by energy equation given by \ref{ener}.

The cosmological framework of the Hot Big Bang in a spatially homogeneous and
isotropic Universe is so widely accepted that it is called the \textit {Standard Hot Big Bang Model}.
This model is supported by a number of observations,
\begin{itemize}
\item
 The relation between distance and recession velocity (Hubble Law) as a consequence
of its metric and also implies that the Universe has a finite age.
\item
 The almost perfect black-body spectrum of the Cosmic Microwave Background,
which is evidence for an extremely hot initial phase of the Universe.
\item
 The excellent match in the observed abundances of light elements and predictions
from primordial nucleosynthesis.
\item
 The evident evolution of the appearance of objects as function of their distance
from us.
\end{itemize}
\subsection{The \lcdm~ Model}
Our current understanding of the components of the Universe is encoded in the Lambda
Cold Dark Matter (\lcdm~) model. In this model we attempt to explain supernova observations in terms of the accelerated expansion of the Universe.
This model is also capable of explaining the observed Cosmic Web and the Cosmic Microwave Background. In the acronym \lcdm,  the term $\Lambda$
refers to the dark energy $(\Omega_\Lambda)$ which is believed to be the driving force behind the accelerated expansion of the Universe at the
present epoch. $\Lambda$ is assumed to have the form of a cosmological constant ($w =-1$). {\it Cold Dark Matter} refers to a model where the dark
matter is explained as being {\it cold},  i.e.,  its velocity was non relativistic at an epoch when it decoupled from other constituents of the Universe. This type of dark matter
is assumed to be non-baryonic,  dissipationless and collisionless. The \lcdm~ model has several parameters from which the most important are shown in
table \ref{tab1}. In this thesis
we base ourselves on the \lcdm~.
\begin{table}[htbp]
\begin{center}
\begin{tabular}{l l c}
  \hline
  \hline
  Parameter & Value & Description \\ \hline
  $H_0$ & $70.4 \pm 2.4 km~s^{-1}~Mpc^{-1}$ & Hubble parameter \\
  $\Omega_m$ & $0.277 \pm 0.029$ & Matter Density \\
  $\Omega_b$ & $0.0459 \pm 0.0028$ & Baryon Density \\
  $\Omega_\Lambda$ & $0.723 \pm 0.029$ & Dark Energy Density \\
  $\rho_{c0}$ & $0.94 \pm 0.07 \times 10^{-26} kg~m^{-3}$ & Critical Density \\
  $t_0$ & $13.72 \pm 0.14 Gyr$ & Age of the Universe \\
  $\sigma_8$ & $0.811 \pm 0.032$ & Galaxy fluctuation amplitude \\
  $n$ & $0.960 \pm 0.014$ & Spectral Index \\
  \hline
\end{tabular}
\caption{Most recent derived values of cosmological parameters \citep[WMAP5 +SDSS;][]{2008arXiv0803.0547K}}
\label{tab1}
\end{center}
\end{table}
\section{\bf  The Gravitation Instability}\label{TGI}
The fact that at the present time we see structures even at scales of hundreds of Megaparsecs
requires an explanation. In order to understand this fact the primordial Universe
is assumed to have been completely homogeneous.  Quantum fluctuation created during inflation led to perturbation to this homogeneous background.
These fluctuations amplified under the influence
of gravitational field,  ultimately resulting in the wealth of structures we can see today pervading
the Universe at different scales. The theoretical framework that describes the
growth of structures from the primordial fluctuations is called the \textit {gravitational instability
theory}.

An integral ingredient of today's standard cosmological model is the assumption that
origin of fluctuations is to be found in the very early universe during the inflationary phase. Shortly after
the Big Bang the Universe entered a phase of extremely rapid expansion. Presumably this
phase may be identified with the $GUT$ transition at $\approx 10^{-34}$ seconds after the Big Bang.
Small quantum fluctuations present in the first instants of the Universe were blown up to
cosmological scales. Not only does this imply a Universe marked by an inhomogeneous
matter and energy distribution,  it also predicts the fluctuations to have the character of a
spatial Gaussian random field. The inhomogeneities in the primordial density field can be conveniently expressed as the fluctuations in density
field superimposed on a uniform and isotropic background.

Consider a density field $\rho({\bf x,t})$. The average density $\bar\rho(t)$ for such a field can be defined by taking average over a constant time
hypersurface. This can be expressed as,
\be
\bar\rho(t) = \frac{\int\rho({\bf x, t}) d^3{\bf x}}{\int d^3{\bf x}}
\ee
The density fluctuation in such a field can now be defined as
\be
\delta\rho({\bf x, t}) = \rho({\bf x, t}) - \bar\rho({\bf t})
\ee
In the linear theory,  we expand the equation of motion around the homogeneous universe. To this end, one commonly introduces a dimensionless density
contrast given by
\be
\delta({\bf x, t})=\frac{\delta\rho({\bf x, t})}{\bar\rho({\bf t})}
\ee
Henceforth,  we shall be denoting $\delta({\bf x, t})$ by $\delta({\bf x})$ and $\delta({\bf t})$ by $\delta$ just for symbolic convenience.
The gravitational acceleration at any position can be described as the contribution from all the matter fluctuations present in the density field,
\be
g({\bf x}) = Ga\bar\rho({\bf t}) \int \frac{\delta({\bf x'})({\bf x'-x})}{|{\bf x'-x}|^3}d^3{\bf x}'
\label{cont2}
\ee
where all the symbols have their usual meaning.

The formation of structures is the result of the gravitational growth of the primordial density
fluctuations. Gravity has an amplifying effect on the initial fluctuations. Any region with a density higher than its surroundings will collapse and
increase its level of density contrast. The increase in density contrast will  reflect in the gravitational
field attracting even more matter into the initial perturbation. The opposite effect
occurs in underdense regions. As matter flows out of them they become less dense.
The gravitational force will be weaker and more mass will escape from the underdense
region. All in all this will result in a runaway process in which any existing perturbation
will be amplified. Overdense regions will collapse until they become bound objects and
underdense regions will expand until they are devoid of matter.

After the epoch of matter radiation equality the Universe is matter dominated and  hence can be assumed to be pressureless to a good approximation.
On cosmological scales one may, to a good approximation, describe the evolution of
the cosmic density field by a set of three coupled differential equations involving the density contrast $\delta$,  the peculier velocity ${\bf v}$ and the gravitational potential $\phi$ :
\begin{itemize}
\item
The continuity equation which ensures mass conservation is given by,
\be
\dot\delta +\nabla.[(1+\delta){\bf v}] =0
\ee
\item
 The Euler equation which is the equation of motion of a fluid element can be expressed as
\be
\dot{\bf v} + \frac{\dot a}{a}{\bf v} +\frac{1}{a}({\bf v}.\nabla){\bf v}= -\frac{1}{a}\nabla\phi
\ee
\item
The Poisson equation relating the distribution of matter and the gravitational field is represented as
\be
\nabla^2\phi=4\pi G \bar\rho a^2 \delta
\ee
\end{itemize}
\section{\bf The Linear Regime}\label{lin}
In the case of small fluctuations $(\delta \ll 1)$ and small streaming motions,  $\delta$ and $v$ can be
computed from linear perturbation theory \citep{1980lssu.book.....P}. In the linear approximation the evolution equation of $\delta$ is given by
\be
\frac{\partial^2 \delta}{\partial t^2}+2\frac{\dot a}{a}\frac{\partial
  \delta}{\partial t}=4 \pi G \bar\rho \delta
  \label{eq0}
\ee
This second order differential equation describes the time evolution for the mass fluctuation
$\delta=\delta\rho/\rho$ for a pressureless fluid. The solution to this differential equation involves
two modes,
\be
\delta=A(x)D_1(t)+B(x)D_2(t)
\label{eq1}
\ee
where $D_1$ and $D_2$ are linearly independent function. They correspond to one growing and one
decaying solution. Usually,  one concentrates on the growing mode because the
decaying solution is damped and becomes subdominant. Taking $D_1(t)$ as the growing mode and $D_2(t)$ as the decaying mode we can simplify equation \ref{eq1} as
\be
\delta=A(x)D_1(t).
\label{eq11}
\ee
For a generic $FRW$ Universe in which we ignore the radiation contribution,  we may find
the following general expression for the growing mode:
\be
D_1(z)=\frac{5\Omega_{m, 0}H_0}{2}H(z)\int_z^{\infty}\frac{1+z'}{H^3(z')}dz'
\ee
where $H(z)$ is the Hubble parameter,  defined as
\be
H(z) = H_0[\Omega_m(1 + z)^3 +\Omega_k(1 + z)^2 +\Omega_\Lambda]^{1/2},
\ee
Here $\Omega_k = 1-\Omega_m-\Omega_\Lambda$. In the Einstein-de Sitter model the expansion parameter varies
as $a \propto t^{2/3}$ and the solution of equation \ref{eq0} is
\be
\delta = At^{2/3}+Bt^{-1}
\ee
\subsection{Cosmic Velocity Flow Perturbations}
In the linear regime (and also $\delta \ll 1$),  using the growing mode solution of $\delta$ (i.e. equation \ref{eq11}),  the  continuity equation
takes the form
\be
\nabla. {\bf v} = - a \dot\delta = -a \delta \frac{\dot{D_1(t)}}{D_1(t)}
\label{cont}
\ee
From the Helmholtz theorem we can express the velocity field as a sum of a divergence free part and an irrotational part. The divergence free part does not contribute
to the evolution of the density contrast $(\delta)$ and decays as $a^{-1}(t)$ \citep{1980lssu.book.....P}. The solution for the curl free part is
given by
\be
{\bf v}({\bf x}) = a \frac{f H}{4 \pi}\int\frac{\bf y-x}{\bf {|y-x|}^3}\delta({\bf y}) d^3{\bf y}
\label{cont1}
\ee
where
\be
f = \frac{a}{\dot{a}}\frac{\dot{D_1(t)}}{D_1(t)} = \frac{1}{H} \frac{\dot{D_1(t)}}{D_1(t)} = \frac{d \log{D_1(t)}}{d \log{a}}
\ee
Comparing equation \ref{cont1} with the equation for acceleration i.e. equation \ref{cont2},   we see that the peculiar velocity can be written as
\be
{\bf v} = \frac{f H}{4 \pi G \bar\rho}{\bf g} = \frac{2}{3}\frac{f}{\Omega_m H}{\bf g}
\ee
\subsection{Growth of Cosmic Structure}
At early times in a matter-dominated Universe the growth of structure closely resembles
that in an Einstein de Sitter Universe at the time $\Omega_m \approx 1$,
\be
D_1(t)\approx a(t) \propto t^{2/3}.
\ee
As the Universe evolves and becomes increasingly empty it enters a nearly free expanding
phase when the scale factor is given by
\be
a_f=\frac{1}{1/\Omega_o - 1}.
\ee
After this time it expands according to
\be
a(t)\approx H_0 t
\ee
The growth of structure in such a scenario freezes out as gravity is no longer able to
counter the fast cosmic expansion. Hence
\be
D_1(t) \approx constant.
\ee
In a $\Lambda$-dominated Universe growth of structures comes to a halt in such a situation. The crucial
transition time is that where dark energy takes over the dynamics, setting the Universe in a phase of
accelerated expansion:
\be
a_{m, \Lambda}=\left(\frac{\Omega_m}{2\Omega_\Lambda}\right)^{1/3}
\ee
In the concordance model this corresponds to $z \approx 0.7$.

This,  however,  does not imply that the
the growth of structures freezes out completely. The growth of structures continues on small scales as long
as they are embedded in an overdense region detached from the general expansion. The
regions in the vicinity of filaments and clusters remain dynamically active and matter still
flows into clusters far beyond the time at which the Universe enters free expansion.
This results in overdense chunks of matter becoming isolated islands in the expanding Universe.
\section{\bf Gaussian Random Fields}
The primordial perturbations in the cosmic matter and energy density are assumed to constitute
a stochastic field of spatially random fluctuations. The density field of the early Universe is
assumed to be a near perfect {\it Gaussian random field}. In addition
to the observed near-Gaussianity of the Cosmic Microwave Background temperature
anisotropies,  the two important rationales behind this expectation are
\ben
\item
The Gaussian nature of quantum fluctuation arising due to inflation and then expanding into macroscopic fluctuation.
\item
The Central Limit Theorem which states that the sum of a sufficiently large number of identically distributed independent random variables each with
finite mean and variance is approximately normally distributed
\een

We can think of the description of a spatial random field in terms of its $n$-point probability distribution
$({\bf PDF})$ $ P^{(n)}\left(\delta_1, \delta_2, \delta_3, \cdots, \delta_n\right)$. The
fluctuations in the primordial density field are
assumed to be Gaussian,  meaning thereby that the $PDF$ is given by
\be
P^{(n)}\left(\delta_1, \delta_2, \delta_3, \cdots, \delta_n\right)d\delta_1, d\delta_2, \cdots, d\delta_n
=\frac{\exp\left[-\frac{1}{2}\delta_i\left(M^{-1}\right)_{i, j}\delta_j\right]}{\left(2\pi\right)^{n/2}\left(detM\right)^{1/2}}
\Pi_{i=1}^{N}d\delta_i
\ee
where $M_{i, j} =\langle\delta_i\delta_j\rangle $ is the covariance matrix. The averaging is performed over ensembles.
Under the assumption of ergodicity,  averages over space approaches averages over
ensembles of Universes. The covariance matrix determines the variance of the distribution,
and the correlation properties of the fluctuation field. For a homogeneous Universe it is given by :
\be
M_{i, j} =\langle\delta_i\delta_j\rangle=\xi\left({\bf x_i-x_j}\right)
\ee
where $\xi({\bf r})$ is the autocorrelation of the density field. For a discrete point
distribution it is usually referred to as  the {\it two-point correlation function} which in the
isotropic case is simply $\xi({\bf r})=\xi(r)$. This reflects the fact that the two point correlation
function only depends on the mutual distance between the points. Phase information is
lost,  limiting our ability to describe the patterns present in the matter distribution.
The statistical properties of a Gaussian random field,  however,  are completely determined by its
two-point correlation function which is the inverse Fourier transform of the \textit{power spectrum}:
\be
\langle\delta({\bf x})\delta({\bf x+r})\rangle=\int\frac{d^3\bf{k}}{\left(2\pi\right)^3}P(k)\exp^{\bf -ikr}
\ee
It also defines the amplitude of density perturbations,
\be
\sigma^2= \frac{1}{(2\pi)}^3\int d^3k P(k) = \int\frac{d\log{k}} {2\pi^2} k^3 P(k),
\ee
where $k^3P(k)$ encapsulates the contribution of fluctuations at wavenumber $k$ to the general
fluctuations field. For a simple power-law power spectrum $P(k) \propto k^n$,  the corresponding
fluctuations on a mass scale are easily shown to be:
\be
{\sigma_m}^2 \propto M^{-(n+3)}
\ee
In other words,  as long as $n > -3$ the fluctuation level is a decreasing function of the
mass scale. Such scenarios are called hierarchical clustering scenarios.

\subsection{The shape of the power spectrum}
The initial shape of the power spectrum is governed by those quantum processes which were responsible for the generation of primordial density
fluctuation. These fluctuations grew to sizes larger than Hubble radius ($cH^{-1}$)during inflation. In the post inflationary era,  these
perturbations re-entered the Hubble radius. The perturbation at the epoch of hubble radius exit determine the nature of perturbation at the Hubble
radius re-entry from Bardeen's gauge invariant formalism \citep{sesh}. In most inflationary models,  the primordial power spectrum is scale invariant or has a weak dependence on $k$ when the corresponding mode enters the Hubble radius. After re-entry,  to the Hubble radius of the expanding Universe,  the fluctuations could start growing. The resulting  power spectrum is of the form given by
\be
P(k) \propto k^n.
\ee
with $n \approx 1$. This is commonly referred to as the Harrison-Zel'dovich spectrum \citep{1970PhRvD...1.2726H}.
This scale-free power spectrum has the property that any perturbation
in the metric or gravitational potential are independent of scale
\be
\frac{d\sigma^2(\phi)}{d\ln k } = constant.
\ee
\citet{1970PhRvD...1.2726H},  \citet{1970A&A.....5...84Z} and \citet{1970ApJ...162..815P}, all pointed out its importance well before inflation was
suggested. The index $n \sim 1$ is now seen as one of the
essential predictions of inflation and has been already observed by $WMAP$. There are
other possibilities with tilted power spectra $n \neq 1$. In this thesis we will restrict ourselves
to the Harrison-Zel'dovich spectrum in a universe with \textit {cold} dark matter.

Once fluctuations have become smaller than the horizon they are
affected by gravity and damping processes. Fluctuations in baryonic matter cannot grow
as a result of the pressure of the coupled baryon-photon fluid,  i.e. as long as they are smaller
than the corresponding Jeans length. The fluctuations in dark matter  hardly grow as
long as the constituents of the Universe is dominated by radiation. Only
after matter takes over as the dynamically dominant component following the epoch of radiation matter equality,
the dark matter perturbations begin to grow. These processes give their
characteristic shape to the \textit {cold dark matter power spectrum}. This information is encoded in the
transfer function
\be
P(k,  z) = A(z)k^nT^2(k,  z).
\ee
where $A(z)$ is a normalization constant determined observationally and $T$ is the transfer
function. We follow the expression for $T_{cdm~}$ given by \citet{1986ApJ...304...15B}:
\be
T_{cdm~} = \frac{\ln (1 + 2.34q)}{2.34q}\left(1 + 3.89q + \left(16.1q\right)^2 + \left(5.46q\right)^3 + \left(6.71q\right)^4\right)^{-1/4},
\ee
where $q \equiv \frac{k}{\Gamma} \, h Mpc^{-1}$ and
\be
\Gamma = \Omega_0 h \exp \left(-\Omega_b\left(1 +\sqrt{2h}/\Omega_0 \right)\right)
\ee
is the shape parameter given by \citet{1995ApJS..100..281S}.

The power spectrum at small scales goes as $k^{-3}$ indicating that asymptotically it is a
hierarchical scenario. On the large scales it remains as the Harrison-Zel'dovich spectrum
set in the inflationary epoch. The horizon scale at the time when matter and radiation densities
were equal is reflected in the power spectrum as the turnover point. This marks the
point when matter overcame radiation in the dominance of the dynamics of the Universe.
\section{\bf The nonlinear Regime}\label{nlin}
The linear regime provides a useful description for the early phases of evolution of the
Universe and it ceases to be valid as the density contrast approaches unity. Since the full
nonlinear solutions are in general too complex to solve analytically,  one must rely on other
alternatives such as solutions for simple configurations and numerical methods. N-body
computer simulations are the most common tool to study the formation and evolution of
structures in the nonlinear regime. They follow the trajectory of particles
sampling the underlying density field.
While the primordial linear density field can be well described as a Gaussian random
field,  in the non linear regime non-gaussianities creep in, making the understanding of the the evolution of density field a lot more complicated. The distribution of
matter in the Universe
at the present time has three important properties that are the result of the processes that
gave it shape:
\begin{enumerate}
\item
Hierarchical Clustering
\item
Anisotropic Collapse into web like structures
\item
Appearance of Voids in the Distribution
\end{enumerate}
\subsection{Hierarchical Clustering}
The fluctuations in a Gaussian random field are fully described by their
power spectrum. It is assumed to have a power-law behaviour $P(k) \propto k^n$
where the relative amplitude between scales is dictated by the index $n$. In order to understand
the role of the index $n$ in the growth of structures it is useful to study a few
simple cases. A density field with power spectrum with index $n = 0$ has same power
at all scales. For such a case however,  the power over a particular scale when it enters the Hubble radius is not the same as that for another scale
when that enters the Hubble radius. Hence for $n=0$,  although the power is same over all scales at a particular time,  it will not be the same for
different scales at the time when the corresponding scales enter the Hubble radius. Hence $n=0$ does not correspond to a scale invariant power
spectrum. It turns out that for scale invariant power spectrum $n=1$ when $P(k)$ is measured for all scales at the same time. For $n=0$,  small-scale
fluctuations will collapse and virialize well before larger scales. Small clumps of matter will aggregate to form larger systems. An index
$n = -2$ will produce an intermediate case where large scale fluctuations will start their
collapse while the small-scales will not yet have fully collapsed. The asymptotic case
where $n = -3$ represents an extreme scenario in which all scales will undergo collapse
at the same time. Hence we can see that only spectra with $n > -3$
leads to a bottom-up structure formation in which small clumps collapse and aggregate
into larger associations. This process of building-up large structures from the merging of
smaller structures is called hierarchical structure formation.
\subsubsection{The Press-Schechter formalism}
\citet{1974ApJ...187..425P} proposed a formalism to compute the average number of objects
that collapsed from the primordial Gaussian density field. They assumed that the dense
objects seen at the present time are a direct result of the peaks in the initial density field.
These small perturbations collapsed spherically under the action of gravity to form selfbound
virilized objects.

In the primordial Gaussian field the probability,  that a given point lies in a region with the density contrast $\delta$  greater than the critical density for collapse $\delta_c$,  is given by
\be
p(\delta > \delta_c|R_f) =\frac{1}{2}\lbrack 1 - erf\left(\frac{\delta_c}{\sqrt{2}\sigma(R_f)}\right) \rbrack
\ee
where $\sigma(R_f )$ is the variance of the density field smoothed on the scale $R_f$. The Press-
Schechter formalism assumes that this probability corresponds to the probability that a
given point has ever been part of a collapsed object of scale $> R_f$. Then,  the comoving
number density of halos of mass $M$ at redshift $z$ is given by
\be
\frac{dn}{dM}(M, z) =\sqrt{\frac{2}{\pi}}\frac{\bar\rho}{M^2}\frac{\delta_c(z)}{\sigma_m}\mid\frac{d \ln \sigma(M)}{d \ln
M}\mid\exp\left(-\frac{{\delta_c(z)}^2}{2\sigma^2(M)}\right)
\label{eq2}
\ee
where $\sigma(M)$ is the variance corresponding to a radius $R_f$ containing
a mass $M$ and $\delta_c(z) = \delta_{c}^{0}/D(z)$ is the critical overdensity linearly extrapolated to the
present time. Here $\delta_{c}^{0} = \delta_c(z=0)$. For an Einstein-de Sitter universe the critical overdensity is $\delta_{c}^{0} = 1.69$.
There are approximations for other models,  in general $\delta_{c}^{0}$ has a weak dependence on $\Omega_m$
\citep{1997ApJ...490..493N}.

One of the limitations of the Press-Schechter formalism is that it assumes  overdense
perturbations to be perfectly spherically symmetric. In reality the situation is more
complex. \citet{1986ApJ...304...15B} extensively studied the statistics of peaks in a random
density field. They showed that peaks in the primordial density field have a degree of
flattening. This departure from a spherical distribution is amplified under the action of
gravity affecting the final collapse of the object.

The original Press-Schechter formalism
also does not properly take into account the cloud-in-cloud problem as it ignores
underdense regions. This is the origin of the contrived factor of 2 in equation \ref{eq2}. An
appropriate description in terms of the excursion set barrier crossing led to the
formulation of the \textit{extended Press-Schechter formalism} by \citet{1991ApJ...379..440B}. Not only
did it provide a powerful enough framework to describe the merging of clumps of matter into even
larger objects \citep{1993MNRAS.262..627L},  but it also allowed a more proper understanding and
description of the mass function of galaxies and haloes given their non spherical shape
\citep{2001MNRAS.323....1S, {2004MNRAS.350.1385S}}. Recently \citet{2004MNRAS.350..517S}
and \citet{2006ApJ...645..783S} provided a viable formalism to describe the hierarchical evolution
of voids and elongated filamentary superclusters.

\subsection{Anisotropic collapse}
The distribution of matter in the Universe is not homogeneous over all scales as is clear from galaxy redshift surveys. The Universe has a variety of structures. The nature of these structures like filaments etc. suggest the gravitation collapse to be anisotropic. Early studies focused  on the anisotropic nature of the gravitational collapse may be found in \citet{1964ApJ...139.1195L} and \citet{1965ApJ...142.1431L}. \citet{1973A&A....27....1I} investigated the evolution of homogeneous ellipsoidal configurations in an
expanding $FRW$ universe and concluded that the predominant final morphologies are flattened
and elongated. One of the most important results of the ellipsoidal collapse model
is that not only gravity sets any overdense perturbation into a runaway collapse but it also has
an amplifying effect on any asphericity present in the initial matter configuration
\citep{1973A&A....27....1I, 1979ApJ...231....1W, 1995ApJ...439..520E, 1996ApJS..103....1B}.

While nearly all these studies address very specific configurations,  the Zel'dovich
formalism clarifies the importance of the anisotropic nature of gravitational collapse for
more generic cosmological circumstances \citep{1970A&A.....5...84Z}. While it formally concerns
a linear Lagrangian formalism it has proven to describe the emergence and development
of structure to weakly nonlinear stages. Not only it
elucidates the first stages of nonlinear clustering but it also has become an essential tool
for setting up the initial conditions used as input for $N$-body computer simulations.
The Zel'dovich formalism is based on the mapping between the initial Lagrangian
position q to a displaced Eulerian position {\bf {x}}. In the weakly non linear regime these two positions are related by
\be
x(t) = q + D(t)\nabla\Phi(q),
\label{eq3}
\ee
where the time dependent function $D(t)$ is the growth rate of linear density perturbations
and the time independent spatial function $\Phi(q)$ is related to the linearly extrapolated
gravitational potential.

Here we concentrate on the anisotropic collapse of a patch of matter. For a particular
structure the force field of the structure hangs together with the flattening of the feature
itself. This induces an anisotropic collapse along the main axes of the structure. Applying
a simple mass conservation relation $\bar\rho d^3q = \rho(x)d^3x$ to equation \ref{eq3}, we get:
\be
\rho(x) = \frac{\bar\rho}{\left[1 - D_+(t)\lambda_1(q)\right] \left[1 - D_+(t)\lambda_2(q)\right] \left[1 -D_+(t)\lambda_3(q)\right]}
\ee
where $\lambda_1,  \lambda_2,  \lambda_3$ are the eigenvalues of the deformation tensor:
\be
\psi_{i, j}=\frac{\partial^2 \Psi}{\partial q_i \partial q_j}
\ee
In order for an object to collapse at least one of the eigenvalues must be positive,  so that
the density $\rho(x)$ diverges as $D_+$ increases.
The Zel'dovich approximation predicts the collapse of matter into planar sheets or
pancakes. The subsequent collapse is determined by the second largest eigenvalue which
produces a filament to finally end up in a spherical clump. This suggest a natural division
of the features of the large scale matter distribution based on their morphology. On the
basis of the eigenvalues we may distinguish three \textit{final configurations}. If
$\lambda_1 > 0$ and $\lambda_2$ and $\lambda_3$ are both less than $0$,  the resulting
configuration is that of a \textit{pancake}. For a \textit{filament} configuration $\lambda_1$ and $\lambda_2$
are positive but $\lambda_3$ is negative. A \textit{clump} configuration is defined by all $\lambda$'s being positive.

Each morphology represents a specific evolutionary state in the gravitational collapse.
In reality the gravitational collapse is not a sequence of single collapses along $\lambda_1,  \lambda_2$ and
$\lambda_3$. Instead it is a more gradual collapse in all three directions. One can then expect the
Universe to contain the three basic morphologies as well as a large number of intermediate
cases. The most conspicuous feature of
the large scale matter distribution is the existence of a pervading filamentary network and
quasi-spherical dense concentrations of matter sitting at the nodes. The planar walls or
pancakes can also be seen as slightly overdense regions located between filaments.
Most of the space is devoid of matter. Large empty regions extend for several Megaparsecs.
These \textit{voids} give the Cosmic Web its characteristic cellular or foamy nature \citep{2002ASSL..276..119V}.
\subsection{The Cosmic Web}

\citet{1996Natur.380..603B} took the analytical description of the hierarchical large-scale
matter distribution to a meaningful description of the nonlocal influences on evolving
matter structures. They coined the word {\it cosmic web} in their study of the physical component of structures of Universe.
Their peak-patch formalism presented a more complete description involving
tidal influences. It provided a basic framework for the Cosmic Web model
for more generic cosmological circumstances of a random density field \citep{1996ApJS..103....1B, 1996ApJS..103...41B, 1996ApJS..103...63B}. The salient feature of finding of \citet{1996Natur.380..603B} was that knowledge of the value of the tidal
field at a few well-chosen locations in some region is sufficient to determine the overall
outline of the web-like pattern in that region.

In the \textit {Cosmic Web Theory} the rare high
peaks corresponding to clusters play a fundamental role. They are the nodes that define
the cosmic web. This relation may be traced back to a simple configuration,  that of a
global quadrupolar matter distribution and the resulting local tidal shear at its central
site. Such a quadrupolar primordial matter distribution will almost by default evolve into
a canonical cluster-filament-cluster configuration which forms the structural basis of
the Cosmic Web.

The Cosmic Web Theory provides a natural explanation to both the elements
that form the Cosmic Web as well as their connectivity properties. This intimate connection
between the \textit{local} force field and the surrounding global matter distribution can be
straightforwardly appreciated on the basis of the {\it constrained random field} study by \citet{1996MNRAS.281...84V}. They,  amongst others,  discussed the
repercussion of a specified constraint on the value of the tidal shear at some specific location. This expression at a particular position is represented by the following expression.
\be
T_{ij}(r,  t) =\frac{3\Omega H^2}{8\pi}\int d^3r' \delta(r', t)\left\{\frac{3(r'_i-r_i)(r'_j-r_j)-{|r'-r|}^2\delta_{ij}}{{|r'-r|}^5}\right\}-\frac{1}
{2}\Omega H^2 \delta(r, t)\delta_{ij}
\label{tt}
\ee
From the expression \ref{tt} of the tidal tensor in terms of the generating density distribution,
we can immediately observe that any \textit{local} value of $T_{ij}$ has \textit{global} repercussions for the
generating density field. Such \textit{global} constraints are in marked contrast to \textit{local} constraints
like the value of the density contrast $\delta$  or the shape of the local matter distribution. One of the major advantages of their {\it constrained random field} construction technique \citep{1987ApJ...323L.103B, 1991ApJ...380L...5H} is that it offers tools for translating locally specified quantities into the corresponding implied global matter distribution for a given structure formation scenario.
\section{\bf Ideas about Galaxy Distribution}\label{gdis}
The \textit{Great Galaxy} view \citep{1922ApJ....55...65A} of the distribution of galaxies depicted the Milky Way as a relatively small
flattened ellipsoidal system. In this model the Sun is supposed to have been located at the center of milky way. The center was supposed to be surrounded by a halo of globular clusters. However,  recognizing the role played by inter stellar absorption,
and also the fact that stars in the Galaxy were orbiting about a distant center,  the sun was placed elsewhere instead of the center.

Another view of Galaxy distribution later confirmed by Edwin P. Hubble \citep{1925PA.....33..252H, {1925ApJ....62..409H}},  was
that there are `field galaxies' largely separated from one another. This view gave rise to the hypothesis of Island
Universe. This hypothesis stated that galaxies are building blocks of the Universe.
In fact,  most galaxies are clustered. The objects which were called nebulae,  at that time,  were in fact
extragalactic system of stars comparable with our own galaxy. The first systematic surveys of the galaxy
distribution were undertaken by Shapley and his collaborators \citep{1938AnHar.106...75S}. It led to the
discovery of numerous galaxy clusters and even groups of galaxy clusters. The clustering together of stars,  galaxies,  and clusters of galaxies in
successively ordered assemblies is normally called a hierarchical or multilevel
clustering \citep{1908ArMAF...4q...1N,  1922ArMAF...16q...1N,  1970Sci...167.1203D}. It has three main consequences. It removes the Olber's paradox
\citep[see e.g.][]{1908ArMAF...4q...1N,  {1922ArMAF...16q...1N}}. The universe retains a primary center and is therefore nonuniform on the largest
cosmic scales.  The total amount of matter is much less than in a uniform universe with the same local density. Hierarchical model of clustering
also assumes that the visible universe is only one of the series of universes nested inside each other.

More recently still there have been a number of attempts to re-incarnate such a universal hierarchy in terms
of fractal models. These models were first proposed by \citet{1907TNW...} and subsequently studied by
\citet{1982fgn..book.....M} and \citet{1987PhyA..144..257P}. Several attempts have been made to construct hierarchical cosmological models. All
these models are,  naturally,  inhomogeneous. These models have preferred position for the observer,  and thus these are unsatisfactory. So the present trend to reconcile fractal models with cosmology is to use the measure of last resort,  and to assume that although the matter distribution in the universe is homogeneous on large scales,  the galaxy distribution can be contrived to be fractal \citep{2001GReGr..33.1699R}. Numerical models of deep samples as well as data from modern redshift surveys contradict this assumption.
\subsection{The cosmological Principle}
The notion that the Earth is not at the center of the Universe is generally referred to
as the Copernican Principle. \citet{1917SPAW.......142E} proposed that on the very largest scales the Universe
should be homogeneous and isotropic.
At that time there could have been no observational support for this assumption. It is a consequence
of the notion that we don't have a special
place in the Universe. Under this assumption Einstein's field equations have a simple solution. Einstein-de
Sitter model of cosmology as well as the famous solution of Einstein's Equation provided by
Robertson and Walker use just this principle.

The first demonstration of homogeneity in the galaxy distribution
was probably the observation by Peebles that the (projected) two-point correlation function estimated from
diverse catalogs probing the galaxy distribution to different depths followed a scaling law that was
consistent with homogeneity. The observations of Cosmic Microwave Background Radiation give evidence of the cosmic isotropy of the Universe.
The $COBE$ satellite all-sky map of the Cosmic Microwave Background Radiation \citep{1992ApJ...396L...1S} is isotropic
to a high degree,  with relative intensity fluctuations only at the level of $10^{-5}$.
\section{\bf Surveys of Cosmic Structures}\label{gsur}
 The first map of the sky came from the Lick survey
of galaxies undertaken by \citet{1967PLO..22.1} using large field plates from the Lick Observatory.
This map revealed widespread clustering and super clustering of galaxies. With each improvement in telescope and associated back end instruments,
we have been able to  probe further into the Universe. One of the  key impetus in understanding
the clustering of galaxies was provided by Palomar Sky survey. Observations were done using a $48''$ Schmidt
telescope.  A catalog of galaxy redshifts,  with information about the clusters to which the galaxies belonged,  was published by \citet{1956AJ.....61...97H}.

These catalogs simply  listed objects as they appeared projected on to the celestial sphere. Only indication
of distance to the object came from its brightness or size. Moreover,  these were subject to human selection effects and hence were not  sufficiently standardized.

What characterizes more recent surveys is the ability to scan photographic plate digitally (e.g: The Cambridge
Automatic Plate Machine APM),  or to create the survey in digital format (e.g: IRAS,  Sloan Survey etc). It is now
far easier to obtain redshifts for large number of objects in these catalogs.  Mapping the Universe this way
provides information about how structured the universe is now at modest redshift. These structures were generated from initial density perturbations in the early Universe. The perturbations led to the anisotropies in the Cosmic Microwave
Background Radiation at the surface of last scattering. The collapse of these tiny fluctuations has given rise to the structures that we observe in the present Universe. Thus observations of Cosmic Microwave Background give us information about the structure of the surface of last scatter.  This information can,  in turn,   serve as the starting point for $N$ body simulations. If we can put these two things (large scale structures and CMBR) together we will have a complete picture of the Universe.

Now we would like to describe briefly some of the recent galaxy redshift surveys that have completed
or are under progress.

\subsection{Cfa and SSRS survey}

The first Cfa survey \citep[][http://tdc-www.harvard.edu/]{1983ApJS...52...89H}  mapped about $2400$
galaxies down to apparent magnitude $m \simeq 14.5$ taken from Zwicky catalog. This survey
was too sparse to show definite structures. The Cfa slice was centered on the Coma cluster,  hence it
was not considered as being representative of the universe as a whole. However,  the breadth of the
slice sampled a far greater volume,  and it was very deep for that time $(\sim 150h^{-1}Mpc)$. Subsequent
surveys like  the following CfA slices and the ESO Southern survey \citep{1991ApJS...75..935D} amply confirmed the
impression given by the CfA slice.

The Southern Sky Redshift Survey  \citep[][http://vizier.u-strasbg.fr/]{1991ApJS...75..935D} was proposed to
complement the original CfA survey. It mapped galaxies in the
southern sky taking redshift of about $2400$ galaxies. The extended SSRS \citep{1998AJ....116....1D}
followed it up with redshifts of about $5400$ galaxies mirroring the Second CfA
survey for the southern sky.

\subsection{The Las Campanas Redshift Survey}
The Las Campanas Redshift Survey \citep{1996ApJ...470..172S} mapped six thin
parallel slices ($1.5^\circ \times 90^\circ$).
It probed the Universe to a depth of about $750 \,h^{-1}$Mpc ($z\approx 0.25$).
It measured redshifts of about 24000 galaxies in these slices.
This was the first deep survey of sufficient volume of the nearby Universe.
The LCRS data can be accessed at http://qold.astro.utoronto.ca/~lin/lcrs.html

\subsection{2dF galaxy redshift survey}
The $2dF$ \citep{2003astro.ph..6581C} used a multi-fiber spectrograph on the $3.9$m Anglo-Australian
Telescope. This survey had a field of view of some $2$ degrees in diameter,  hence the name of
the survey.  The redshifts measurement was carried out on  some $250, 000$ galaxies located in
extended regions around the north and south Galactic poles. The
source catalog is a revised APM survey. The galaxies in the
survey go down to  magnitude $b_J = 19.45$.  The median
redshift of the sample is $z = 0.11$ and redshifts extend to about
$z \simeq 0.3$. The survey is already complete,  and the data can be
downloaded from http://www2.aao.gov.au/2dFGRS/
\subsection{Sloan Digital Sky Survey}
 The Sloan Digital Sky Survey (SDSS) \citep{2000AJ....120.1579Y,  2002AJ....123..485S} is the
largest galaxy redshift survey to date. It employs a specially designed $2.5$ m telescope
with a $3^\circ$ field of view. It uses  a mosaic CCD camera,  and dual fiber-fed spectrograph,  to
obtain five band (u, g, r, i, z) digital photometry.
The spectroscopic information is obtained over full range of optical wavelengths. The main
spectroscopic galaxy sample of the SDSS \citep{2002AJ....124.1810S} includes objects having
Petrosian magnitude of $r < 17.77$ after correction for Galactic extinction. It
is designed to measure a million galaxy redshifts
over $\sim 10^4$ square degrees of sky. The sixth major public release of SDSS data
\citep[SDSS DR6;][www.sdss.org/dr6 and www.cas.sdss.org/dr6/en]{2008ApJS..175..297A}
in June,  $2007$ includes $8520$ square degrees imaging and $6860$ square degrees of spectroscopy. As of now
the spectroscopic data includes $1163520$ spectra with $792680 $ galaxy redshift. The
survey area covers a single contiguous region in the Northern Galactic Cap and three
non-contiguous region in the Southern Galactic Cap. The SDSS surveys to a depth
that has been probed previously by earlier surveys like LCRS,  however the volume
covered by SDSS is enormously greater. The solid angle coverage of the SDSS is almost
$14$ times that of the LCRS. For a detailed discussion about SDSS we refer the reader to section \ref{Data} of this thesis.
\section{\bf Goals and outline of this thesis}\label{got}
The main goal of this thesis is to understand the nature of clustering of matter over
large scales in the Universe.
There are various methods for the statistical characterization of large scale structures. The traditional approaches
include the {\it two point correlation function,  counts in cells,  nearest neighbor approximation and N point correlation function}.
The approach we use in this thesis for this purpose is the {\bf multifractal analysis} of simulated distribution of points as well as of galaxy
distributions from galaxy redshift surveys. Multifractal analysis is a useful tool in this case because the large scale distribution of matter has a
scaling behaviour over a range of scales.  Galaxy distributions also exhibit self similarity  on small scales
\citep{1987PhyA..144..257P, 1988ApJ...332L...1J}.

Fractals have been invoked to describe many physical phenomena which exhibit self-similarity
\citep{1982fgn..book.....M}. A multi-fractal is an extension of the concept of a fractal. It includes the possibility that the self similar behaviour
of particle distributions may be different in different density environments. In order to give a complete statistical information about the
point distribution the multifractal analysis characterizes scaling properties of moments at all levels. One of the advantages of using this technique
over the traditional approaches is that it does not require {\it apriori} information about the average density of the Universe. This enables us to
use this approach
in finding the scale at which the matter distribution in the Universe attains homogeneity. It means we are interested in finding
the scale above which the cosmological principle can be assumed to be valid and the Friedman-Robertson-Walker-Lemaitre (FLRW) metric is a correct description of the Universe.


Chapter 2 describes various statistical methods used in the analysis of  distribution of galaxies. We start with the standard tool of two point
correlation function of the distribution and discuss its merits and demerits.  N-point correlation functions and the counts in cells statistics of
the number of particles in the distribution are discussed in order to calculate higher moments of the distribution. Fractal analysis as an
alternative to the two point correlation function has been extensively described. Different definitions of fractal dimension (e.g Minkowski -
Bouligand Dimension) have been discussed which are useful for deterministic as well as statistical fractal distributions.

Chapter 3 deals with the calculation of Minkowski- Bouligand fractal dimension ($D_q$) for both Homogeneous and weakly clustered distribution of
points. We have described the relation between $D_q$ and the probability distribution function of a distribution. We have investigated how the
computed dimension changes with the number of particles in the distribution. The  fractal dimension has also been calculated  for a general
mathematical distribution in which the particles are weakly clustered.  We also describe the individual contribution of finite number and clustering
to the Minkowski Bouligand Dimension and derive an analytical expression to quantify the deviation of $D_q$ from Euclidean dimension  due to these two contributions.

In chapter 4 the application of our model of calculating Minkowski Bouligand Dimension developed in chapter $3$ has been discussed. For this purpose
various distribution of points have been considered. To test the correctness of our model the application to {\it Multinomial Multifractal}
distribution has been studied.  We have also discussed the application of  our model to the concordance model of cosmology. We describe the scale at
which the unbiased distribution of $L_*$ type galaxies is homogenously distributed. Application of our model to biased distribution of Large
Redshift Galaxies (LRG) has also been discussed. Contribution of clustering term to the Minkowski-Bouligand Dimension has been discussed for the
distribution of points having a feature (like the Baryon Acoustic Oscillation) in the correlation function.

Chapter 5 tests the large scale homogeneity of the galaxy distribution in the Sloan Digital Sky Survey Data Release One (SDSS-DR1) using  volume
limited subsamples extracted from two equatorial strips. The two dimensional multifractal analysis of the  galaxy distribution projected  on the
equatorial plane has been studied. The galaxy distribution has also been compared with the distribution generated from random catalog and also from
N-Body simulations. The effect of bias to the scale of homogeneity of the galaxy distribution has also been discussed in this chapter.

Chapter 6 gives a summary of the thesis along with the future scope of our work.

\chapter{Statistical tools to analyze Distribution of Galaxy}\label{chap2}
\section{\bf Introduction}\label{Introduction}
One of the goals of modern cosmology is to understand and quantify the nature
of large scale matter distribution of the Universe. An accurate empirical description of large scale clustering of matter, derived from systematic
observations of visible matter in the Universe, is   essential to achieve this goal. Efforts in this directions have vastly improved due to better
instruments that have been available in recent times for astronomical data acquisition as well as better statistical techniques of
analysis of the acquired data. It is as a result of these efforts  that an enormous amount of data about the observable universe has been
accumulated in the form of the now well-known {\it redshift surveys}, and some widely accepted conclusions  drawn from these data have created a
certain confidence in many researchers that an accurate description of the large scale matter distribution is just about being achieved.

In statistical analysis of galaxy distribution, we are not interested in the number of
galaxies in a particular region of the sky but we are rather interested only in the average properties of  number distribution of galaxies. We
are {\it e.g.} interested in knowing whether or not distribution of galaxies is clumpy, and if so, how we can quantify the nature of clumpiness.
In the literature \citep{2002sgd..book.....M}, various statistical methods of analysis of galaxy clustering have been discussed. The broad feature
of all these methods is
to discuss the nature of clumpiness of the galaxy distribution. Among all these methods the historical favorites have been variants of two
point correlation function (see equation \ref{eq21}). This function measures the excess probability, relative to a Poisson distribution, of finding
an object near another object. Bok's statistic (the dispersion of the counts $N$ in cells), is an integral over two point correlation function.
Zwicky's index of clumpiness is the ratio of variance of $N$ to what would be expected for a uniform random distribution. From the two point
correlation function of the counts of galaxies in the Lick survey, Limber showed that there is a linear integral equation relating the angular
correlation function to the corresponding spatial correlation function. Neyman and Scott devised a {\it priori} statistical model of clustering
and then adjusted the parameters to fit model statistics to estimates from data. A recent program in a similar vein is called the halo model.
Recently there have been precise estimates of two point correlation function from redshift surveys like Two Degree Field Galaxy Redshift Survey
(2dFGRS) and the Sloan Digital Sky Survey (SDSS).

The Fourier or spherical harmonic transform of the two point correlation function is the power spectrum (see equation \ref{eq221}). It is the
description of clustering in terms of wavenumbers $k$ that separates the effects of different scales. Other descriptors of statistics have been the
$N$ point correlation function of the distribution of points, moments and counts in cells, void probability function and nearest neighbor  distances
etc. However all these methods are based on the idea of eventual homogenization of the matter distribution within the sample size itself. A group of researcher feels that
this idea of homogenization is flawed and the distribution of matter in the Universe is intrinsically inhomogeneous to largest observed scales and,
perhaps, indefinitively.

The debate of homogeneous versus non homogeneous distribution of matter has
taken a new vigor with the arrival of a new method for describing the clustering of galaxies.
This method is based on the ideas of a new geometrical perspective for the description
of irregular patterns in nature. We generally refer to it as \textit{the fractal geometry}. In this chapter we intend to show the basic idea behind
this geometrical approach.

The plan of this chapter is as follows. In section \ref{Standard Analysis} we  briefly present the basic tools used in statistical analysis of
the large scale distribution of galaxies, its estimations, difficulties and answers given to these difficulties. The subsection \ref{Higher
Order} describes Higher order statistics of the distribution. Section \ref{other statistical} describes various other methods of statistical
analysis. Section \ref{Fractals}  presents a brief, but general, introduction to fractals, which emphasizes their empirical side and
applications. The discussion on various fractal dimensions along with their merits (and demerits) follow in section \ref{Other Frac Dim} and
\ref{Prob}. We conclude the chapter with a small discussion on Lacunarity of the point distribution in section \ref{lacunarity}.
\section{\bf The Standard Correlation Function Analysis}\label{Standard Analysis}
The standard statistical analysis assumes that the objects under
discussion (galaxies) can be regarded as point particles. These particles are
assumed to be distributed homogeneously on a sufficiently large scale within the sample boundaries. This means that
we can meaningfully assign an average number density to the
distribution. Therefore, we can characterize the galaxy distribution
in terms of the extent of the departures from uniformity on various
scales. The correlation function as introduced by \citet{1980lssu.book.....P} is basically the statistical tool that permits the
quantitative study of this departure from homogeneity.

Consider a set of  $N$ galaxies contained in a volume $V$. The average number density of galaxies is defined by
$\bar{n} = N/V$. It implies that we have to go on an average a distance of ${ ( \bar{n} )}^{-1/3}$ from a given galaxy before another is
encountered. This means that local departures from uniformity can be
described if we specify the distance we actually go from any particular
galaxy before encountering another. This will sometimes be larger than
average, but sometimes less. Specifying this distance in each case is
equivalent to giving the locations of all galaxies. This is an awkward
way of doing things and does not solve the problem. What we require is
a statistical description giving the probability of finding the nearest
neighbor galaxy within a certain distance.

As we know the probability of finding a galaxy closer than, say, 50 kpc
to the Milky Way is zero, and at a distance greater than this value
is one. This sort of probability information is not useful to us.
What is necessary is some sort of average. We can view the
actual universe to be a particular realization of some statistical
distribution of galaxies. The departure from
randomness due to clustering of these galaxies is expressed by the fact that the
average separation of galaxies over the statistical ensemble of this separation is less than
${ ( \bar{n} ) }^{-1/3}$.

For a completely random and homogeneous distribution of galaxies, the
probability $dP_1$ of finding a galaxy in an infinitesimal volume
$dV_1$ is proportional to $dV_1$ and to $\bar{n}$, and is independent of
position. So we have
\[
  dP_1 = \frac{ \bar{n} }{N} dV_1, \nonumber
\]
where $N$ is the total number of galaxies in the sample. The sample space is
divided into cells of volumes $dV_1$ and we count the ratio of those cells which
contain a galaxy to the total number. The probability of finding two
galaxies in a cell is of order ${(dV_1)}^2$, and so can be ignored in
the limit $dV_1 \to 0$. It is important to state once more that this
procedure only makes sense if the galaxies are distributed randomly
on some scale less than that of the sample.

Similarly  the joint 
probability $dP_{12}$ of finding galaxies in volumes $dV_1$ and $dV_2$ at positions $\vec{r}_1$, $\vec{r}_2$ respectively is just the product of
probabilities of finding each of the
galaxies, i.e.
\[
  dP_{12} \propto dP_1dP_2, \nonumber
\]
This is because in a random distribution the positions of galaxies are
uncorrelated. On the other hand, if the galaxies are correlated we would have a
departure from the random distribution. In that case the joint
probability is different from a simple product. The {\it two-point
correlation function} $\xi (\vec{r}_1, \vec{r}_2)$ is by definition a
function which determines this difference from a random distribution. So
we have
\be
 dP_{12} = { \left( \frac{ \bar{n} }{N} \right) }^2
           \left[ 1+ \xi (\vec{r}_1, \vec{r}_2) \right] dV_1 dV_2
 \label{eq21}
\ee
as the probability of finding a pair of galaxies in volumes $dV_1$,
$dV_2$ at positions $\vec{r}_1$, $\vec{r}_2$. Obviously, the assumption
of randomness on sufficiently large scales means that $\xi (\vec{r}_1,
\vec{r}_2)$ must tend to zero if $\left| \vec{r}_1 - \vec{r}_2 \right|$
is sufficiently large. In addition, the assumption of homogeneity and isotropy
implies that $\xi$ cannot depend on the location of the galaxy pair, but only
on the distance $\left| \vec{r}_1 - \vec{r}_2 \right|$ that separates
them, as the probability must be independent of the location of the
first galaxy. If $\xi$ is positive we have an excess probability over
a random distribution and, therefore, clustering. If $\xi$ is negative
we have anti-clustering. Obviously $\xi > -1$.

The two-point
correlation function can be generalized to define $n$-point correlation
functions, which are functions of $n-1$ relative distances, but in
practice computations have not been carried out beyond the four-point
correlation function.

It is a common practice to replace the description above using point
particles by a continuum description. So if galaxies are thought to be
the constituent parts of a fluid with variable density $n(\vec{r})$, and
if the averaging over a volume $V$ is carried out over scales large
compared to the scale of clustering, we have
\be
 \frac{1}{V} \int_V n(\vec{r}) dV = \bar{n},
 \label{eq212A}
\ee
where $dV$ is an element of volume at $\vec{r}$. The joint probability
of finding a galaxy in $dV_1$ at $\vec{r} + \vec{r}_1$ and in $dV_2$ at
$\vec{r} + \vec{r}_2$ is given by
\[
  { \left( \frac{1}{N} \right) }^2 n(\vec{r} + \vec{r}_1)
  n(\vec{r} + \vec{r}_2) dV_1 dV_2.
\]
Averaging this equation over the sample gives
\be
  dP_{12} =  \frac{1}{N^2 V} \int_V  n(\vec{r} + \vec{r}_1) n(\vec{r} +
  \vec{r}_2) dV dV_1 dV_2.
\label{p12}
\ee
Now if we compare the equation \ref{p12} with equation (\ref{eq21})
we obtain
\be
   { \bar{n}}^2 \left[ 1+ \xi(\vec{\tau}) \right] = \frac{1}{V} \int_V
   n(\vec{R}) n(\vec{R}+\vec{\tau}) dV,
 \label{eq22}
\ee
where $\vec{\tau} = \vec{r}_2 - \vec{r}_1$, $\vec{R} = \vec{r}
+ \vec{r}_1$ and $dV$ is the volume element at $\vec{R}$.

It is worth mentioning that in statistical mechanics the correlation function normally used is $g(r)= 1+ \xi(r)$
which is called the radial distribution function. Statisticians call this quantity the pair correlation function. The number of galaxies, on
average, lying between $r$ and $r+dr$ is $4 \pi r^{2}ng(r)$ with $n$ being the average number density.

Related to the correlation function is the so-called {\it power
spectrum} of the distribution, defined by the Fourier transform of
the correlation function.
\be
   \xi(r)=\frac{1}{2\pi^2}\int dk~k^2~P(k)\frac{sin(kr)}{kr}
 \label{eq221}
\ee
The scale or wavelength $\lambda$ of a fluctuation is related to
the wavenumber $k$ by $k=\frac{2\pi}{\lambda}$. As explained in chapter \ref{chap1} the power spectrum describes the way that large, intermediate
and small structures combine to produce the observed distribution of luminous matter.
It is also possible to define an {\it angular correlation function} which will express the probability of
finding a pair of galaxies separated by a certain angle, and this is
the  appropriate function to studying catalogs of galaxies which
contain only information on the positions of galaxies on the
celestial sphere. It means that the angular correlation function is used to study the projected galactic
distribution when the galaxy distances (i.e. the redshift information) are not available. Further
details about these two functions can be found at various places in the literature \citep{1980lssu.book.....P,2002sgd..book.....M}.
Finally, for the sake of easy comparison with
other works it is useful to write equation (\ref{eq22}) in a
slightly different notation:
\be
 \xi(r) = \frac{ \langle n(\vec{r}_0) n(\vec{r}_0+\vec{r}) \rangle }
               { { \langle n \rangle }^2} - 1.
 \label{eq203}
\ee

The usual interpretation of the correlation function obtained from the
data is as follows: when $\xi~\gg~1$ the system is strongly correlated
and for the region when $\xi~\ll~1$ the system has small correlation.
From direct calculations from catalogs it was found that at small
values of $r$ the function $\xi(r)$ can be characterized by a power
law \citep{1987PhyA..144..257P,1988ApJ...333L...9D}:
\be
 \xi(r) = \left(\frac{r}{r_0}\right)^{- \gamma}~ \approx A r^{- \gamma}, \ \ \ \ (\gamma \approx 1.7),
 \label{eq215}
\ee
where $A$ is a constant. This power law behavior holds for galaxies and
clusters of galaxies. The distance $r_0$ at which $\xi~=~1$ is called
the {\it correlation length}, and this implies that the system becomes
essentially homogeneous for lengths appreciably larger than this
characteristic length. This also implies that there should be no
appreciable overdensities (superclusters) or underdensities (voids)
extending over distances appreciably larger than $r_0$.

In the calculation of two point correlation function we
have assumed an average value for the density of matter on the scales well within the sample size. In
practice however, we do not have a statistical ensemble from which the average
value can be derived. So what we can do is to take a
spatial average over the visible universe, or as much of it as has been cataloged,
in place of an ensemble average. This only makes sense
if the departure from homogeneity occurs on a scale smaller than the depth of the sample,
so that the sample will statistically reflect
the properties of the universe as a whole. In other words, we need to
have a {\it fair sample} of the Universe in order to fulfill this
program. This fair sample ought to be homogeneous, by
assumption. If, for some reason, the sample we have is not a fair sample in the above sense, we can not construct an average density of the Universe
with this sample. In that case this whole program breaks down.
\subsection{ Estimators of Two point Correlation Function}
The two-point correlation function $\xi(r)$ can be estimated
in several ways from a given galaxy sample. At
small distances, nearly all the estimators provide very similar
performance. However at large distances, their performance
is not equivalent any more and some of them could be biased.
Considering the galaxy distribution as a point process,
the two-point correlation function at a given distance $r$ is
estimated by counting and averaging the number of neighbors
each galaxy has within a given scale. It is clear that the
boundaries of the sample have to be taken into account, because as
no galaxies are observed beyond the boundaries, the number
of neighbors is systematically underestimated at larger distances.
If we do not make any assumption regarding the kind
of point process that we are dealing with, the only solution is
to use the so-called minus-estimators, the kind of estimators
favored by Pietronero and co-workers \citep{1998PhR...293...61S}. In this estimation the averages of the number
of neighbors at a given distance are taken omitting those galaxies lying closer than $r$ to the
sample boundary. Thus at large scales only a small fraction of the
galaxies in the sample enter in the estimation. This increases
the variance. To make full use of the surveyed galaxies, the
estimator has to incorporate an edge-correction.The most
widely used estimators in cosmology are the Davis and Peebles
estimator \citep[$\xi_{DP}$,][]{1983ApJ...267..465D}, the Hamilton estimator \citep[$\xi_{HAM}$,][]{1993ApJ...417...19H} and the
Landy-Szalay estimator \citep[$\xi_{LS}$,][]{Landy:1993yu}. Here we provide their formulae.

Consider a complete galaxy sample in a given volume
with $N$ objects. A Poisson catalog generated by  a binomial process with
$N_{rd}$ points has also to be produced within the same volume.
Then the three estimators are represented by
\be
\xi_{DP}(r)=\frac{N_{rd}}{N}\frac{DD(r)}{DR(r)}-1
\ee

\be
\xi_{HAM}(r)=\frac{DD(r).RR(r)}{[DR(r)]^2}-1
\ee

\be
\xi_{LS}(r)=1+\left({\frac{N_{rd}}{N}}\right)^2\frac{DD(r)}{DR(r)} -2\frac{N_{rd}}{N}\frac{DD(r)}{DR(r)}.
\ee
Here $DD(r)$ is the number of pairs of galaxies with separation
within the interval $r-\frac{dr}{2}$ to $r + \frac{dr}{2}$ and $RR(r)$ is the number of pairs with separation
in the same interval in the Poisson catalog. In order to calculate $DR(r)$ we use a combination of points in the galaxy sample as well as in Poisson
catalog. From a point in the poisson catalog, we find the the number of galaxies that are within $r-\frac{dr}{2}$ to $r + \frac{dr}{2}$ of the
point. It is this number that we call $DR(r)$.  At large scales the performance of the Hamilton and Landy-Szalay estimators
has been proved to be better.

Now that we have obtained the explicit form of the two-point correlation function, it is important to
emphasize at this point two essential aspects of this method. First that this analysis fits very
well in the standard Friedmannian cosmology which {\it assumes} spatial
homogeneity, but it does not take into consideration any effect due to
the curvature of the spacetime. In fact, this method overlooks this
problem altogether under the assumption that the scales under study are
relatively small. However, it is important to keep in mind that beyond some scales
curvature effects become significant. Secondly, if Universe is inhomogeneous, this
analysis is {\it inapplicable}. Moreover, since this analysis starts
by assuming the homogeneity of the distribution within the sample size, {\it it does not offer
any kind of test for the hypothesis itself}. In other words, {\it this
correlation analysis cannot disprove the homogeneous hypothesis}.
\subsection{Difficulties of the Standard Analysis}\label{Difficulties Standard
Analysis}
The first puzzling aspect found using the method just described
is the difference
in the amplitude $A$ of the observed correlation function ( see equation \ref{eq215})
when measured for galaxies and clusters of galaxies. While the exponent
$\gamma$ is approximately $1.7$ in both cases, for galaxies
$A_G~\simeq~20$ and for clusters $A_C~\simeq~360$. Its value for
superclusters of galaxies is found to be $A_{SC}~\simeq~1000-1500$.
 The correlation length
was found to be $r_0~\simeq~5$~h$^{-1}$ Mpc for galaxies and
$r_0~\simeq~25$~h$^{-1}$ Mpc for clusters.

Since $A_C~\simeq~18A_G$, clusters appear to be much more correlated than galaxies. This dicripency in the value of $A$ is puzzling because cluster
themselves are made of galaxies. Similarly superclusters will then appear to
be more correlated than clusters. From the interpretation of $\xi(r)$
described above, the galaxy distribution becomes homogeneous at the
distance $\simeq$ 10-15 h$^{-1}$ Mpc where $\xi(r)$ is found to become zero,
while clusters and superclusters are actually observed at much larger
distances.

The second problem of the standard analysis has to do with the
homogeneity assumption itself and the possibility of achieving a
fair sample, which should not be confused with a homogeneous sample as
the standard analysis usually does. A fair sample is one in which there exists
enough points from where we are able to derive some unambiguous statistical
properties of the entire distribution. Improvements in
astronomical detection techniques, in particular the new sensors and
automation, have enabled astronomers to obtain a large amount of galaxy redshift
measurements per night.  With these improvements, it is now possible to map the
distribution of galaxies in three dimensions. The picture that emerges from
these surveys appears to be far from the expected homogeneity.  We can clearly see
clusters of galaxies, voids and superclusters appearing at all scales, with no
clear homogenization of the distribution. The first `slice' of the universe shown
by \citet{1986ApJ...302L...1D} confirmed this inhomogeneity with very
clear pictures. Inhomogeneities are more perceptible in the observed slices of the Universe when we compare with a randomly
generated distribution. Of course, with the modern galaxy redshift surveys it may
be possible to demonstrate that the survey is a fair sample
of the universe by showing that the values of $\bar\rho$ derived from
sub-samples of the survey are consistent with each other, or that the value of
$\bar\rho$ computed at different scales converges to a definite value at
scales much smaller than the size of the survey. However, to verify and hence validate
the cosmological principle, it is useful to consider a statistical test which does
not presuppose the premise being tested.

Some people think that third difficulty of the standard analysis is related to
the correlation length \citep[see e.g. ][]{1986MNRAS.219..457E,1988ApJ...333L...9D,1987Ap&SS.137..101C,1988A&A...200L..32C,1997ApL&C..36...65P}. They found that the correlation
length $r_0$ {\it increases with the sample size}. However, this interpretation is also not supported by data from modern galaxy redshift survey
\citep[e.g.][]{1990ApJ...357...50M}.
\subsection{Higher Order Correlation Function}\label{Higher Order}
The two-point correlation function is not the only method to quantify clustering. It is merely the first of an infinite hierarchy of
such descriptors describing the distribution of galaxies
taken $N$ at a time. Two quite different distributions can have
the same two-point correlation function. In particular, the fact
that a point distribution generated by any random walk (e.g.,
as a $L{\acute{e}}vy$ flight as proposed by Mandelbrot ($1975$)) has the
correct two-point correlation function does not convey much information
unless other statistical measures of clustering are tested.

The present day galaxy distribution is manifestly not a gaussian random
process: there is, for example, no symmetry about the mean density. Even this fact alone tells us that there is more to
clustering of galaxies than two point correlation function. The higher order correlation functions provide a much more detailed description of
galaxy clustering probing the low and high count tails of the distribution. The simplest higher order correlation function is the $3$-point
correlation function. It is defined in terms of probability of finding three points inside the infinitesimal volume elements $dV_1$, $dV_2$ and
$dV_3$ respectively placed at the vertices
of a triangle with sides $r_{12}, r_{23}$,and $r_{31}$. This probability, of occurrence of three points nearby, is given by
\be
dP_{123} = { \left( \frac{ \bar{n} }{N} \right) }^3
           \left[ 1+ \xi(r_{12}) +\xi(r_{23})+\xi(r_{31}) + \zeta(r_{12,r_{23},r_{31}}) \right] dV_1 dV_2 dV_3
           \label{3pcf}
\ee
where $\zeta(r_{12},r_{23},r_{31})$ is the reduced or connected three point correlation function, while the full three-point correlation
function is given by the sum of last four terms in the square brackets. The reduced three point correlation function  appears to be simply
related to the two-point function through a Kirkwood-like relationship:
\be
\zeta(r_{12}, r_{23}, r_{31}) = Q[\xi(r_{12})\xi(r_{23}) + \xi(r_{23})\xi(r_{31}) + \xi(r_{31})\xi(r_{12})]
\label{eq204}
\ee
where $Q \sim 1$ is a constant, and the equality is due
to the usual assumption of homogeneity and isotropy. This
scaling law is called the hierarchical model in cosmology,
and it agrees rather well with observations. The full
Kirkwood law would require an additional
term on the right-hand side of this equation, proportional to
$\xi(r_{12})\xi(r_{23})\xi(r_{31})$.
Equation $\ref{eq204}$ can be generalized to any order $n$ by the expression
\be
\xi_n(r_1,\cdots,r_n)=\sum_{t=1}^{T(n)}Q_{n,t}\sum_{L_{n,t}}\Pi^{n-1}\xi(r_{ij}).
\label{eq205}
\ee
where $Q_{n,t}$ are known as structure constants. $T(n)$ are the number of distinct structures formed by linking of $n$ galaxies and $L_{n,t}$
is the number of possible relabelings for each distinct structure. A simplification of equation $\ref{eq205}$ is the scale invariant model
proposed by \citet{1989A&A...220....1B}
\[
\xi_n(kr_1,\cdots,kr_n) = k^{-(n-1)\gamma}\xi_n(r_1,\cdots,r_n)
\]

\section{\bf Other Statistical Measures}\label{other statistical}
\subsection{Moments of Counts in Cells and Void Probability}
The probability that a randomly placed cell $A$  of volume $V(A)$ contains exactly $N$ objects of the point process is denoted by $P(N,V(A))$. For a Poisson process with intensity $\lambda$, these quantities are completely known
\be
P(N,V(A))= \frac{(\lambda V(A))^N}{N!}exp(-\lambda V(A)).
\ee
When $N=0$, the above quantity quantifies a region of space which has no particles in the cell. Such regions of space is called a {\it void} and the corresponding probability $P(0,V(A))$ is known as the emptiness or void probability. The moments of order $n$ of the counts are defined by
\be
\mu^n(A^n) = E(\phi(A^n)) = \sum_{N=0}^{\infty}N^n P(N,V(A)).
\ee
N-point volume averaged correlation function can now be related to the $n^{th}$ moment of the counts in cells by
\be
\bar\xi_n(V) = \frac{\mu^n(A^n)}{{\bar{N}}^n}
\label{co-corr}
\ee
As we have seen previously in subsection \ref{Higher Order} that higher-order correlation functions depend on a large number of arguments,
indicating that they are rather difficult to estimate. However, since counts in cells are easy to estimate, they
are frequently used to study higher order correlations in the galaxy distribution using equation \ref{co-corr}.
\subsection{Nearest Neighbor Distances}
In this statistical method the distribution function $G(r)$ is chosen in a way so that $G(r)$ is the probability that the distance between a
randomly chosen galaxy and its nearest neighbor is less than or equal to $r$. If the measure of this function $G(r)$ for the given distribution
of galaxies is greater than its counterpart for a Poisson distribution, the galaxy distribution is said to be clustered. Likewise, for a
regular distribution  the measure of $G(r)$ is smaller than that for a Poisson distribution.
This is because a point in a regular distribution is, on average, farther away  from its nearest neighbor than in Poisson distribution.
\subsection{Need of fractal Hypothesis}
The problems of combinatorial explosions of terms in $N$ point correlation
function together with the power law behavior of $\xi(r)$ clearly pose difficulties in their calculation. While many solutions to these issues have
been proposed they usually deal with each of these issues separately. As we shall see, the fractal hypothesis, on the other hand, deals with all
these problems as a whole and offers an explanation to each of them within the fractal picture. We, however, do not intend to claim that the fractal
hypothesis is the only possible explanation to these problems, whether considering them together or separately. From now on in this thesis we shall
take the point of view that fractals offer an attractively simple description of the large scale distribution of galaxies. Therefore the analysis
offered by them deserves a deep, serious and unprejudiced investigation.

From its basis, the fractal hypothesis in many ways represents a radical
departure from the orthodox traditional view of an {\it observationally} homogeneous universe. By using tools of fractal analysis we will
investigate if the distribution of large scale structures {\it actually} behaves as a fractal to arbitrary large scales, or if there is a
transition to homogeneity.
Fractal analysis is going to be used as the main statistical tool, to quantify the clustering in the distribution of large scale structures, in this
thesis. Keeping this in mind Section \ref{Fractals} introduces a brief background on fractals necessary in
this thesis.
\section{\bf On the ``Definition'' of Fractals}\label{Fractals}
Fractal geometry deals with the objects which are highly irregular and can not be handled by the tools of differential geometry. A geometric object
can in general be described in terms of its topological dimension which is an integer that defines the number of coordinates needed to specify the
geometric object. A {\it fractal} is defined to be a set of points for which the Hausdorff-Besicovitch dimension strictly exceeds the topological
dimension. We shall discuss later the Hausdorff dimension, but the important point here is that this original tentative definition is very abstract.
It is often too difficult to be used in practice. This definition  also excludes some sets which ought to be regarded as fractals (e.g. space
filling (Hilbert) curve). 

Loosely speaking a fractal is a shape that tends to have a {\it scaling} property, implying that degree of its
irregularity and/or fragmentation is identical at all scales. Time and again the
definition of fractal has been modified by arguing that a single definition of fractal would be restrictive and, perhaps,
it would be best to consider fractals as a collection of techniques and methods applicable
in the study of the irregular, broken and self-similar geometrical patterns.

It seems best to
regard a fractal as a set that has properties
such as those described below : when we refer to a set as a fractal, we will
typically keep in mind that this set has a fine structure,
i.e., one has to look for detail on arbitrarily small scales. It is
too irregular to be described in traditional geometrical language, both locally and
globally. This set which we call a fractal, often has some form of
self-similarity, perhaps  approximate or statistical. The `fractal dimension' of
such a set (defined in some way) is greater than its topological
dimension. And in most cases of interest this set of points is defined in a very
simple way, perhaps recursively.

With the above description of fractals, we have
kept open the possibility that a given fractal shape can be
characterized by more than one definition of fractal dimension, and
they do not necessarily need to coincide with each other, although
they have in common the property of being able to take fractional
values. Therefore, an important aspect of studying a fractal
structure (once it is characterized as such by, say, at least being
recognized as self-similar in some way) is the choice of a
definition for fractal dimension that best applies to, or is derived
from, the case in study.

Fractal geometry has been considered a revolution in the way we are able
to mathematically represent and study figures, sets and functions. In
the past, sets or functions that are not sufficiently smooth or regular
tended to be ignored as ``pathological''. Nowadays, there is a realization that a lot can and is worth being said
about non-smooth sets. Another interesting points is that the
irregular and broken sets provide a much better representation of many
phenomena than do figures of classical geometry.
\subsection{Application of fractals}\label{App Fractals}
The concept of scale invariance is of key importance in the
characterization of many physical system. It has long been
recognized that scale invariance are usually associated with the
complexity displayed by a given structure, for which the differentiable
geometry based techniques are completely inadequate. A classic example
is that of Brownian motion, where the concepts of non differentiable manifolds were used to describe the random motions of particles in liquid or
gaseous medium. Although  the concept
of non differentiable geometry has been subsequently used in many physical
and mathematical application, the concept of fractal object has been
explicitly  introduced and formalized only recently  by Mandelbrot.
A description in terms of fractal gives a good representation of
wide spectrum of phenomena, not only in physics, but also in
biology, geology etc.

The application of fractal techniques to the study of chaotic dynamical system has been quite fruitful. Fractal techniques have been used to
describe geometric structures which have completely unpredictable trajectories in
the configuration space. Such structures, which are generally termed as \textit{strange attractor}, can not be represented by means of usual
geometric tools. Fractal techniques have been extensively applied in the statistical characterization of the study of turbulence.

In this thesis we are going to apply the fractal techniques to statistically describe the distribution of gravitationally evolved Large Scale
Structures of the Universe. Despite the great difference existing between the dissipative dynamics of fully developed turbulence and the non-
dissipative gravitational dynamics, several common aspects can be identified. First of all, both the Navier-Stokes equation of fluidodynamics and
the BBGKY equation which describe the gravitational dynamics, do not contain intrinsic scales. Further numerical simulation of both non linear
gravity and the turbulent flows are seen to generate small scale coherent structures arising from a large scale smooth background. On the ground of
these similarities, we feel that the fractal description, that is so successful in describing the statistics of dissipative eddies of turbulent
flows, to be equally useful in describing the statistics of gravitational clustering. The following discussion starts on the mathematical aspects
associated with fractals, but gradually there is a growing emphasis on applications.
\subsection{The Hausdorff Dimension}\label{Haus Dim}
An important step in the understanding of fractal dimensions is for one
to be introduced to the {\it Hausdorff-Besicovitch dimension} (often known
simply as {\it Hausdorff dimension} \citep{fal90}). It can take non-integer values and was found to coincide with
many other definitions. In obtaining the dimension that bears his name, Hausdorff used the
idea of defining measures using covers of point sets (first proposed by C.\
Carath\'{e}odory). We shall offer an {\it illustration} of
the Hausdorff measure whose final result is the same as achieved by
the formal mathematical proof.

\begin{figure}[thb]
  \centerline{\epsffile{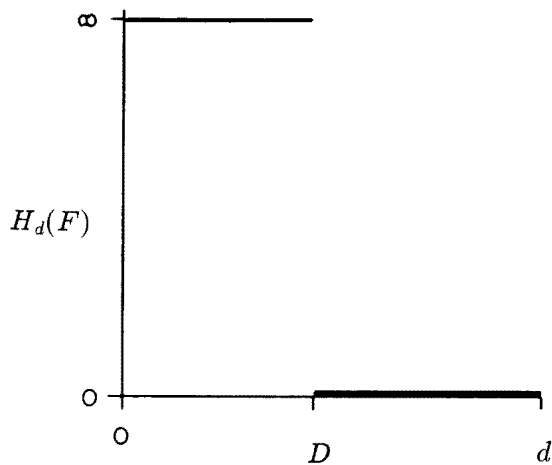}}
  \caption[\sf Graph of the Hausdorff $d$-measure $H_d(F)$ against $d$ for a
           set $F$ ]{\sf Graph of $H_d(F)$ against $d$
           for a set $F$. The Hausdorff dimension $D$ is the critical value
           of $d$ at which the jump of Hausdorff measure from $\infty$ to $0$ occurs. It can be an integer (for line, plane and sphere etc.) as well
           as a fractional number(e.g. for irregular shapes and clustered point set) }
           \label{2-3}
\end{figure}

Consider a set $F$ of points. In order to give a measure of the size of this set
in space we have taken a test function $h(\delta) = \gamma (d) \delta^d$, where $\gamma (d)=[\Gamma (1/2)]^d/\Gamma(1+d/2)$. For lines, squares and
cubes we have the geometrical factor $\gamma (d)$ equal to $ 2,\pi$ and $\frac{4}{3}\pi$ respectively. Here $\delta$ is the length of the line
segment needed to cover the point set completely. Hausdorff proposed a {\it measure} $H_d(F) = \sum h(\delta)$ of such a point set.
In general we find that, as $\delta \to 0$, the measure $H_d(F)$ is either infinite or zero depending on the value of $d$, the dimension of the
measure. The Hausdorff dimension $D$ of the set $F$ is the {\it
critical dimension} for which the measure $H_d(F)$ jumps from infinity to zero (see figure \ref{2-3}):
\be
  H_d(F) = \sum \gamma (d) \delta^d = \gamma (d) N(\delta) \delta^d
           \mathop\rightarrow_{\delta \to 0}
           \left\{ \begin{array}{ll}
                      0,      &  d > D, \\
                      \infty, &  d < D.
                   \end{array}
           \right.
  \label{eq206}
\ee
The quantity $H_d(F)$ is called the $d$-measure of the set and its
value for $d=D$ is often finite, but may be zero or infinity. It is
the position of the jump in $H_d$ as function of $d$ that is
important. Note that this definition means the Hausdorff dimension
$D$ is a local property in the sense that it measures properties of
sets of points in the limit of a vanishing diameter of size $\delta$
of the test function used to cover the set. It also follows that the
$D$ may depend on position.

The familiar cases are $D=1$ for lines, $D=2$ for planes and
surfaces and $D=3$ for spheres and other finite volumes. There are
many sets, however, for which the Hausdorff dimension is non
integer and is said to be fractal. In other words, because the jump
of the measure $H_d(F)$ can happen at non integer values of $d$,
when $H_d(F)$ is calculated for irregular and broken sets the value
$D$ where the jump actually occurs is usually non integer.
\subsection{Fractal Dimension of orthogonal projections and  intersections}
\label{orthogonal}
We briefly present the properties of orthogonal projections and
intersections of fractal structures. This discussion is
useful in the interpretation of angular and one dimensional (pencil beams)
catalogs.

Orthogonal projections preserve the sizes of objects in the perpendicular direction.
If an object of fractal dimension $D$, embedded in a  space of dimension $d=3$, is projected on a plane
(of dimension $d'=2$) it is possible to show that
the projection has dimension $D'$ such that
\be
\label{epro1}
D'=\left\{ \begin{array}{ll}
                      D,      &  \mbox{if} \; \; D<d'=2 \\
                      d', &  \mbox{if} \; \; D>d'=2
                   \end{array}
           \right.
\ee
This explains, for example, why clouds which have fractal dimension
$D \approx 2.5$, give rise to a compact shadow of dimension $D'=2$.
The angular projection represents a more complex problem
due to the mix of very different length scales. Nevertheless
the theorem given by Equation \ref{epro1} can be extended
to the case of angular projections in the limit of small angles.

We discuss now a different but related problem. We investigate the properties
of the structure that comes out from the intersection of a fractal with dimension $D$,
embedded in the $d=3$ Euclidean space, with an object of dimension $D'$ ? The later can
be for example a line ($D'=1$ - schematically a pencil beam survey), a plane ($D'=2$) or a
random distribution ($D'=3$). Then the {\it co-dimension} of the intersection is equal
to the sum of the {\it co-dimensions} of the two intersecting structures. We can represent this as
\be
d-D_I =(d-D)+(d-D')
\ee
where $D_I$ is the fractal dimension of the intersection set.
Hence we can write for the dimension $D_I$:
\be
\label{add1}
D_I = D + D' - d
\ee
If $D_{I} \le 0$, in
the intersection it
is not possible to recover any correlated signal.
Hence, e.g., the intersection of a stochastic fractal with a
random distribution has the same dimension $D_I=D$ of the
original structure. Such a property is useful in the discussion of
surveys in which a random sampling has been applied.
\section{\bf Other Fractal Dimensions}\label{Other Frac Dim}
The illustration of the Hausdorff dimension shown previously
may be a good description  from a mathematical point of view, but it is hard
to get an intuitive feel about significance of the fractal dimension from it. Moreover, we do
not have a clear picture of what this fractional value of dimension
means. In order to try to answer these questions let us see  different
definitions of fractal dimension and some examples.
\subsection{Similarity Dimension}\label {similarity dim}
The fractals we discuss may be considered to be sets of points
embedded in space. This space has the usual topological dimension
which we are used to, and from a physicist's point of view it
coincides with degrees of freedom defined by the number of
independent variables. So the location of a point on a given line is
determined by one real number and a set of two independent real
numbers is needed to define a plane. If we define dimension by
the number of degrees of freedom in this way, we are in a position to
consider a $d$-dimensional space for any non-negative integer $d$. In
fact, in mechanics it is conventional to consider the motion of $m$ particles in 3 dimensions
as being the motion of one particle in a $6m$-dimensional space if
we take each particle's position and momentum as independent.

The dimension defined by degrees of freedom seems very natural but
contains a serious flaw. In practice we can have a curve (e.g. a Peano Curve) that
folds so wildly that it nearly `fills' a plane. In this way we are able to
define the position of any point on the plane by a single real
number. Hence the degree of freedom, or the dimension, of {\it this} plane
becomes 1, which contradicts the empirical value 2. In order to
explain such structures we have to define a new dimension based on
similarity.

\begin{figure}[p]
  \centerline{\epsffile{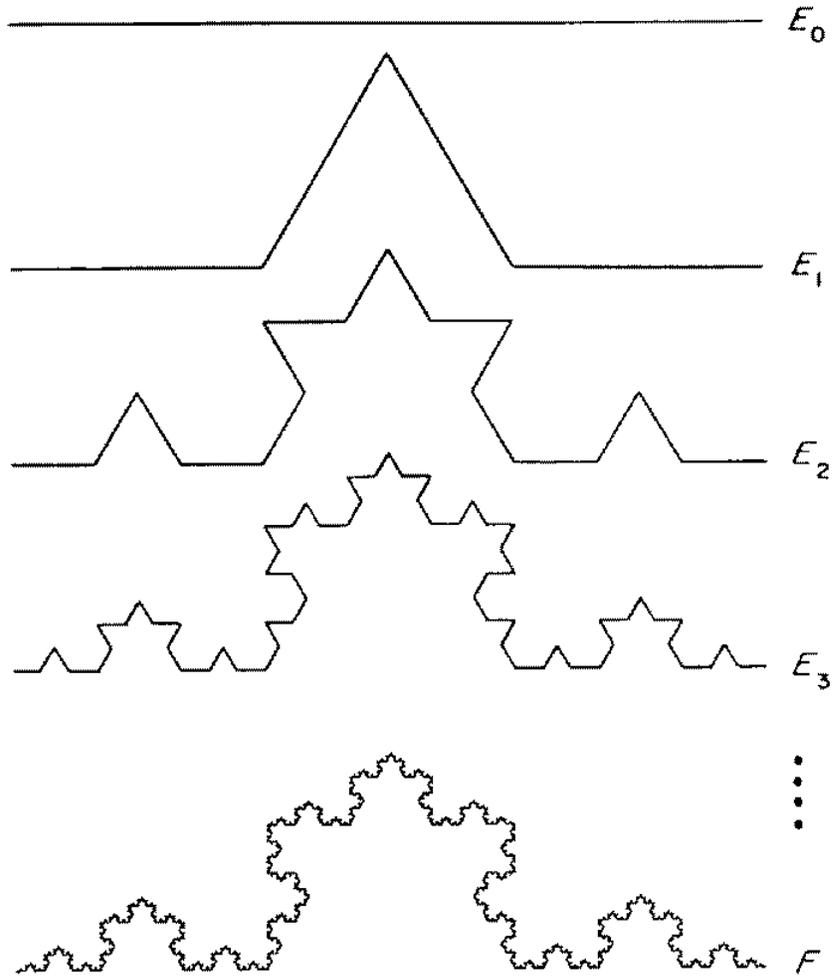}}
  \caption[\sf Construction of the von Koch curve]{\sf
           Construction of the von Koch curve $F$. At each stage,
           the middle third of each interval is replaced by the
           other two sides of an equilateral triangle. We can see that the length of the curve $E_k$ goes to infinity as $k$ tends to infinity. The von Koch curve occupies almost zero area implying that geometric measures like length and area are not well defined for fractal structures.}
           \label{5_fractal}
\end{figure}

Let us consider dividing a unit line segment into $N$ parts. As we decrease the size (say $\delta$) of each part, the $N$ will correspondingly increase. This results in $N(\delta) \propto 1/\delta$, implying for a straight line segment of unit length we have  $N \delta^1 = 1$. Similarly, if we divide
a unit square into $N$ similar parts, each one is scaled by a factor
$\delta = 1/N^{1/2}$ (if $N=4$ the square is scaled by half the side length); so $N
\delta^2 =1$. Now if a unit cube is divided in $N$ parts, each
scaled by $\delta = 1/N^{1/3}$ (again if $N=8$ the cube is scaled by
half the side length), we have $N \delta^3 =1$. Note that the
exponents of $\delta$ correspond to the space dimensions in each
case. Generalizing this discussion we may say that for an object of
$N$ parts, each scaled down from the whole by a ratio $\delta$, the
relation $N \delta^D =1$ defines the {\it similarity dimension} $D$
of the set as
\be
   D= \lim_{\delta \to 0}\frac{d \log N}{d \log 1/\delta}.
   \label{eq207}
   \ee
The calculation of similarity dimension of sets can be better viewed
with the example of a strictly self similar fractal called "Koch
Curve". Figure \ref{5_fractal} shows the construction of the von Koch
curve, and
any of its segments of unit length is composed of 4 sub-segments each of
which is scaled down by a factor 1/3 from its parent. Therefore, its
similarity dimension is $D=~\log 4~/~\log 3~\cong~1.26$. This
non-integer dimension, greater than one but less than two, reflects the
properties of the curve. It somehow fills more space than a simple line
($D=1$), but less than a Euclidean area of the plane ($D=2$). Figure
\ref{5_fractal} also shows that the von Koch curve has a finite structure
which is reflected in irregularities at all scales; nonetheless, this
intricate structure stems from a basically simple construction. Whilst
it is reasonable to call it a curve, it is too irregular to have
tangents in the classical sense. A simple calculation on the von Koch
curve (Figure \ref{5_fractal}) shows that at $k^{th}$ step $E_k$ is of length ${ \left( \frac{4}{3} \right) }^k$;
letting $k$ tend to infinity implies that the curve $F$ has infinite length. On
the other hand, curve $F$ occupies almost zero area in the plane. So neither length
nor area provides a very useful description of the size of the von Koch curve $F$.

After this discussion we start to have a better idea of what those
fractal dimensions mean. Roughly, a fractal dimension provides a description
of how much space a set fills. It is a measure of the prominence of
the irregularities of a set when viewed at very small scales. We can therefore
expect that a shape (or point set)  with a high fractal dimension will be more complicated (or clustered)
than another shape with a lower fractal dimension. The Hausdorff dimension
described previously can be seen as a generalization of this
similarity dimension. Unfortunately, similarity dimension is only
meaningful for a small class of strictly self-similar sets. For non
self similar fractals we need to introduce more general measures
like box counting dimensions discussed below.
\subsection{Box Counting Dimension}\label{box-cou-dim} The
Hausdorff and similarity dimensions defined so far provide
definitions of fractal dimension for deterministic fractals, i.e.,
classical fractal sets in a mathematically idealized way. Although
some of these classical fractals can be used to model physical
structures, what is necessary now is to discuss structures that are statistically self similar,
which are encountered in natural phenomena. Hence, we need to apply
as far as possible the mathematical concepts and tools developed so
far in the study of such statistically self similar fractal structures.
One such tool is called the box counting dimension. In this approach the irregular curve or the distribution of particles is covered
with a set of boxes of size $\delta$ and the number of boxes are counted which contain the part of the fractal. This size $\delta$ is varied over a
range and the resulting number of boxes required to cover the
distribution of points gives the number $N(\delta)$. Obviously $N(\delta)$ will increase as the size $\delta$ decreases. If we proceed this way and find
$N(\delta)$ for smaller values of $\delta$, we are able to
plot a graph of $N(\delta)$ versus $\delta$, for different grid
sizes. Now it follows from equation (\ref{eq206}) that
asymptotically in the limit of small $\delta$ the following equation
is valid:
\be
 N( \delta ) \propto \frac{1}{\delta^D}.
 \label{eq23}
\ee
So the fractal dimension $D$ of the distribution can be determined
by finding the slope of $\log N(\delta)$ plotted as a function of
$\log \delta$. We get the expression for the box counting dimension $D_b$ as
\be
 D_b := \lim_{ \delta \to 0} \frac{d \log N(\delta)}{d \log(1/\delta)}.
\ee
If the limit does not exist then one must talk about the upper box
dimension and the lower box dimension which correspond to the upper
limit and lower limit respectively in the expression above. In other
words, the box-counting dimension is strictly defined only if the upper
and lower box dimensions are equal. The upper box dimension is
sometimes called the entropy dimension, Kolmogorov dimension,
Kolmogorov capacity or upper Minkowski dimension, while the lower
box dimension is also called the lower Minkowski dimension.
Box counting dimension $D_b$ is, in essence, a scaling rule comparing how a pattern's detail changes with the scale at which it is considered.
According to equation (\ref{eq23}), for a space filling
distribution we expect that $N(\delta)$ should decrease as $\delta^{-3}$, so
that fractal dimension for such a distribution is equal to the dimension of the ambient space {\it i.e.} $D_b=3$. In a similar way for a filamentary
structure it is $N(\delta) \sim \delta^{-1}$, while for a planer point distribution $N(\delta) \sim
\delta^{-2}$, with resulting box counting dimensions $D_b=1$ and $D_b=2$
respectively. In more general cases, non integer dimensions can also
be expected. By means of equation (\ref{eq23}) we can obtain the
expression for the length $L$ of irregular curve as
\[
 L = N(\delta)~ \x ~\delta  \propto \delta^{1-D},
\]
which shows its explicit dependence on the yardstick chosen.

The box counting dimension proposes a systematic measurement which
applies to any structure in the plane, and can be readily adapted
for structures in the space. It is perhaps the most commonly used
method of calculating dimensions. Its advantage lies in the easy
and automatic computability provided by the method, as it is
straightforward to count boxes and maintain statistics allowing
dimension calculation. The program can be carried out for shapes
with and without self-similarity and, moreover, the objects may be
embedded in higher dimensional spaces.

Note that the above two definitions of dimension (i.e. Hausdorff and
Box counting) deal with the number of required coverings.These
definitions have no regard to the number of points contained inside
each of the covering boxes. In this sense, such dimensions depend on
the `shape' of the distribution. In this way they provide a purely
geometrical description, while no information is given about the
clumpiness, as correlation functions do. In order to extend the
description in terms of fractal dimensions, so as to include the
clustering properties of a distribution, we need to introduce a
probability measure, so that adequate information about the
clustering of the distribution is available.

\section{\bf Probability and Probability Measure}\label{Prob}
The probability of an event is not just a property of an individual experiment. It is  a joint property of the number of different possible
outcomes of the experiment performed under similar conditions. If an experiment is performed $N$ times, and a certain outcome $A$ occurs in $M$ of
these cases,  the ratio $M/N$ approaches a limiting value as $N \rightarrow \infty$. This ratio is defined as the probability $P(A)$ of $A$. In
other words probability of an event is a measure of how likely the event is to occur when the random experiment within a sample space is run. In a
sample space $S$, a measure $d\mu$ is said to be a probability
measure if $d\mu(A) \ge 0$ for any event $A$ within the sample space.
The probability measure has also got to satisfy $d\mu(S) = 1$. While
analyzing the clustering properties of distribution of points, the coarse grained probability, $p_i(r)$, in terms of the probability measure $d\mu$
is given by
\be
{\it p}_i(r) = \int_{\Lambda_i}d\mu(x)
 \label{eq24}
\ee
It provides the measure of "mass" contained inside
the hypercube $\Lambda_i$ of side $r$, with ${\it i} = 1, 2,\cdots,N (r)$. Accordingly, the set
${\it P}_r= \{{\it p}_i; i = 1,\cdots,N(r)\}$ is the probability
distribution over the $N(r)$ different states. In this way the information
content of the distribution can be defined as
\be
{\it J}(r,{\it P}(r)) = \log_{2}N(r)
+\sum_{i=1}^{N(r)}{\it p}_i\log_2{\it p}_i
 \label{eq25}
\ee
This is characterized in terms of Information Dimension of the distribution which we discuss below.
\subsection{Information Dimension}\label{info dim}
For a homogenous distribution, all the boxes are expected to be
equally populated, that is, all the states are equally probable
(maximum entropy configuration). Correspondingly, the quantity $J
(r, P(r))$ vanishes, thus indicating the absence of any information
carried by unclustered structures. Conversely, the maximum
information content is obtained when one single state has unity
probability, while it is vanishing for all the other states (minimum
entropy configuration). In this case, ${\it J}(r,{\it P}(r)) =\log_2 N(r)$,
while in general $0 \leq {\it J}(r, {\it P}(r)) \leq \log_2 N(r)$.We define
the Shannon information (or entropy),
\be
{\it I}(r,{\it P}(r)) = -\sum_{i=1}^{N(r)}{\it p}_i\log_2{\it p}_i
 \label{eq26}
\ee
as the difference between the maximum information content and the
actual information provided by the ${\it P}_r$ distribution.
Therefore, the information dimension,
\be
{\it D}_{{\it I}} = \lim_{r\to
0}\frac{{\it I}(r,{\it P}(r))}{\log_2(1/r)} =-\lim_{r\to 0}\frac{\langle \log{\it p_i}\rangle}{\log \frac{1}{r}}
 \label{eq27}
\ee
is related to the rate of information loss as the resolution scale
increases.
\subsection{Correlation Dimension} \label{corr dim}
A further important characterization of the scale-invariant
properties of a fractal set is given in terms of the correlation
dimension, originally introduced by Grassberger \& Procaccia .  For
a given point ${\bf  x_i}$ belonging to ${\bf  \it A}$, let
\be
{\it C}_i(r) = \frac{1}{N}\sum_{j=1 \neq i}^{N}\Theta(r-|{\bf x_i-x_j}|)=
\frac{n_i(r)}{N}
 \label{eq28}
\ee
is the probability of finding $n_i(< r)$ points out of the
N points of the set within a distance $r$ from ${\bf x_i}$. In equation
\ref{eq28}, $\Theta$ is the Heaviside  step function with the following property:
\be
\th(x) = \left\{ \begin{array}{ll}
                      1,      &  x > 0, \\
                      0, &  x < 0.
                   \end{array}
           \right.
\label{eq281}
\ee
 We, then, introduce the correlation integral
\be
{\it C}(r) = \frac{1}{N}\lim_{N \to \infty}\sum_{i=1}^{N}{\it C}_i(r)
 \label{eq29}
\ee
whose scaling in the limit $r \to 0 $ defines the
correlation dimension, $D_\nu$, according to
\be
{\it C}(r) \sim r^{D_\nu}
 \label{eq210}
\ee
Note that for a structure that behaves like a fractal at all the
scales it is not possible to define an average density, since it
turns out to depend on the dimension of the fractal itself. In fact,
since equation \ref{eq210} gives the scaling of the number of
neighbors, the density around the $i^{th}$ point will scale as
$r^{3-D_\nu}$, and, thus, unless $D_\nu = 3$, it decreases for
increasing scales. Note that this kind of behavior is not expected
for the distribution of cosmic structures, which, on grounds of the
Cosmological Principle, should reach homogeneity at sufficiently
large scales. However, we can define fractal dimensions in a finite
scale range, while taking homogeneity at large scales. In this case,
following the definition of the 2-point correlation function given
in subsection \ref{Standard Analysis}, it is easy to see that it is
related to the correlation integral of equation \ref{eq29}
according to
\be
{\it C}(r) =\int_{0}^{r}d^3r'[1+\xi(r')] =
\bar{N}\left[1+(\frac{r_c}{r})^\gamma \right]
 \label{eq211}
\ee
Here, $\bar{N}=\frac{4}{3}\pi r^3 \bar{n}$ is the number of neighbors
within a radius $r$ expected for a homogeneous distribution, while the
clustering scale $r_c$ is related to the correlation length $r_o$ as
$r_c =\left[3/(3- \gamma )\right]^{1/\gamma}r_o$. Thus, according to
the definition \ref{eq210} of correlation dimension, the
observed power law shape of the 2-point correlation function implies
that at $r \ll r_c$ the galaxy distribution behaves like a fractal
with $D_\nu \neq 3$, while assuming large scale homogeneity gives
$D_\nu = 3$ at $r \gg r_c$.

As we have seen in subsection \ref{Higher Order}, a complete
statistical description of a given points distribution requires the
knowledge of correlations or moments of any order. In a similar way,
a complete characterization of the scaling properties
of a fractal set should require the introduction of a hierarchy of
scaling indices, that generalize those already introduced  and that
account for the scaling of correlation functions of different
orders. This will be realized in the following subsection by introducing the
concept of the multifractal spectrum of generalized dimensions.
\subsection{ The Generalized Dimension}
The various definitions of fractal dimension that we have introduced
represent particular cases of a continuous sequence of scaling
indices, known as multifractal spectrum of generalized dimensions
\citep[see e.g.][]{bor}. A first definition can be given in
terms of the generalized Hausdorff dimensions, which represents the
extension of the classical Hausdorff dimension of equation
(\ref{eq206}). Consider ${\it p_i}$ to be the measure associated with a given
set $\Lambda_i$ (as defined by equation \ref{eq24}). We can introduce a partition function $\Gamma(q,\tau)$ as
\be
  \Gamma(q,\tau) = \left\{
                        \begin{array}{ll}
                      \lim_{r\to 0}\inf_{\Gamma^r_A}\sum_i\frac{{\it {p}_i}^q}{{r_i}^\tau},~~ \tau \leq 0,~~ q\leq 1, \\
                      \lim_{r\to 0}\sup_{\Gamma^r_A}\sum_i\frac{{\it {p}_i}^q}{{r_i}^\tau},~~ \tau \geq 0,~~ q\geq 1.
                   \end{array}
           \right.
  \label{eq217}
\ee
For each value of q, the respective $ \tau(\equiv \tau_q)$ is defined as the
unique value which makes $\Gamma(q, \tau )$ a finite constant. Then,
the generalized Hausdorff dimensions are defined as
\be
D(q)=\frac{\tau(q)}{(q-1)}
  \label{eq218}
\ee
For $q=1$, $D(1)=\lim_{q \to 1} D(q)$. From this definition, it is easy to recognize that Hausdorff Dimension, $D_H = D(0) =
-\tau(0)$.

A further set of scaling indices is given by the {\it Renyi dimensions}.
Let us consider a covering, of a point set, formed by $N(r)$ cells of the same
size $r$. Then, if $n_i$ is the number of points in the cell $i$,
the probability ${\it p}_i = n_i(r)/{\sum_jn_j}$ is the measure
associated to the $i^{th}$ box. The Renyi dimensions are defined as
\be
D_q=\frac{1}{q-1}\lim_{r\to 0} \frac{\log \sum_i
{\it {p}_i}^q}{\log r}
  \label{eq219}
\ee
In this case, the capacity dimension corresponds to the $q = 0$
case, while the information dimension is recovered in the limit $q
\to 1$. In general, it can be proved that $D(q) \leq  D_q$,
while in most cases of practical application the two definitions
(\ref{eq218}) and (\ref{eq219}) of generalized dimensions
can be considered as completely equivalent.

A slightly different definition is represented by the
Minkowski-Bouligand dimensions. In this case, the covering of the
fractal set is obtained by means of spheres of radius $r$, that are
centered each at a point belonging to the fractal. If $n_i(< r)$ is
the number of points within $r$ from the $i^{th}$ point, the
Minkowski-Bouligand dimensions are defined as
\be
D'_q=\lim_{N\to \infty}\frac{1}{N^2}\lim_{r\to
0}\frac{1}{q-1} \frac{\log \sum_i {n_i}^{q-1}}{\log r}
  \label{eq220}
\ee
and generalize the correlation dimension of equation \ref{eq210}
. Further, it can be proved that the Renyi and Minkowski-Bouligand
dimensions are completely equivalent \citep{1980ApJ...236..351D}.

An important class of fractals is represented by self-similar
monofractals. These fractal sets are characterized by the fact that
every part of the set represents an exact replica of the whole set
(in a statistical sense), so that the scaling properties are the
same around each point. For these fractals $D_q = D_H$ for any q, so
that a single dimension gives a complete characterization of the
whole set. More complex fractal sets are represented by
multifractals. In this case, the entire spectrum of
generalized fractal dimensions $D_q$ is required to describe the
local character of the scaling properties. For a multifractal, it
can be shown that $D_q \leq D_{q'}$ for $q \geq q'$. According to the definitions
(\ref{eq217}) of the $\Gamma$ partition function and
(\ref{eq219}) of Renyi dimensions, in the case $q \gg 0$ the
summations are dominated by the densest regions in the set, while
for $q \ll 0$ the least dense regions give the largest contribution.
In this sense, for positive q's the generalized dimensions provide
information about the scaling properties of the distribution inside
the regions of high density, as correlation functions do, while for
$q \ll 0$ they account for the scaling inside the underdense
regions, thus providing a comprehensive statistical description of
the entire point distribution.
We are going to characterize both homogeneous as well as slightly clustered distribution of points using Minkowski-Bouligand Dimension in chapter \ref{chap3}.
\section{\bf Lacunarity}
\label{lacunarity}
So far we have quantified fractal structures by their dimension.
That this is not a sufficient characterization can be illustrated by the fact that we can have two points sets
(one deterministic and one  stochastic Cantor set) with the same
fractal dimension $D$ but with different {\it morphological properties}.
In order to distinguish such sets,  Mandelbrot has
introduced the concept of {\it lacunarity}  $F$ as
\be
\label{e322}
Nr(\lambda >\Lambda) = F \Lambda^{-D}
\ee
where $ Nr(\lambda >    \Lambda)$
is the number of voids with a size
 $ \lambda  > \Lambda$.
The scaling behavior of  $ Nr(\lambda)$  is the same for both Cantor sets.
However the lacunarity $F$, i.e. the prefactor of the distribution, takes
different values for the two Cantor sets.

In order to define lacunarity for random fractals
we need a probabilistic form
of Equation \ref{e322}. This can be done by introducing $P(\lambda)$,
which is the conditional probability that, given a box of size
 $\epsilon$
containing points of the set,
this box is neighbored by a void of size $\lambda > \Lambda$.
Lacunarity is defined as the prefactor of the void distribution
\be
\label{e323}
P(\lambda >   \Lambda) = F \Lambda^{-D} \; .
\ee
It is easy to show that in the case
of deterministic fractals this
definition gives the same value of the lacunarity defined in
Equation \ref{e322}.
Lacunarity plays a very important role
in the characterization of voids distribution in the available galaxy
 catalogs. 

\chapter{Fractal Dimensions of a Homogeneous and Weakly Clustered Distribution}\label{chap3}
Homogeneity and isotropy of the universe at {\sl sufficiently} large scales is
a fundamental premise on which modern cosmology is based.
Fractal dimensions of matter distribution is a parameter that can be used to
test the hypothesis of homogeneity.
In this method, galaxies are used as tracers of the distribution of matter and
samples derived from various galaxy redshift surveys are used to
determine the scale of homogeneity in the Universe.
Ideally, for homogeneity, the distribution should be a mono-fractal with the
fractal dimension equal to the ambient dimension.
While this ideal definition is true for infinitely large point sets, this may
not be realized, as in practice, we have only a finite point set.
The correct benchmark for realistic data sets is a homogeneous distribution of
a finite number of points and this should be used in place of the
mathematically defined fractal dimension for infinite number of points $(D)$
as a requirement for approach towards homogeneity.
We derive the expected fractal dimension for a homogeneous distribution of
a finite number of points.
We show that for sufficiently large data sets the expected fractal
dimension approaches $D$ in the absence of clustering.
It is also important to take the weak, but non-zero amplitude of clustering at
very large scales into account. In this chapter\footnote{This chapter is based on {\it Fractal Dimensions of a Weakly Clustered Distribution and the
Scale of Homogeneity} (J.S. Bagla, Jaswant Yadav \& T.R. Seshadri), Monthly Notices of Royal Astronomical Society {\bf 390}, 829} we also compute the expected fractal dimension for a finite
point set that is weakly clustered. Clustering introduces departures in the Fractal dimensions from $D$ and in
most situations the departures are small if the amplitude of clustering is
small. Features in the two point correlation function, like those introduced by
Baryon Acoustic Oscillations (BAO) can lead to non-trivial variations in the
Fractal dimensions where the amplitude of clustering and deviations from $D$
are no longer related in a monotonic manner. We show that the contribution of clustering and finite numbers to the fractal dimension is given by two
separate terms in the expression.

\section{\bf Introduction}
We expect the Universe to be homogeneous and isotropic on the largest scales.
Indeed, one of the fundamental postulates in cosmology is that the Universe is
spatially homogeneous and isotropic.
It is this postulate, generally known as the Cosmological Principle
(CP)\citep{1917SPAW.......142E}, that allows us to describe the geometry of the space time over large scales in terms of the Friedman-Robertson-Walker-Lemaitre (FLRW) metric.
The standard approach to cosmology assumes that the universe can be modeled
as a perturbed FLRW universe.
The large scale structures (LSS) in the universe are believed to have been
formed due to the collapse of small inhomogeneities present in the early
Universe \citep{1980lssu.book.....P,1999coph.book.....P,2002tagc.book.....P,2002PhR...367....1B}.
Thus it is of paramount importance to test whether the observed distribution
of galaxies approaches a homogeneous distribution at large scales.

The primary aim of galaxy surveys \citep{2001MNRAS.328.1039C,2000AJ....120.1579Y,1996ApJ...470..172S} is to
determine the distribution of matter in our Universe.
Redshift surveys of galaxies have revealed that the universe consists of a
hierarchy of structures starting from groups and clusters of galaxies to
superclusters and interconnected network of filaments spread across the
observed Universe \citep{2007arXiv0708.1441V,2000PhRvL..85.5515C,1986ApJ...302L...1D,2002AJ....123...20K}.

Fractal dimensions can be used as an indicator to test whether or not the
distribution of galaxies approaches homogeneity.
One of the reasons that make the Fractal dimensions an attractive option is
that one does not require the assumption that an average density of matter in the sample region is the same as that of the Universe as a whole \citep{1982fgn..book.....M}. In other words one does not have to assume that the matter distribution in the Universe has achieved homogeneity within
the sampled region itself \citep{2002sgd..book.....M}.
Ideally one would like to work with volume limited samples in order to avoid
corrections due to a varying selection function.
Redshift surveys of galaxies can be used to construct such sub-samples from
the full magnitude limited sample but this typically leads to a sub-sample
that has a much smaller number of galaxies as compared to the full sample.
This limitation was found to be too restrictive for the earliest surveys and
corrections for the varying selection function were attempted in order to
determine the scale of homogeneity: \citep[for example see][]{bharad1}
With the large surveys available today, this limitation is no longer very
serious.
Fractal dimensions are computed for the given sample or sub-sample and the
scale beyond which the fractal dimension is close to the physical dimension of
the sample is identified as the scale of homogeneity.
We expect that at scales larger than the scale of homogeneity, any fluctuation
in density is small enough to be ignored.
Thus at larger scales, CP can be assumed to be valid and it is at these scales
that the FLRW metric is a correct description of the Universe.

Fractal Dimension is defined in the mathematically rigorous way only for an
infinite set of points.
Given that the observational samples are finite, there is a need to understand
the relation between the fractal dimension and the physical dimension for such
samples.
In this chapter, we compute the expected fractal dimension for a finite
distribution of points \citep[see e.g.][]{1993PhRvE..47.3879B,1994PhRvE..49.4907B}
The early work on these effects has focused on small scales where the
amplitude of clustering is large.
In this work, we calculate fractal distribution for a uniform distribution, as
well as for a weakly clustered distribution of a finite number of points.
This is of interest at larger scale where fractal dimensions are used as a
tool to find the scale of homogeneity.

catalogs of different extra-galactic objects have been studied using various
statistical methods.
One of the important tools in this direction has been the use of two point
correlation function $\xi(r)$ \citep{1980lssu.book.....P} and its Fourier
transform the power spectrum $P(k)$.
We have precise estimates of $\xi(r)$ \citep{2007MNRAS.378.1196K,2007MNRAS.381..573R,2007A&A...472...29D}
and the
power spectrum  $P(k)$
\citep{2005MNRAS.362..505C,2007ApJ...657..645P}
from different galaxy surveys.
Different measurements appear to be consistent with one another once
differences in selection function are accounted for
\citep{2006astro.ph.11178C} \citep[but also see][]{2008MNRAS.385..830S}
On small scales the two point correlation function is  found to be well
described by  the form
\begin{equation}
  \xi(r)=\left(\frac{r_0}{r}\right)^{\g}
\label{eq:1}
\end{equation}
where  $\g =1.75\pm 0.03 $ and $ r_0=6.1\pm 0.2$~$h^{-1}$Mpc for the SDSS
\citep{zevi} and $\g =1.67\pm 0.03 $ and $ r_0=5.05\pm
0.26$~$h^{-1}$Mpc for the 2dFGRS \citep{haw}.
Recent galaxy surveys have reassured us that the power law behavior for
$\xi(r)$ does not extend to arbitrarily large scales.
The breakdown  of this behavior occurs at  $r> 16 h^{-1} {\rm Mpc} $ for SDSS
and at $r> 20 h^{-1} {\rm Mpc} $ for 2dFGRS,
which appears consistent with the distribution of
galaxies being homogeneous at large scales.
A note of caution here is that though the $\xi(r)$ determined from
redshift surveys appear to be consistent with the universe being homogeneous at
large scales in that $\left|\xi(r)\right| \ll 1$ at large $r$, it does not
actually imply that the universe is necessarily homogeneous.
This is because the two point correlation function given by,

\begin{equation}
\xi(r)=<\delta(x+r)\delta(x)>
\label{eq:2}
\end{equation}
where
\begin{equation}
\delta(x) = \frac{\rho(x) - \bar\rho}{\bar\rho}
\label{eq:3}
\end {equation}

presupposes that galaxy distribution that we
are analyzing is homogeneous on the scales within our survey region. This is
implicit in the fact that $\bar\rho$, which is assumed to be the spatial
average density of matter in the universe, is computed by averaging the
density from within the survey volume.
Of course, it may be possible to demonstrate that the survey is a fair sample
of the universe by showing that the values of $\bar\rho$ derived from
sub-samples of the survey are consistent with each other, or that the value of
$\bar\rho$ computed at different scales converges to a definite value at
scales much smaller than the size of the survey.
However, to verify and hence validate the cosmological principle, it is useful
to consider a statistical test which does not presuppose the premise being
tested, In other words the survey region need not be assumed to represent a fair sample of the whole Universe.
In this chapter we consider one such test, the ``multi-fractal analysis''
and apply it to distribution of particles in random as well as clustered
distributions.

Fractal dimension, which is generally a fractional number, is
characterized by the scaling exponent. A single exponent is sufficient to characterize a monofractal distribution. However, in many physical situations, we need to use a set with an invariant measure characterized by a whole spectrum of scaling exponents, instead of a single number.
Such a system is called a multi-fractal and we need to do a multifractal analysis of a point set to study the system.

In this  chapter we calculate the fractal dimension for a distribution
of finite number of points which are distributed homogeneously as well as for those which are weakly clustered.
For this purpose we use the multi-fractal analysis to study the scaling behavior of uniform as well as weakly clustered distributions in turn
finding the relationship between the fractal dimension and the two point correlation
function. We find deviations of fractal dimension $D_q$ from the $D$ arising
due to a finite number of points for a random distribution with uniform
density, these deviations arise due to discreteness.
In this case we can relate the deviation (of $D_q$ from $D$) to the number
density of points.
We further show that for a distribution of points with weak clustering, there
is an additional deviation of $D_q$ from $D$.
This deviation can be related to the two point correlation and the intuitive
relation between the amplitude of clustering and deviation from homogeneity can be quantified.
We then apply the derived relation to cosmology and compute the expected
deviations in a model that fits most observations.

A brief outline of this chapter is as follows. In \S\ref{fractaldimension} we describe briefly the method of calculation of fractal dimension from
various measures.  \S\ref{hodi} contains the expression of Minkowski Bouligand dimension for a homogeneous distribution of points. The fractal
dimension of a weakly clustered distribution has been calculated in \S{\ref{wcd}}. The appendices \ref{apphomo} and \ref{appweak} contain the
detailed derivation of the expression for fractal dimension.
\section{\bf Fractal Dimensions}\label{fractaldimension}
Fractal dimension is a basic characterization of any point
distribution.
There are many different methods that can be used to calculate the fractal
dimension.
Box counting dimension of fractal distribution is defined in terms of non
empty boxes $N(r)$ of radius $r$ required to cover the distribution. Cosider a distribution of points in a region. We start by covering this distribution with certain number of spheres $N(r)$ of radius $r$. As the radius of the sphere increases, we note, that the number of the spheres required to cover the distribution of point decreases. Hence their is a scaling relation between the number of spheres and the radius of the sphere. This relation can be represented as
\begin{equation}
 N(r) \propto r^{-D_b}
\label{eq:4}
\end{equation}
where  $D_b$ is defined to be the box counting dimension of the distribution of particles.
One of the difficulties with such an analysis is that it does not depend on
the number of particles inside the boxes and rather depends only on the number
of boxes. As such it provides limited information about the degree of clumpiness of the
distribution and is a purely geometrical measure.
To get more detailed information on clustering of the distribution we need to use higher order moments of the distribution. The simplest of these
moments is the correlation dimension.
Instead of using the formal definition of correlation dimension, which demands
that the number of points in the distribution approach infinity, we choose a
`working definition' which can be applied to a distribution of a finite number
of points.

We consider a distribution of $N$ points. Out of these $N$ points we choose $M$ points and put spheres centered on these $M$ points. Let the number of points lying inside the sphere of radius $r$ centered on $i^{th}$ point be $n_i(r)$. Correlation integral for such a system of points is then given by
\begin{equation}
C_2(r)=\frac{1}{NM}\sum_{i=1}^{M}n_i(r)
\label{eq:5}
\end{equation}
In general the number of points and spheres are different as one cannot use
points near the edge of the sample where a sphere of radius $r$ is not
completely inside the sample.

The term
$n_i(r)$ denotes the number of particles within a distance $r$ from a
particle at the point $i$ :
\begin{equation}
n_i(r)=\sum_{j=1}^{N}\th(r-\mid \x_i-\x_j \mid)
\label{eq:6}
\end{equation}
where $\x_i$ is the position coordinate of the $i^{th}$ particle.  $\th(x)$ is the  Heaviside function given by equation \ref{eq281}.

An alternative way of defining correlation integral $C_2(r)$ is in terms of the probability distribution of particles. In this method we define
$C_2$ in terms of the probability of finding particles in a sphere of radius
$r$ centered on one of the particles in the distribution. Let $P(n;r,N)$ represent the probability of finding $n$ neighbours inside a radius $r$ to any particle in a distribution of $N$ particles. Then  correlation integral $C_2$ can be defined as,
\begin{equation}
C_2(r)=\frac{1}{N}\sum\limits_{n=0}^{N}n P(n;r,N)
\label{eq:7}
\end{equation}
For a homogeneous distribution of points, the probability for any point to
fall within a sphere is proportional to the ratio of the volume of
the sphere to the total volume of the sample.
In such a case $C_2$ reduces to the product of the volume of a sphere of
radius $r$ and the total number of particles, divided by the total volume.
As the total number and total volume are fixed quantities, $C_2$ for a
homogeneous distribution of points scales as $r^{D}$ at sufficiently large
scales, where as before $D$ is the dimension of the ambient space in which the particles are distributed.

For a general point distribution the  power law scaling of correlation integral i.e. $C_2(r) \propto r^{D_2}$
defines the correlation dimension $D_2$ of the distribution.
\begin{equation}
D_2(r)=\frac{\partial \log C_2(r)}{\partial \log r}
\label{eq:78}
\end{equation}
Depending on the scaling of $C_2$, the value of correlation dimension $D_2$
can vary with scale $r$.
For the special case of a homogeneous distribution, we see that $D_2(r) =
D$ at sufficiently large scales and this matches the intuitive expectation
that the correlation dimension of a homogeneous distribution of points should
equal the mathematically defined fractal dimension for infinite number of
points.

We see from equation \ref{eq:7} that the correlation integral is defined in terms of
probability of finding $n$ point out of a distribution of $N$ points
within a distance $r$. This makes it a measure of one of the moments
of the distribution. We need all the moments of the distribution to
completely characterize the system statistically. The multifractal
analysis used here does this with the generalized dimension $D_q$,
the Minkowski-Bouligand dimension, which is defined for an arbitrary
$q$ and typically computed for a range of values. It is different
from Renyi dimension only in the aspect that in this case the
spheres of radius $r$ have been centered at the point belonging to
the fractal whereas in Renyi dimension the sphere need not be
centered on the particle in the distribution
\citep{bor}. However, it can be proved that these two
definitions are completely equivalent \citep{1980ApJ...236..351D}.

The definition of generalized dimension $D_q$ is a generalization of
the correlation dimension $D_2$. In stead of taking the first moment of the distribution as in case of correlation integral, we need to take the ${(q-1)}^{th}$ moment in order to define the generalized integral $C_q(r)$. Consider $n_i(r)$ to be  the number of neighbours within $r$ of the particle $i$. The correlation integral for such a distribution of particles can now be generalized to define $C_q(r)$ as
\be
C_q(r) = \frac{1}{NM}\sum_{i=1}^{M}n_i^{q-1}(r)
\label{cqre}
\ee
Alternatively, in terms of probability distribution function (equation \ref{eq:7}) this can be written as
\bq
C_q(r) &=& \frac{1}{N}\sum\limits_{n=0}^{N}n^{q-1} P_c(n) \nonumber \\
 &=& \frac{1}{N} \langle \mathcal{N}^{q-1} \rangle_c
\label{eq:8}
\eq
The subscript $c$ denotes that spheres have been
centered on points within the distribution. Thus the $q^{th}$ order generalized integral (i.e. $C_q(r)$) is related to the $(q-1)^{th}$ order moment of the distribution. Equation \ref{eq:8} can now be used to define the Minkowski-Bouligand dimension as
\begin{equation}
D_q=\frac{1}{q-1}\frac{d\log{C_q(r)}}{d\log{r}}
\label{eq:9}
\end{equation}
As is obvious from equation \ref{eq:9}, the generalized dimension corresponds to the correlation dimension
for $q=2$. For the case  $q=1$ we can see from equation (\ref{eq:8})
that $C_q(r)$ does not contain any information about the number of
neighbours of the particle which has been taken as center. If we place the center of the sphere randomly, as is done in calculating the Renyi Dimension , then q=1  corresponds to  box counting dimension.
 For $q>2$ the contribution to $C_q$ comes from a range of correlation function ranging from $2$ point correlation function to $q$ point correlation function.  If the distribution of points is a monofractal then we have $D_q = D_2$ for all $q$ and at all scales. However, a multifractal distribution of points can only be described by the full spectrum of $D_q$.

By construction , the positive values of $q$ in the generalized integral (see equation \ref{cqre}) give more weightage to regions with a high number density whereas the negative values of $q$ give more weightage to underdense regions. Thus we may interpret $D_q$  for  $q \gg 0$ as characterizing the scaling behavior of the galaxy distribution in the high density regions like clusters
whereas $q \ll 0$ characterizes the scaling in voids. In the situation where the galaxy distribution is homogeneous and isotropic
on large scales, we intuitively expect $D_q \simeq D=3$ independent of
the value of $q$ at the relevant scales.
\section{\bf Fractal Dimension for a Homogeneous Distribution}\label{hodi}
In our analysis, we first compute the expected values for $C_q$ and $D_q$ for
a homogeneous distribution in a finite volume.
The volume $V_{tot}$ over which the points are distributed is taken to be much
larger than volume of spheres ($V$).
The points are distributed randomly and hence we can use a set of points generated from the Binomial distribution.
The conditional probability of finding $n$ points in a sphere of volume $V$ centered on a
point, if $V_{tot}$ contains $N$ particles is:
\begin{equation}
P_c(n) =  \left({N-1\atop n-1}\right) p^{n-1} \left(1-p\right)^{N-n}
\label{eq:11}
\end{equation}
where $p$ is the probability that a given point (out of $N$) is located in a
randomly placed sphere. This $p$, however, is not the same as the probability of finding only one particle in such a sphere. If we place a sphere of volume $V$ inside a distribution which is contained in
volume $V_{tot}$ then $p = \frac{V}{V_{tot}}$.
. Recalling that we have taken $V_{tot} \gg V$, we shall assume in our calculations that $p \ll 1$.
The above expression \ref{eq:11} follows from the fact that  with the sphere centered on one point, this
point is already in the sphere and we need to compute the probability of $n-1$
points out of $N-1$ being in the sphere.
For comparison, the probability of finding $n$ particles in a randomly placed
sphere of volume $V$ is:
\begin{equation}
P(n) =  \left({N\atop n}\right) p^n \left(1-p\right)^{N-n}
\end{equation}
The average number of points in a randomly placed sphere which we define by $\bar{N}$ is $Np$ and
we assume that this is much larger than unity.
Thus we work in the limit where $1 \ll Np \ll N$.

In order to calculate the generalized correlation integral we need to calculate the higher order moments of the distribution. Moments of the
distribution for spheres centered at points (denoted by$ \langle \mathcal{N}^m \rangle_c$ ) can be related to moments for randomly placed spheres
(denoted by $\langle \mathcal{N}^m\rangle$). We start with the expression for the conditional moment given by
\bq
\langle \mathcal{N}^m \rangle_c &=& \sum n^m P_c(n) \nonumber \\
        &=& \sum n^m \left({N-1 \atop n-1}\right) p^{n-1} \left(1-p\right)^{N-n} \nonumber \\
&=& \sum \left(n - 1 + 1\right)^m \left({N-1\atop n-1}\right) p^{n-1}
\left(1-p\right)^{N-n} \nonumber \\
&\simeq & \sum \left(n - 1\right)^m  \left({N-1\atop n-1}\right)
  p^{n-1} \left(1-p\right)^{N-n}  \nonumber \\
&& + \sum m \left(n-1\right)^{m-1} \left({N-1\atop n-1}\right) p^{n-1}
\left(1-p\right)^{N-n}
\label{rel}
\eq
In the limit $1 \ll Np \ll N$ we can replace $N-1$ by $N$ in the above expression \ref{rel}. So we get the following approximation
\bq
\langle \mathcal{N}^m \rangle_c &\simeq& \langle \mathcal{N}^m\rangle + m\langle \mathcal{N}^{m-1} \rangle
\label{rel1}
\eq
It is once again stressed that subscript $c$ on the angle brackets denotes that the average is for spheres
centered on points within the distribution. A specific application of the above expression is to compute the average
number of points in a spherical region. The average number of points in a sphere centered at a point is $ 1 +
(N-1)p \simeq Np + 1 = \bar{N} + 1$. $\bar{N}$, here, is the average number of points in the randomly placed sphere in the binomial distribution.
The difference between the two expressions for average number of particles arises due to fluctuations that are present in an uncorrelated
distribution of points.

The generalized correlation integral can now be expressed in terms of the
moments of this probability distribution.
In the limit $1 \ll Np \ll N$ we can write down a
leading order expression for the generalized correlation integral for $q > 1$
as:
\begin{equation}
NC_q(r) \simeq {\bar{N}}^{q-1} + \frac{(q-1)(q-2)}{2} {\bar{N}}^{q-2} +
(q-1){\bar{N}}^{q-2} + \cdots
\label{genin}
\end{equation}
(See Appendix \ref{apphomo} for a detailed discussion on how we arrived at this
expression.) Here we have ignored terms that are
of lower order in $\bar{N}$ and terms of the same order in $\bar{N}$ with
powers of $p$ multiplying it.

The Minkowski-Bouligand dimension corresponding to this generalized integral(\ref{genin}) has contribution from the fluctuations in the distribution due to discreetness and also from preferential placing of the spheres on the points in the distribution. We can quantify these contributions by:
\begin{equation}
D_q(r) = D - \frac{\left( q - 2 \right)}{2} \frac{D }{\bar{N}} - \frac{D
}{\bar{N}}
\label{eqn:homogen}
\end{equation}
to the same order as for equation \ref{genin}.
The last two terms in the intermediate expression for $D_q(r)$ have a
different origin: the first of the two terms arises due to fluctuations
present in a random distribution and the second term arises due to the spheres
being centered at points within the distribution and this leads to weak
clustering.
A few points of significance are:
\begin{itemize}
\item
We do not expect $D_q(r)$ to coincide with the $D$ even if the distribution of points is homogeneous. Thus the benchmark for a sample of points is
not $D$ but $D_q(r)$ given by expression \ref{eqn:homogen}, and if the Minkowski-Bouligand dimension for a distribution of points coincides with
$D_q(r)$ then it may be considered as a homogeneous distribution of points.
\item
For $q > 1$, the correction due to a finite size sample always leads to a smaller value for
$D_q(r)$ than the $D$.
\item
The correction is small if $\bar{N} \gg 1$, as expected. The correction arises primarily due to discreteness and has been discussed by
\citet{bor}.
The major advantage of our approach is that we are able to derive an
analytic expression for the correction.
\end{itemize}
\section{\bf Fractal dimension of a Weakly Clustered Distribution}\label{wcd}
In a weakly clustered distributions of points, the counts of number of neighbours, for spheres whose centers are randomly placed and for
those whose centers are placed on the points in the distribution, differ by a significant amount.
Also there is no simple way of relating the two and hence we cannot use the approach we followed in section \ref{hodi} for estimating the
generalized correlation integral.

In order to make further progress, we note that we can always define an
average density for a distribution of a finite number of points in a finite
volume. This allows us to go a step further and also define $n-$point correlation
functions. It is well known that this can be used to relate the generalized correlation
integral with $n-$point correlation functions \citep[ e.g. see][]{bor}. We shall show in Appendix \ref{appweak} that it is possible to considerably simplify this relation, between the generalized correlation integral and two point correlation function, in the limit of weak clustering. We can show that the correlation integral may be written as follows (see Appendix \ref{appweak} for details).
\begin{eqnarray}
NC_q(r) &\simeq& {\bar{N}}^{q-1}
\left( 1 +
  \frac{\left(q-1\right)\left(q-2\right)}{2{\bar{N}}} +
  \frac{q(q-1)}{2}\bar{\xi}\right.          \nonumber \\
&&+ \left. \mathcal{O}\left({\bar\xi}^2\right) +
\mathcal{O}\left(\frac{\bar\xi}{\bar{N}}\right)
+ \mathcal{O}\left(\frac{1}{{\bar{N}}^2}\right)\right)
\label{geninwe}
\end{eqnarray}
Here we have used the assumption that $|\bar\xi| \ll 1$ ({\it weak clustering limit}) and that higher powers
of $\bar\xi$ as well as higher order correlation functions can be ignored when
compared to terms of order $\bar\xi$ and $1/{\bar{N}}$.
This assumption is over and above the limit $1 \ll Np \ll N$.
The first two terms on the right hand side of equation \ref{geninwe} are same as the
first two terms in the expression \ref{genin} for $C_q$ that we derived for a homogeneous
distribution of points.
The third term encapsulates the contribution of clustering.
This differs from the last term in the corresponding expression for a
homogeneous distribution as in that case the ``clustering'' is only due to
spheres being centered at points whereas in this case the locations of every
pair of points has a weak correlation.
It is worth noting that the highest order term of order
$\mathcal{O}\left({\bar\xi}^2\right)$ has a factor
$\mathcal{O}\left(q^3\right)$ and hence can become important for sufficiently
large $q$. {\it This may be quantified by stating that $q\bar\xi \ll 1$ is the more
relevant small parameter for this perturbative expansion.}

The Minkowski-Bouligand dimension for such a system can now be expressed in
the form
\begin{eqnarray}
 D_q(r) &\simeq& D -
\frac{D \left( q - 2 \right)}{2\bar{N}} +
\frac{q}{2} \frac{\partial{\bar\xi}}{\partial{\log{r}}}   \nonumber \\
&=& D  -
\frac{D \left( q - 2 \right)}{2\bar{N}} -
\frac{Dq}{2} \left(\bar\xi(r) - \xi(r) \right) \nonumber  \\
&=& D - \left(\Delta{D_q}\right)_{\bar{N}} -
  \left(\Delta{D_q}\right)_{clus}
\label{eqn:clus}
\end{eqnarray}
It is interesting to see that the departure of $D_q$ from $D$ due to a
finite sample and weak clustering is given by distinct terms at the leading
order.
This expression allows us to compute $D_q$ for a distribution of points if the
number density and $\bar\xi$ are known.

Recall that $D$ is the mathematically defined fractal dimension for an
infinite set of points with a homogeneous distribution.
We have already noted some aspects of the correction due to a finite number of
points in the previous section, here we would like to highlight aspects of
corrections due to clustering.
\begin{itemize}
\item
For hierarchical clustering, the last two terms in equation \ref{eqn:clus} have the same sign and lead to a
smaller value for $D_q$ as compared to $D$.
\item
Unless the correlation function has a feature at some scale, smaller
correlation corresponds to a smaller correction to the Minkowski-Bouligand
dimension.
The expression given above quantifies this intuitive expectation.
\item
Note that for $q=2$, the expression given here is exact.  For this case, the
contribution of clustering has also been discussed by \citet{1998MNRAS.298.1212M}.
\item
If the correlation function has a feature then it is possible to have a small
correction term $\left(\Delta{D_q}\right)_{clus}$ for a relatively large
$\xi$.
The relation between $\xi$ and $\left(\Delta{D_q}\right)_{clus}$ is not longer
one to one.
\end{itemize}

Our results as given in equation \ref{eqn:homogen} and \ref{eqn:clus} are completely general and apply to any distribution of points with weak clustering.
In chapter \ref{chap4} we shall be using the example of galaxy clustering in concordance model of cosmology to illustrate our calculations.

\newpage
\begin{appendices}
\chapter{}
\section{\bf Derivation of $D_q$ for Homogeneous Distribution}\label{apphomo}
In a homogeneous distribution of points the probability for a point to be found in a sphere of volume $V$ enclosed
within a total volume of $V_{tot}$ is $p=V/V_{tot}$.
In an uncorrelated distribution of points, the probability of finding a point is independent of the location of other points in the distribution  and hence the probability of $n$ out of $N$
points falling in a sphere of volume $V$ is:
\begin{equation}
  P(n,N) = \left({N\atop n}\right)p^n (1-p)^{N-n}
  \label{appa1}
\end{equation}
The distribution function determined by the probability function $P(n,N)$ is
called a Binomial distribution.
As discussed in the text, the probability distribution for occupation number
of spheres centered at points (i.e. $P_c(n,N)$), as given by equation \ref{eq:11}) is different from the one given in equation \ref{appa1}. However,
moments of the distribution in both these cases can be related in the limit $N \gg Np \gg 1$ as shown by expression \ref{rel1}.
We will be interested in the description being accurate to first order in
$1/\bar{N}$.

We start with the moment generating function for the Binomial distribution given by
\begin{equation}
G(t)=\sum_{n=0}^N e^{tn}\left({N\atop n}\right)p^n(1-p)^{N-n}  = (pe^t+1-p)^N
\end{equation}
The $m^{th}$ moment of this distribution can then be calculated by differentiating
$G$, $m$ times with respect to $t$, and then setting $t$ to zero.
The $m^{th}$ derivative of $G(t)$, at $t=0$ can be written as:
\begin{equation}
G^{(m)}(t)\left|_{t=0}\right. = \sum_{l=1}^{m}H_{m,l}\frac{N!}{(N-l)!}p^l
\end{equation}
where $H$ satisfies the following recurrence relation
\begin{equation}
H_{m,l}=l H_{m-1,l} + H_{m-1,l-1}
\label{rec}
\end{equation}
with  the assumption that $H_{1,1}=1$ and $H_{m,l}=0$ for $l>m$ and $l<1$.
It can now be shown that for all $l$,
\[
H_{l,l}=1
\]
The $m^{th}$ moment of the distribution is thus given by:
\begin{eqnarray}
\langle \mathcal{N}^m \rangle &=& G^{(m)}(t)\left|_{t=0} \right. \nonumber \\
 &=& \sum_{l=1}^{m} H_{m,l}\frac{N!}{(N-l)!}p^l \nonumber
\end{eqnarray}
On the face of it this expression has a large number of terms for $m \gg 1$
and is difficult to analyze.
But if we assume that $p \ll 1$ and $\bar{N} = Np \gg 1$ then we can rewrite
the expression in the following form:
\begin{eqnarray}
\langle \mathcal{N}^m \rangle &=& H_{m,m}\frac{N!}{(N-m)!}p^{m}
  +H_{m,m-1}\frac{N!}{(N-m+1)!}p^{m-1} + \cdots \nonumber \\
 &=& \bar{N}^{m} + \mathcal{O}(p\bar{N}^{m-1})
+ H_{m,m-1}\bar{N}^{m-1} +  \mathcal{O}(p\bar{N}^{m-2}) +
\mathcal{O}(\bar{N}^{m-2}) \nonumber \\
&\simeq& \bar{N}^{m}+\frac{m\left(m-1\right)}{2}\bar{N}^{m-1}
\label{appa}
\end{eqnarray}
Here we have retained terms up to $\mathcal{O}(\bar{N}^{m-1})$ and have
dropped all other lower order terms. We have also used the recurrence relation \ref{rec} and find that $H_{m,m-1}=
m(m-1)/2$.

We can now write the expression for the generalized correlation integral with the help of equation \ref{eq:8}, \ref{rel1} and \ref{appa}  as:
\begin{eqnarray}
NC_q(r) &=&  \langle \mathcal{N}^{q-1} \rangle + (q-1)\langle \mathcal{N}^{q-2}
\rangle \nonumber \\
&\simeq& {\bar{N}}^{q-1} + \frac{(q-1)(q-2)}{2} {\bar{N}}^{q-2}
+ (q-1){\bar{N}}^{q-2} + \cdots
\label{eq:app1}
\end{eqnarray}

The Minkowski-Bouligand dimension for a homogeneous distribution is then given by
\begin{eqnarray}
D_q(r)&=&\frac{1}{q-1}\frac{\partial \log C_q(r)}{\partial \log r} \nonumber \\
 &\simeq& \frac{1}{q-1}\frac{\partial }{\partial \log r}
 \log\left[\bar{N}^{q-1}\left(1 +
     \frac{\left(q-1\right)\left(q-2\right)}{2\bar{N}}
   \right)
+ \frac{\left(q-1\right)}{\bar{N}} \right]
 \nonumber \\
&\simeq& D\left(1-\frac{(q-2)}{2\bar{N}} - \frac{1}{\bar{N}}\right)
\label{eq:app3}
\end{eqnarray}
where $D$ is the dimension of the space in which particles are
distributed. This is the relation we have used in equation \ref{eqn:homogen} to describe a homogeneous distribution of points.
In this calculation, we have again made use of the fact that $\bar{N} \gg 1$
and that it scales as the $D^{th}$ power of scale $r$ for a random
distribution such that
\be
D = \frac{\partial  \log \bar{N} }{\partial \log r} \nonumber
\ee
\section{\bf Derivation of $D_q$ for a Weakly Clustered Distribution}\label{appweak}
In this section we will derive the form of the correlation integral for a
weakly clustered distribution of points.
Consider a sphere of volume $V$ contained within the sample of volume
$V_{tot}$.
We follow the approach given in \S{36} of \citet{1980lssu.book.....P} for
estimating the correlation integral.
In order to estimate the correlation integral, we divide the sphere into
small elements such that each element contains at most one point.
This is a useful construct because in such a case we have $n_i^m=n_i$ for all $m \geq 0$, where $n_i$ is
the number of points occupying the $i^{th}$ infinitesimal volume element.
If the occupancy of the $i^{th}$ volume element is $n_i$ then we have the total number of particle in this volume given by:
\begin{equation}
\mathcal{N} = \sum\limits_i n_i
\end{equation}
We can also define the mean count of such a distribution as:
\begin{equation}
\left\langle \mathcal{N} \right\rangle = \left\langle \sum\limits_i n_i
\right\rangle = \bar{N}
\end{equation}
Knowing the mean, the $m^{th}$ moment of the distribution is given by:
\begin{equation}
\left\langle {\mathcal{N}}^m \right\rangle = \left\langle \left(\sum\limits_i
    n_i \right)^m
\right\rangle
\end{equation}
Since we are not interested in number of galaxies in a region, but are rather interested in the number of galaxies around a galaxy, we have to take care that the spheres that we throw in the sample are centered on the points of the sample. If the sphere is centered at a point in the distribution then the average number of points, and hence the different order moments of the distribution, are
denoted as $<\mathcal{N}^m>_c$. This is what we are interested in for the
purpose of computing the correlation integral.
\begin{equation}
\left\langle \mathcal{N}^m \right\rangle_c = \left\langle \left(\sum\limits_i
    n_i  \right)^m \right\rangle_c
\end{equation}
Averaging the sum raised to a positive integral power will lead to averaging
of terms of type $n_i n_j$, $n_i n_j n_k$, etc. and the expression for such
terms involves $n$-point correlation functions, $n$ being the number of terms
being multiplied.
With this insight, we can write
\begin{eqnarray}
\left\langle \mathcal{N}^m \right\rangle_c  &=& \left\langle \sum n_1^m
\right\rangle_c + m \left\langle \sum n_1^{m-1} n_m
\right\rangle_c   + \cdots
+ \frac{m(m-1)}{2} \left\langle
  \sum n_1^2 n_3 \ldots n_m \right\rangle_c  \nonumber \\
&& + \left\langle
  \sum n_1 n_2 n_3 \ldots n_m \right\rangle_c \nonumber \\
&=& \left\langle \sum n_1
\right\rangle_c + m \left\langle \sum n_1 n_m
\right\rangle_c   + \cdots
+ \frac{m(m-1)}{2} \left\langle
  \sum n_1 n_3 \ldots n_m \right\rangle_c \nonumber \\
&& + \left\langle
  \sum n_1 n_2 n_3 \ldots n_m \right\rangle_c
\label{eq:app4}
\end{eqnarray}
Here the terms in the expansion correspond to all indices being equal for the first term,
only one of the indices differing from the rest for the second term and so
on. The last term in this series is for all the $m$ indices being different.
We have shifted the notation in order to write down the explicit form for
arbitrary $m$. It is interesting to see that for $2^{nd}$ order moment of the distribution, the expression \ref{eq:app4} reduces to
\begin{eqnarray}
    \langle\mathcal{N}^2\rangle_c &=& \langle\sum\limits_i n_i^2\rangle_c
    +\langle\sum\limits_{i,j} n_i n_j\rangle_c \nonumber \\
    &=& \langle\sum\limits_i n_i\rangle_c
    +\langle\sum\limits_{i,j} n_i n_j\rangle_c \nonumber \\
    &=& \bar{n}V + \bar{n}\int{\xi(r)\delta{v}}
    +{\bar{n}}^2\int{\delta{v_1}\delta{v_2}(1+\xi(r_1)+\xi(r_2)+\xi(r_{12})+\zeta(r_1,r_2))}
    \nonumber
  \end{eqnarray}
   here $\zeta$ is the reduced three point correlation function given by equation \ref{3pcf}. $V$ and $\delta v$ represent the total sample volume and the volume of the sphere that has been thrown in the sample. $\bar{n}$ is the average number density (i.e. $\bar{n}=\bar{N}/V$).

For a weakly clustered set of points with statistical isotropy and
homogeneity, we can safely assume that on large scales the magnitude of the two point
correlation function is small compared to unity, and higher order correlation
functions are even smaller. Keeping this in mind we get the
  second moment of this weakly clustered distribution as
  \begin{eqnarray}
    \langle\mathcal{N}^2\rangle_c &=& \bar{n}V + {\bar{n}}^2V^2+\bar{n}\int{\xi(r)\delta{v}}
    +3{\bar{n}}^2\int{\delta{v_1}\delta{v_2}(\xi(r))} \nonumber \\
    &=& {\bar{N}}^2 \left(1 +\frac{3}{V}\int \xi(r)  \delta v +\frac{1}{\bar{N} V}\int  \xi(r) \delta v +\frac{1}{\bar N} \right)
    \label{n2c}
  \end{eqnarray}
Further, we continue to use the assumption that $\bar{N} \gg 1$ and hence we
need to retain only terms of the highest and the next highest order in this
parameter.
Thus we have two small parameters in the problem: $\xi$ and $1/\bar{N}$ and
our task is to compute the leading order terms in $\left\langle \mathcal{N}^m
\right\rangle_c$. Here $\xi$ is the two point correlation function.

It can be shown that the leading order contribution to $\left\langle \mathcal{N}^m
\right\rangle_c$ comes from the last term in the series in Equation (\ref{eq:app4}), and the next to leading order
contribution is from the last two terms. We should note that these terms also contain several terms that are smaller
than the leading and next to leading order within them.

The foremost contribution comes from the uncorrelated component of the last term, i.e., ${\bar{n}}^m \int \delta v_1 \delta v_2 \ldots \delta v_m = {\bar{N}}^m$. The integral here is over $m$ independent volumes and $\bar{n}$ is the average number density.  The corresponding term for $m=2$ is the first term in expression \ref{n2c}.

The next contribution comes from components of this term that include the
effect of pairwise correlations. As there are $m$ distinct points in the last term of expression \ref{eq:app4}, the number of distinct pairs is $m(m+1)/2$. It implies that the contributing term due to the effect of  pairwise correlation has the form:
\begin{equation}
{\bar{N}}^m \frac{m(m+1)}{2} \bar\xi(r)
\end{equation}
where $r$ is the radius of the sphere with volume $\delta v$ and $\bar\xi$ is given
by:
\begin{equation}
\bar\xi(r) = \frac{3}{r^3} \int\limits_0^r x^2 \xi(x) dx ~~~~~~~ .
\end{equation}
The corresponding term for $m=2$ is the second term in expression \ref{n2c}. It can be shown that all other components of the last term involve higher order correlation functions, or higher powers of $\xi$.

Further, it can be shown that the contributions that contain only a single
power of $\xi$ from other terms in the series in Equation (\ref{eq:app4}) contain a
lower power of $\bar{N}$ {\it i.e.} it is of ${\bar{N}}^{m'} \bar\xi$ form where $m'$ ranges from $0$ to $m-1$.
Lastly, the only other term that we need to take into account comes from the penultimate term in the series in Equation (\ref{eq:app4}). The uncorrelated component of this term is:
\begin{equation}
\frac{m(m-1)}{2} {\bar{N}}^{m-1}
\end{equation}
Thus we have for the $m^{th}$ moment of the counts of neighbors:
\begin{eqnarray}
\left\langle \mathcal{N}^m \right\rangle_c &=& \bar{N}^m + \frac{m(m+1)}{2}
\bar{N}^m \bar{\xi}+\frac{m(m-1)}{2} \bar{N}^{m-1}  \nonumber \\
&& + \bar{N}^m \left(\mathcal{O}\left({\bar\xi}^2\right) +
  \mathcal{O}\left(\frac{\bar\xi}{\bar{N}}\right)
+ \mathcal{O}\left(\frac{1}{{\bar{N}}^2}\right) \right) \nonumber \\
&\simeq&
\bar{N}^m\left(1+\frac{m(m+1)}{2}\bar{\xi}+\frac{m(m-1)}{2\bar{N}}\right)
\label{eq:appb2}
\end{eqnarray}
The largest term of order $\mathcal{O}\left({\bar\xi}^2\right)$ arises from
the contribution of correlated triangles in the last term of
Equation (\ref{eq:app4}).
The number of triangles scales as $m^3$ and hence can become important for
sufficiently large $m$.
This may be codified by stating that $m\bar\xi \ll 1$ is the more relevant
small parameter.

Hence the correlation integral (\ref{eq:8}) can be written as
\begin{equation}
NC_q(r) \simeq \bar{N}^{q-1}\left(1+\frac{q(q-1)}{2}\bar{\xi}
  +\frac{(q-1)(q-2)}{2\bar{N}}\right)
\label{eq:appb3}
\end{equation}
From this we can calculate the Minkowski-Bouligand dimension using equation
\ref{eq:app3} as
\begin{eqnarray}
 D_q(r) &=& \frac{1}{(q-1)} \frac{\partial\log C_q(r)}{\partial\log
   r}      \nonumber \\
&\simeq& D -
\frac{D \left( q - 2 \right)}{2\bar{N}} +
\frac{q}{2} \frac{\partial{\bar\xi}}{\partial{\log{r}}}   \nonumber \\
&=& D  -
\frac{D \left( q - 2 \right)}{2\bar{N}}   - \frac{D q}{2} \left(\bar\xi(r) - \xi(r) \right)
\end{eqnarray}
This is the required expression for the fractal dimension of a weakly clustered distribution and has been used in text as equation \ref{eqn:clus}.
\end{appendices} 

\chapter{Fractal Dimension as a measure of Homogeneity}\label{chap4}
There are various observational probes which test the cosmological principle.
The near isotropicity of Cosmic Microwave background Radiations (CMBR) tell us that our space time locally is very well described by $FRW$ metric.
Similarly the CMBR anisotropies at large angular scales ($~10^\circ$) imply that density fluctuations in the distribution of matter in the
Universe is very small on large scales ($\delta\rho/\rho \sim 10^{-4}$ on $1000 \,h^{-1}Mpc$). The absence of big voids in the distribution of Lyman
$\alpha$ absorbers is also consistent with the Universe being Homogeneous on large scales. In this chapter\footnote{This chapter is based on {\it Fractal Dimensions of a Weakly Clustered Distribution and the Scale of Homogeneity} (J.S. Bagla, Jaswant Yadav \& T.R. Seshadri), Monthly Notices of Royal Astronomical Society {\bf 390}, 829} we have tested the large scale homogeneity of the
homogeneous as well as slightly clustered distribution of points using {\it multifractal analysis} developed in chapter $3$. The distribution is
said to be homogeneous when the multifractal dimension of the point distribution is same for all moments and is equal to the ambient dimension of the space in which the points
are distributed. We have also applied our analysis to the distribution of various types ($L_*$ as well as $LRG$) of galaxies in the concordance
model of cosmology. We see that in the concordance model, the fractal dimension makes a rapid transition to values close to $3$ at scales between
$40$ and $100$~Mpc.
\section{Introduction}
Various groups have used the concept of fractals to analyze catalogues of
extra-galactic objects.
See \citet{2005RvMP...76.1211J} for an excellent review of quantitative
measures used for describing distributions of points.
Based on the scale invariance of galaxy clustering,
\citet{1987PhyA..144..257P} suggested that the distribution of galaxies is a
fractal to arbitrarily large  scales.
In a later analysis of different samples of galaxies \citet{coleman} obtained results consistent with this argument.
On the other hand \citet{bor} showed that the distribution is a
fractal only on small scales and on large scales there is a transition to
homogeneity.
If the distribution of galaxies is found to be a fractal then the average
number of galaxies in a volume of radius $r$ centred on a galaxy should scale
as $r^{d}$, where $d$ is the fractal dimension.
Hence, the number density of neighboring galaxies would go as $\rho = r^{d -
  D}$ in a $D$ dimensional distribution.
This, when calculated for higher values of $r$ will show a decrease compared to its value for lower scales.
This effect led \citet{1998PhR...293...61S} to believe that the value of
correlation length $r_0$ (see equation \ref{eq:1}) increases with the increase in size
of the sample.
However, this interpretation is not supported by volume limited samples of
various galaxy redshift surveys \citep{1996ApJ...472..452B, martinez}.

A number of authors
\citep[{e.g.}][]{cappi, haton, best,amen, baris}
have shown the distribution of galaxies to be a mono-fractal up to the largest
scales that they were able to analyze.
On the other hand homogeneity has been seen at large scale in other analysis \citep[see {\it e.g.}][]{gujo,bharad1, 1999Sci...284..445M,kur,pan,hog}.
The best argument in favour of large scale homogeneity stems from the near
isotropy of radio sources or background radiation in projection on the sky \citep{wu}.

We have developed a model (equation \ref{eqn:homogen} and \ref{eqn:clus}) which predicts the scale of homogeneity for a homogeneous distribution of points as well as for a slightly clustered distribution of points. In order to test our model we have applied it to a multifractal multinomial distribution of points for which the fractal dimension is analytically known. We have also applied our model to the concordance model of cosmology. In this case we have considered unbiased (e.g. $L_*$ galaxies) as well as biased (e.g. Large Redshift Galaxies (LRG)) tracers of the underlying dark matter distribution, to test the scales at which these distributions attain homogeneity. For this purpose we have taken the large scale two point correlation function for a flat \lcdm~ model with a power law initial power spectrum that best fits the observations. We have used this correlation function to see how the clustering in the distribution of galaxies is going to affect the scale of homogeneity. We have also studied the effect of the bump in the correlation function on the behaviour of the fractal dimension of a clustered galaxy distribution.

A brief outline of this chapter is as follows. In section \ref{mmd} we discuss the application of our model to the multifractal multinomial distribution. The discussion about the scale of homogeneity of different types of galaxies follows in section \ref{discussion4}. We end this chapter with a list of conclusions in section \ref{conclusion4}.
\section{\bf Multifractal Multinomial Distribution}\label{mmd}
We have applied our method to the multinomial multifractal model
discussed in literature \citep[See e.g.][]{2002sgd..book.....M}.
The set of points for this model can be generated by starting with a
square and dividing it into four parts.
We assign a probability $\{f_i\}$ to each of these subsquares
$\left(\sum\limits_{i=1}^{4}f_i=1\right)$. At the second step each of the small squares is again subdivided into $4$ parts thus getting 16 squares. The probability attached to each one of these 16 squares is the product of the probability of the individual square (one of the $f_i$'s) multiplied by the probability of its parent square. This construction can be continued iteratively by dividing each smaller square further and assigning probability by multiplying the corresponding number ${f_i}$ by all its ancestors.
We performed this construction to $L=8$ levels, thus getting a $256^2$ lattice
with the corresponding probability measures associated with each pixel of lattice. As an illustration we have performed several realisations of this process for different choices of the initial parameters :
\begin{table}[h]
\begin{center}
\begin{tabular}{l l l l l}
  \hline
  Model & $f_1$ & $f_2$ & $f_3$ & $f_4$ \\ \hline
  $I$ & $0.25$ & $0.25$ &$0.25$ &$0.25$  \\
$II$ & $0.23$ & $0.27$ &$0.25$ &$0.25$  \\
$III$ & $0.15$ & $0.20$ &$0.30$ &$0.35$  \\
$IV$ & $0.05$ & $0.50$ &$0.35$ &$0.15$  \\
  \hline
\end{tabular}
\caption{Generation of various multinomial multifractal Distribution}
\label{tab2}
\end{center}
\end{table}

\begin{figure}
\begin{center}
\includegraphics[height=7.2truein,width=9.2truein]{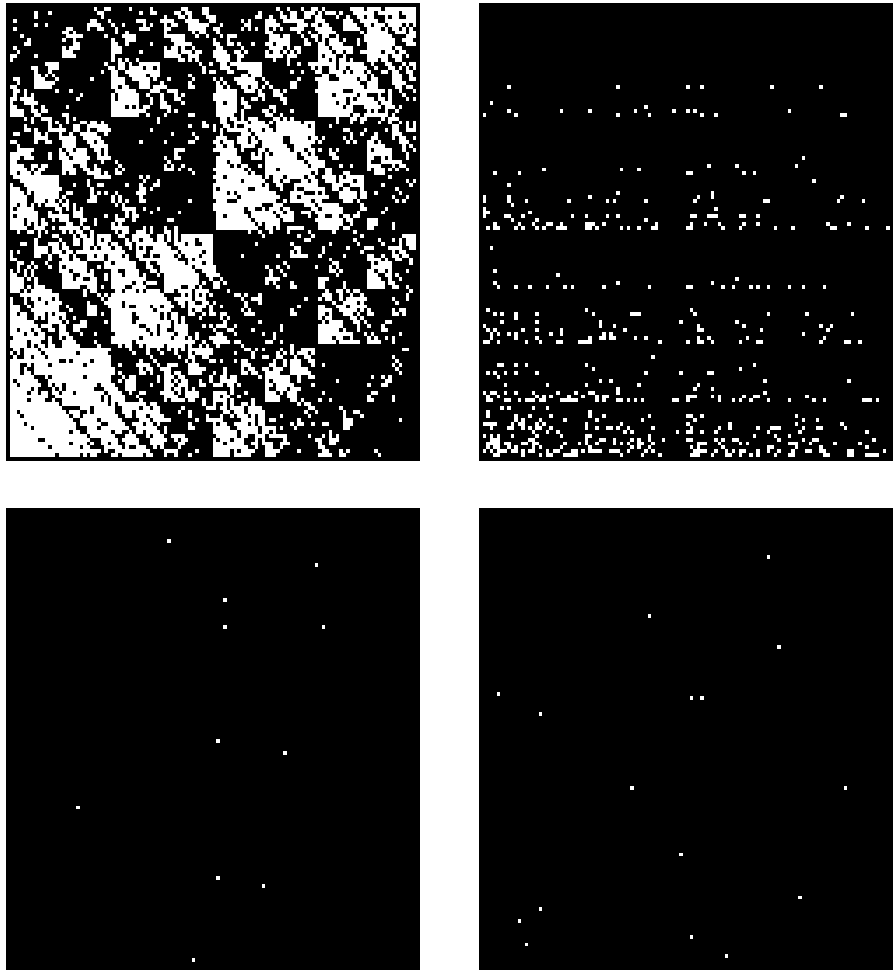}
\caption[\sf Multifractal Multiplicative cascade distribution of points]{\sf Multifractal multiplicative cascades with parameters given in table \ref{tab2}. Models are $I$, $II$, $III$ and $IV$ from bottom to top and left to right}
\label{multip}
\end{center}
\end{figure}

These realisations are shown in figure \ref{multip} on page \pageref{multip}.
For such models we have an analytical expression for the generalized
dimension \citep{fal90,1990ApJ...357...50M} given by
\begin{equation}
D_q=\frac{1}{q-1}\log_2\left(\sum\limits_{i=1,{f_i}\neq 0}^{4} {f_i}^q \right)
\label{eq38}
\end{equation}
As equation \ref{eq38} is an analytical expression, it can be used to check whether our model for finite number and
correlation work correctly or not.

\begin{figure}
\begin{center}
\includegraphics[width=5.2truein]{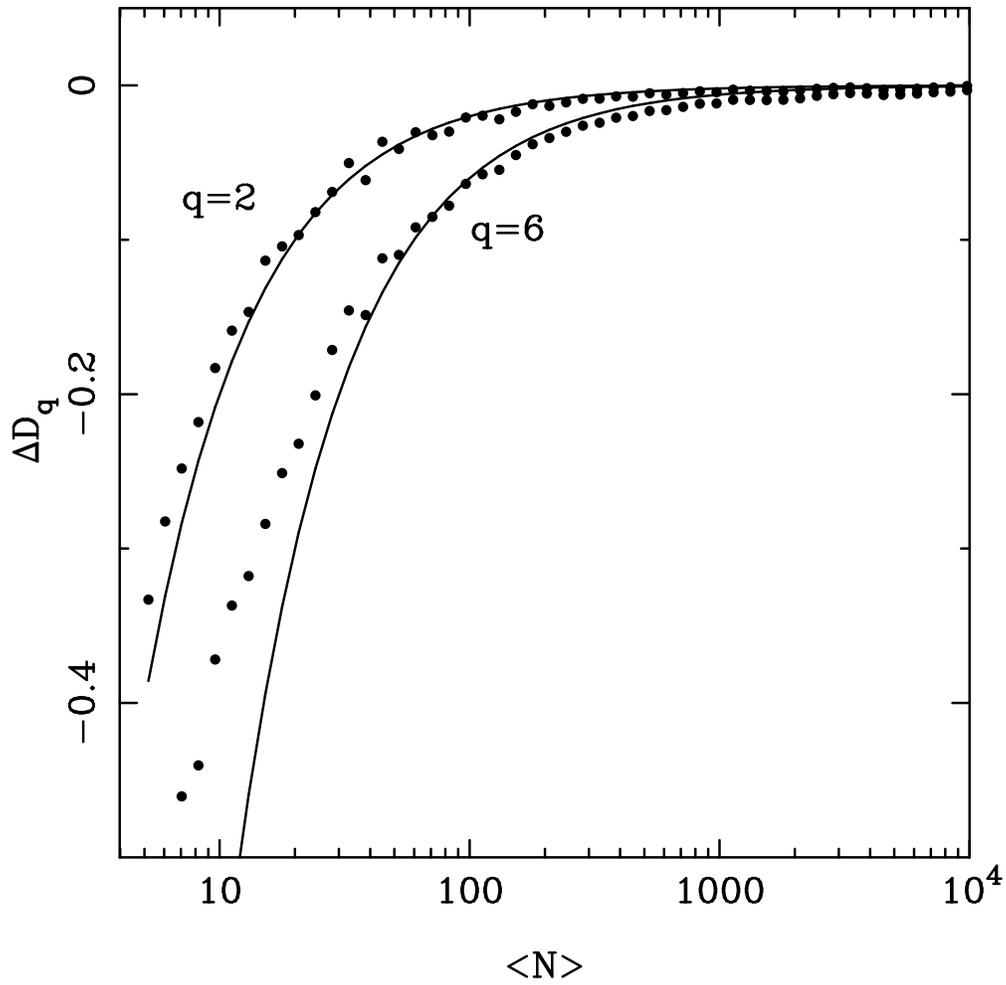}
\caption[\sf Comparison of our model with observed fractal dimension for a random 	
Distribution of points]{\sf Our model 	
is compared with the observed Fractal dimensions for a
  random distribution of points in the special case of the multinomial model (model $I$).
  $\Delta D_q$ is shown as a function of $\langle N \rangle \equiv \bar{N}$
  for $q=2$ and $6$ for this distribution.  $\Delta D_q$ measured from a
  realization are plotted as points, and our model is shown as a curve.}
\label{fig1_chap4}
\end{center}
\end{figure}

We have calculated the generalized dimension for this model taking four
different combination of ${f_i}$.
In one of the cases all four ${f_i}$'s are $0.25$ so that the distribution is
homogeneous.
In this case the expected $D_q = 2$ for all $q$ using the above expression.
Our model in this case gives a scale dependent correction to this due to a
finite number of particles.
Figure \ref{fig1_chap4} on page \pageref{fig1_chap4} shows $\Delta D_q =D_q - D$, as a function of $\langle N \rangle \equiv
\bar{N}$ for $q=2$ and $6$ for this distribution.
$\Delta D_q$ measured from a realization are plotted as points, and our model
is shown as a curve.
It is clear that for $\bar{N} \leq 10^3$, there is a visible deviation of
$D_q$ from the expected value and that our model correctly estimates this
deviation.

\begin{figure}
\begin{center}
\includegraphics[width=5.2truein]{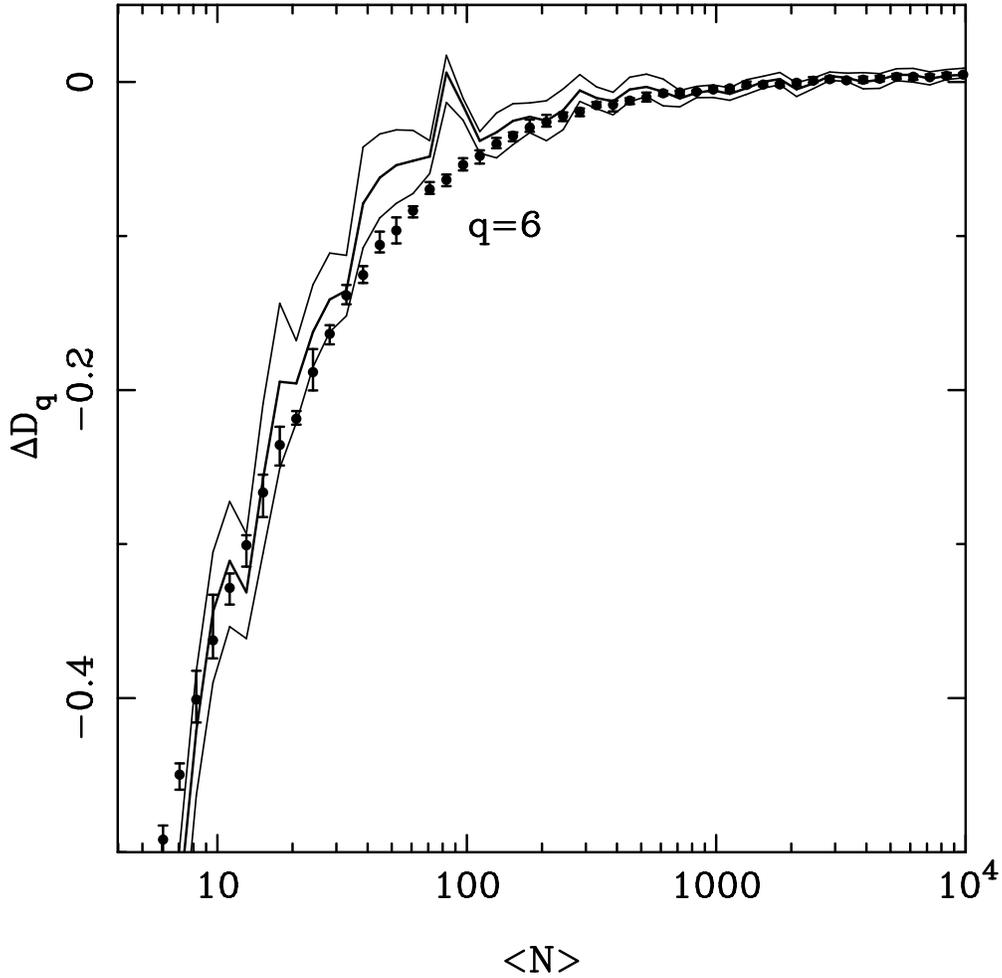}
\caption[\sf Comparison of our model with observed fractal dimension for a weakly clustered  	
distribution Distribution of points]{\sf Our model is compared with the observed Fractal dimensions for a
multinomial fractal with $f_i = {0.23,0.27,0.25,0.25}$ (model $II$).  $\Delta D_q \equiv
D_q - D_{q\, exp} $ is shown as a function of $\langle N \rangle$ for $q=6$.
We have plotted $\Delta D_q$ measured in the five realizations as points with
error bars.  The error bars mark the extreme values of $\Delta D_q$ seen in
these realizations whereas the central point marks the average value.
Predictions of our model based on correlation function measured in these
realizations is shown as a thick line.  Thick line corresponds to the average
value of $\xi$ and $\bar\xi$ measured in simulations, and thin lines mark the
predictions of our model based on extreme values seen in these simulations.}
\label{fig2_chap4}
\end{center}
\end{figure}

We now simulate a weakly clustered distribution of points by taking the values of $f_i$ slightly different from one another. More specifically, we present here, an example where the ${f_i}$'s are close to $0.25$ but not
exactly equal to $0.25$. This construction provides us a slightly clustered distribution.
We used parent $f_i$ to be $0.23,0.27,0.25$ and $0.25$ and generated five realizations of
this fractal. In this case, the expected $D_q = 1.986$ for $q=6$.
As this differs from $D=2$, the difference in our model must come from
clustering present in this fractal.
Figure \ref{fig2_chap4} on page \pageref{fig2_chap4} shows $\Delta D_q \equiv D_q - D_{q\, exp} $ as a function of
$\langle N \rangle$ for $q=6$, where $D_{q\, exp} $ follows from Equation \ref{eq38}.

We have plotted $\Delta D_q$ measured in the five realizations as points with
error bars. The error bars mark the extreme values of $\Delta D_q$ seen in these
realizations whereas the central point marks the average value.
Predictions of our model (Equation \ref{eqn:clus}) based on correlation function measured in
these realizations is shown as a thick line.
This line corresponds to the average value of $\xi$ and $\bar\xi$ measured in
simulations, and thin lines mark the predictions of our model based on extreme
values seen in these simulations.
At $\langle N \rangle \ll 100$, where the effect of a finite number is
dominant, our model matches the measured $\Delta D_q$ very well.
At $\langle N \rangle \gg 100$ where the effect of clustering is dominant we
again find a good match between the model and measured values.
It is significant that at very large $\langle N \rangle$, the deviation of $D_q$ from $D=2$ is correctly accounted for in our model.

However, there appears to be a mismatch in the transition region around
$\langle N \rangle \simeq 100$.
On inspection, we find that $\xi - \bar\xi$ has an oscillatory behavior up to
this scale and the discrepancy corresponds to the last oscillation.
At the scale of maximum discrepancy, $\xi - \bar\xi \simeq 0.05$ and perhaps
we cannot ignore values of this order.

In summary we can say that our model works very well for the multinomial model
and we find that the correction due to clustering as well as a finite number
of points matches with the observed behavior of $D_q$.

\section{\bf Discussion}\label{discussion4}
\begin{figure}
\begin{center}
\includegraphics[width=5.2truein]{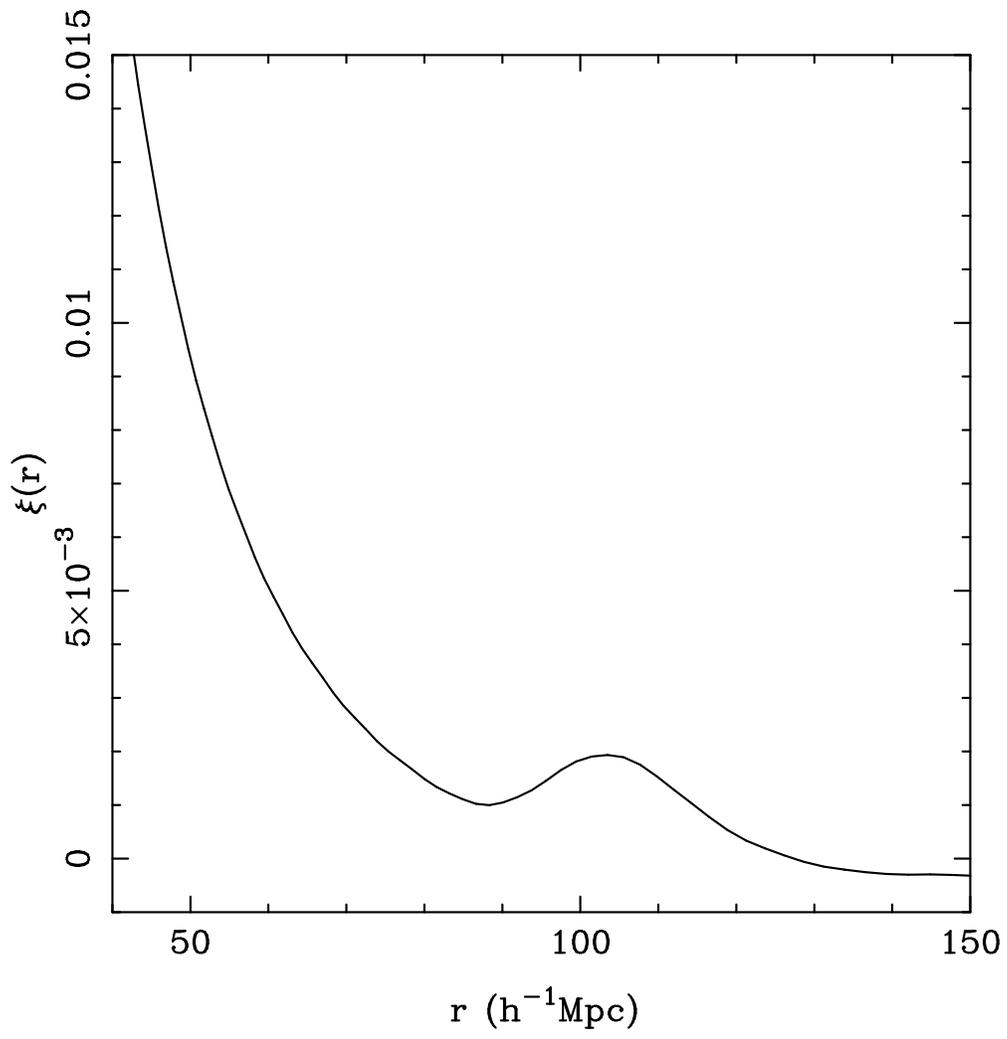}
\caption[\sf Linearly extrapolated two point correlation function for the best fit model for WMAP-3]
{\sf The linearly extrapolated two point correlation
  function is shown as a function of scale for the best fit model for WMAP-3
  (see section \ref{discussion4} for details).  This has been used, for calculation of $(\Delta
  D_q)_{clus}$.}
  \label{fig3_chap4}
  \end{center}
\end{figure}
The expressions of $D_q$ derived in chapter \ref{chap3} in section \ref{hodi} and \ref{wcd} have a rich structure and we illustrate some of the features here. We would also like to discuss the application to the concordance model here.
The two point correlation function for the model that fits best the WMAP-3
data \citep{2007ApJS..170..377S} is shown in Figure \ref{fig3_chap4} on page \pageref{fig3_chap4}.
We have used the flat $\Lambda$CDM model with a power law initial power
spectrum that best fits the WMAP-3 data here. Parameters of the model used here are: $H_0=73$~km/s/Mpc, $\Omega_b h^2 = 0.0223$, $\Omega_c h^2 = 0.105$, $n_s=0.96$ and $\tau=0.088$.
For this model, $\sigma_8=0.76$. The two point correlation has been shown at large scales where the clustering
can be assumed to be weak. The most prominent feature here is the peak near $100$ Mpc.
This peak is caused by baryon acoustic oscillations (BAO) prior to
decoupling \citep[see, e.g.,][]{1998ApJ...496..605E}.
Apart from this peak, the two point correlation function declines from small
scales towards larger scales at length scales shown here.
\begin{figure}
\begin{center}
\includegraphics[width=5.2truein]{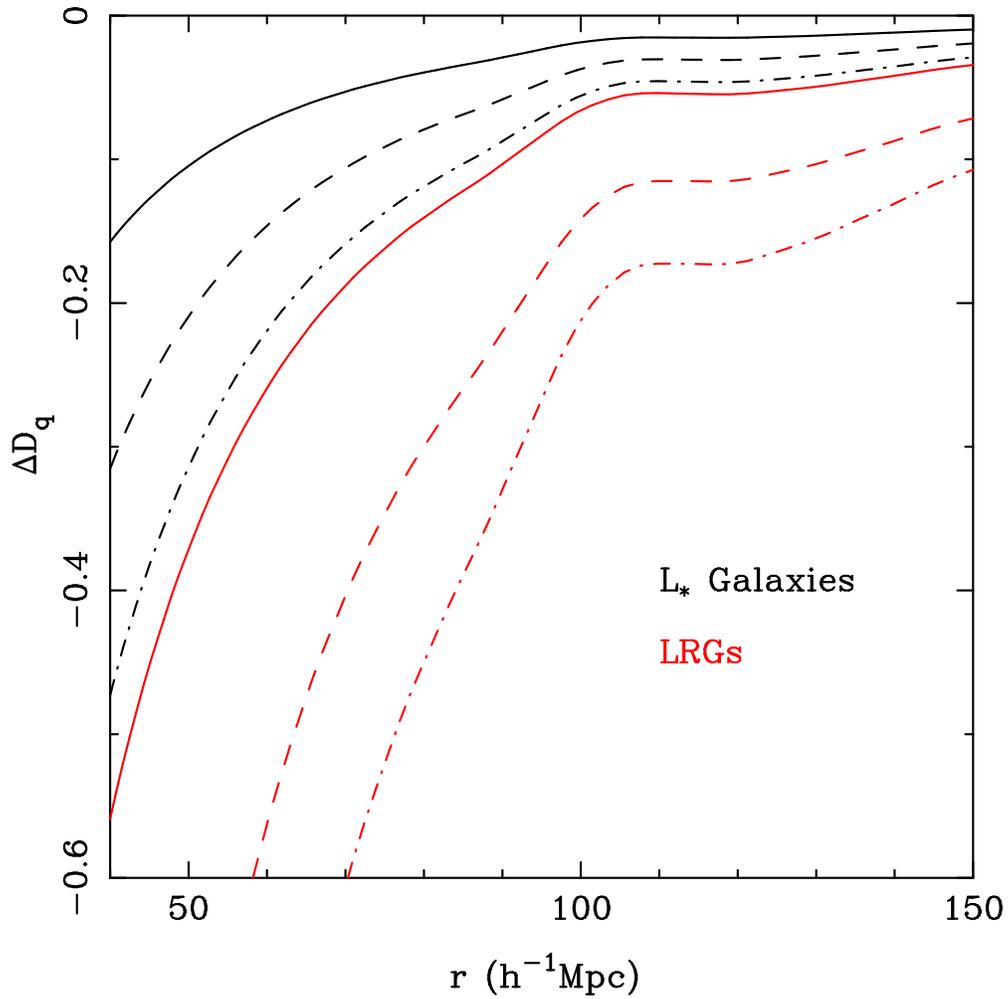}
\caption[\sf Estimated deviation of the Minkowski-Bouligand dimension from the
  physical dimension]{\sf Estimated deviation of the Minkowski-Bouligand dimension from the
  physical dimension is shown here for two types of populations.  In black we
  have plotted $\Delta D_q$ for an unbiased sample of points, distributed in
  redshift space with the real space correlation function as shown in
  Figure \ref{fig1_chap4}.  The solid curve shows $\Delta D_2$, whereas $\Delta D_4$ and
  $\Delta D_6$ are shown with a dashed curve and a dot-dashed curve
  respectively.  Curves in red correspond to an LRG like population with a
  number density of $5 \times 10^{-5}$~h$^{-3}$Mpc$^3$ and a linear bias of
  $2$.}
  \label{fig4_chap4}
\end{center}
\end{figure}
All observations of galaxies are carried out in redshift space.
Therefore we must use the correlation function in redshift space.
At large scales, redshift space distortions caused by infall lead to an
enhancement of the two point correlation function.
The enhancement is mainly along the line of sight but the angle averaged
two point correlation function is also amplified by some amount \citep{1987MNRAS.227....1K}.

Further, we must also take into account the {\sl bias} in the distribution of
galaxies while using the correlation function shown in Figure \ref{fig3_chap4}.
This has been discussed by many authors \citep[ e.g.][]
{kais,1986ApJ...304...15B,1994ApJ...431..477B,1996ApJ...461L..65F,mo,1998MNRAS.297..251B, 1998MNRAS.299..417B, dekel}. At large scales, we may
assume that the linear bias factor $b$ is sufficient for describing the redshift space distortions and clustering.

Lastly, we should mention that we are working with the linearly extrapolated
correlation function at these scales even though there is some evidence that
perturbative effects lead to a slight shift in the location of the peak in
$\xi$ \citep[For example, see][]{2008PhRvD..77d3525S}.
The only change caused by such a shift in the location of the peak is to in
turn shift the scale where there appears to be a transition from large values
of $\Delta D_q$ towards small and constant values.
As the shift does not alter our key conclusions, we will ignore such effects
in the following discussion.

\begin{figure}
\begin{center}
\includegraphics[width=5.2truein]{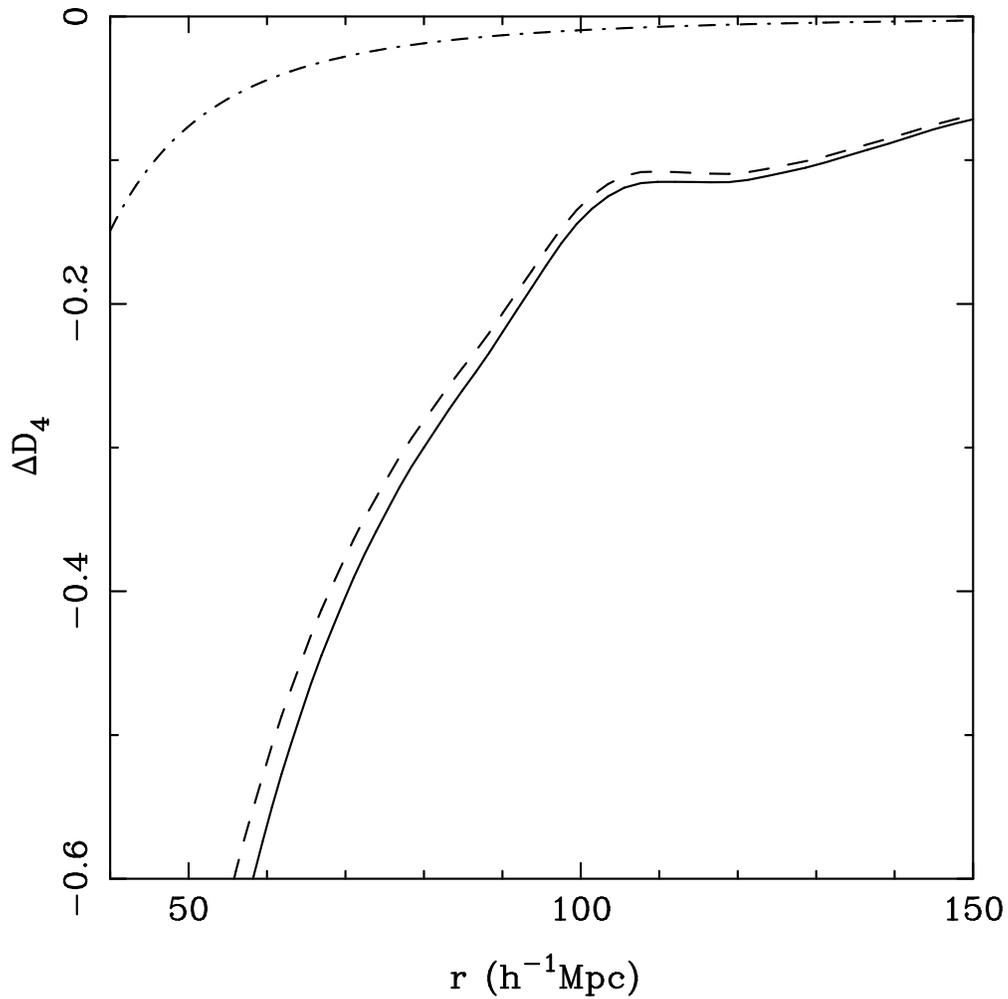}
\caption[\sf Components of $\Delta D_q$ for an LRG like population of galaxies for $q=4$]
{\sf This figure shows the components of $\Delta D_q$ for an LRG like
  population of galaxies for $q=4$.  This value of $q$ was chosen as the
  contribution of a finite number of galaxies does not vanish in this
  case. The solid line shows $\Delta D_4$, the dashed line shows the
  contribution of clustering to $\Delta D_q$ and the dot-dashed line is the
  correction due to a finite number of galaxies.  Clearly, the correction due
  to clustering is the dominant reason for departure of $D_q$ from $D$.}
  \label{fig5_chap4}
  \end{center}
\end{figure}

We plot the expected departure of $D_q$ from $D$ for an unbiased sample
of galaxies in Figure \ref{fig4_chap4} on page \pageref{fig4_chap4}.
We assumed that typical (L*) galaxies have an average number density of
$0.02$~h$^3$Mpc$^{-3}$ \citep{1980lssu.book.....P}and a bias factor of unity.
$\Delta D_q$ for such a population is shown as a function of scale by black
curves for $q=2$, $4$ and $6$.
Red curves show the same quantity for a sample of galaxies similar to Luminous
Red Galaxies (LRGs).
We used a bias factor $b=2$ and a number density of $5 \times
10^{-5}$~h$^3$Mpc$^{-3}$ that is representative of such a population \citep{2007ApJ...657..645P}.
$\Delta D_q$ is negative at all scales shown here, as expected from the
expression (see Eqn.(\ref{eqn:clus})) for hierarchical clustering.
The behavior of $\Delta D_q$ as a function of scale has two distinct
regimes on either side of $100$~h$^{-1}$Mpc.
The magnitude of $\Delta D_q$ decreases rapidly as we go from smaller scales towards
$100$~h$^{-1}$Mpc. At scales larger than $100$~h$^{-1}$Mpc, magnitude of $\Delta D_q$ either stays constant or
decreases at a very slow rate. The behavior of $\Delta D_q$ around $100$~h$^{-1}$Mpc is dictated largely by
the BAO peak in $\xi$ at this scale. Although there is no peak in $\bar\xi$, $\partial \bar\xi / \partial\log r$ which is given by
$-0.5 D (\bar\xi(r) - \xi(r))$ has a minima and a maxima near the scale of the peak in $\xi(r)$.
This results in a corresponding minima and maxima for $\Delta D_q$ as the
contribution of a finite number of galaxies is subdominant at such large
scales. We illustrate this in Figure \ref{fig5_chap4} on page \pageref{fig5_chap4} where $\Delta D_4$ is plotted for an LRG like
sample, and the two contributions (from a finite sample and weak clustering)
are also shown. If $\xi$ has a power law form then there are no extrema for $\partial
\bar\xi / \partial\log r$ and the magnitude of both $\xi$ and $\Delta D_q$
becomes progressively smaller as we get to larger scales.
There is a one to one relation between $\xi$ and $\Delta D_q$ for a given
model of this type. However, a feature like the peak introduced by BAO leads to the non-trivial
behavior illustrated in Figure \ref{fig4_chap4}.
Here we find that $D_q$ can be smaller at scales with a larger $\xi$.
For example, the scale with the local maxima of $\xi$ is very close to the
scale with the local minima of $D_q$.
The intuitive correspondence of a small $\xi$ implying a smaller deviation of
$D_q$ from $D$ does not apply in this case.

The difference between the unbiased galaxy population, and an LRG like sample
is stark. The LRG like sample has a Minkowski-Bouligand dimension that differs from
$D=3$ by a significant amount. The main reason for this difference is the high bias factor associated with
the LRG population, although a smaller number density also makes some
difference. Different clustering properties for different types of galaxies imply that
these will have not have the same Minkowski-Bouligand dimension.
This has no impact on determination of the scale of homogeneity for the
universe, where we must use unbiased tracers.

The calculations presented in the section \ref{hodi} and \ref{wcd} allow us to estimate the
offset of the Minkowski-Bouligand dimension from the physical dimension due to
weak clustering and a finite sample.
This has to be accompanied by a calculation of the dispersion in the expected
values \citep{1999MNRAS.310..428S,2000MNRAS.313..711C}. The natural estimate for the scale of homogeneity is the scale where the
offset of the Minkowski-Bouligand dimension from the physical dimension
becomes smaller than the dispersion in a sufficiently large survey.
Given that the offset is dominated by the effect of clustering, we have
$\Delta D_q \simeq 0.5 D q (\xi - \bar\xi) \sim q (\xi - \bar\xi)$.
The offset scales with $q$.
Further, it is apparent that the dispersion in $\Delta D_q$ must also scale
with $q$.
This implies that the requirement of dispersion being greater than the offset
leads to the same scale for all $q$.
This is a very satisfying feature of this approach in that the scale of
homogeneity does not depend on the choice of $q$ as long as the effect of a
finite number of points is subdominant.

Alternatively, we may argue that the scale of homogeneity should be identified
with the scale above which the variation of $\Delta D_q$ is very small.
While this is an acceptable prescription for typical galaxies where $\Delta
D_q \leq 0.06$ at scales above $100$~h$^{-1}$Mpc, it does not appear
reasonable for an LRG like population.
The scale of homogeneity for the latter population is clearly much larger than
$100$~h$^{-1}$Mpc.

\section{\bf Conclusions}\label{conclusion4}

We have studied the problem of the expected value of the
Minkowski-Bouligand dimension for a finite distribution of points.
For this purpose, we have studied a homogeneous distribution as well as a
weakly clustered distribution.
In our study, $q/\bar{N}$ and $q\bar\xi$ are taken to be
small parameters and the deviation of $D_q$ from $D$ is estimated in terms of
these quantities.
In both cases we find that the expected values of the Minkowski-Bouligand
dimension $D_q$ are different from  $D$ for the distribution of points.
For generic distributions, the value of $D_q$ is less than the dimension $D$.
We have derived an expression for $D_q$ in terms of the correlation function
and the number density in the limit of weak clustering. We see that the dimension of the ambient space (i.e. $D$) is not the correct benchmark for
defining a homogeneous distribution and instead the minkowski bouligand dimension of the distribution should match with equation \ref{eqn:homogen}
for the distribution to be homogeneous.

We find that $\Delta D_q = D_q - D$ is non-zero at all scales for
unbiased tracers of mass in the concordance model in cosmology.
For this model $\Delta D_q$ is a large negative number at small scales but it
rapidly approaches zero at larger scales.
$\Delta D_q$ is a very slowly varying function of scale above
$100$~h$^{-1}$Mpc and hence this may be tentatively identified as the scale of
homogeneity for this model.
A more quantitative approach requires us to estimate not only the systematic
offset $\Delta D_q$ but the dispersion in this quantity.
The scale of homogeneity can then be identified as the scale where the offset
is smaller than the expected dispersion.

Although we have used the example of galaxy clustering for illustrating our
calculations, the results as given in Equation (\ref{eqn:homogen}) and
Equation (\ref{eqn:clus}) are completely general and apply to any distribution of
points with weak departures from homogeneity. A detailed derivation of the relations presented here, with verification using
mock distributions of points will be presented in a future publication, where we also expect to highlight other applications.

We have obtained the scale of homogeneity of the galaxy distribution in the volume limited samples obtained from first data release of Sloan Digital Sky Survey. We will discuss this in detail in chapter \ref{chap5}.
\chapter{Testing homogeneity on large scales in the Sloan Digital Sky Survey}\label{chap5}
One of the most important assumption on which modern cosmology is based is the Cosmological Principle.
It states that the Universe is homogeneous and isotropic on large
scales. We\footnote{This chapter is based on {\it Testing homogeneity on large scales in Sloan Digital Sky Survey Data Release One}, (Jaswant Yadav, S. Bharadwaj, B. Pandey \& T.R. Seshadri), Monthly Notices of Royal Astronomical Society, 2005, {\bf 364}, 601} have tested the large scale homogeneity of the galaxy distribution in the
Sloan Digital Sky Survey Data Release One (SDSS-DR1) using
volume limited subsamples extracted from the two equatorial strips. These strips are nearly two dimensional (2D).
The galaxy distribution is projected  on the equatorial plane and
we have carried  out a 2-dimensional  multi-fractal analysis  by counting the number of galaxies
inside circles of different radii $r$ in the range $5 \, h^{-1} {\rm Mpc}$ to $150 \,
h^{-1} {\rm Mpc}$. The circles  have been centred on galaxies. Different moments of the
count-in-cells have been analyzed to identify a range of length-scales
($60-70 \, h^{-1} {\rm Mpc}$ to $150 h^{-1} {\rm Mpc}$ ) where the moments
show a power law scaling behavior.  It has helped us to determine the scaling
exponent which gives the spectrum of generalized dimension $D_q$.
If the galaxy distribution is homogeneous, $D_q$ does not vary with $q$ and
is equal to the Euclidean dimension which in our case is 2.
We find that $D_q$ varies in the range
$1.7$ to $2.2$. We also constructed mock data from random, homogeneous
point distributions and from \lcdm~ N-body simulations with bias $b=1, 1.6$
and $2$,  and analyzed these in exactly the same way. The values of
$D_q$ in the random distribution and the unbiased simulations show
much smaller variations  and these are not consistent with the actual
data.  The biased simulations, however, show larger variations in $D_q$ and
these are  consistent with both the random and the actual data.
Interpreting the actual data as a realization of a biased \lcdm~
universe, we conclude that the galaxy distribution is homogeneous on
scales larger than  $60-70 \, h^{-1} {\rm Mpc}$.
\section{\bf Introduction}
The collapse of small scale inhomogeneities created due to some quantum process in the early Universe gave rise to Large Scale Structures that we observe in the Universe at present epoch. The primary aim of all  galaxy
redshift surveys which have completed or are still underway is to determine the distribution of large scale structures in the universe. Though the galaxy distribution  exhibits a large
variety of structures starting from groups and clusters, extending to
superclusters  and an interconnected network of filaments which appears to
extend across the whole  universe, we expect the galaxy distribution
to be homogeneous on large scales. The assumption  that the universe is
homogeneous and isotropic on large scales is known as the
Cosmological Principle and this is one of the fundamental  pillars of
cosmology today. In addition to determining the large scale structures,
galaxy redshift surveys can also be used to verify if the galaxy
distribution does indeed become homogeneous on large scales and
thereby  validate  the Cosmological Principle.
Further, the galaxy redshift surveys can also be used to investigate the scales at which this
transition to homogeneity takes place.
In this chapter we test using multifractal analysis whether the galaxy distribution in the SDSS-DR1 \citep{abaz}  is {\em actually}   homogeneous on large scales and if so what is the scale at which the transition to homogeneity takes place.

A large variety  of methods have been developed and used to quantify
the galaxy distribution in redshift surveys. Prominent among these
methods are  the two-point correlation function $\xi(r)$  \citep{1980lssu.book.....P} and
its Fourier transform the power spectrum $P(k)$. There now exist very precise estimates of
$\xi(r)$ \citep[e.g.][]{zevi,haw}  and the
power spectrum  $P(k)$ \citep[e.g.][]{perci,teg2} determined from different large redshift surveys.
As has been pointed out in chapter \ref{chap3},  the two point correlation function is  found to be
well described by a power law (equation \ref{eq:1}) on small scales.

The power law behavior of $\xi(r)$  suggests a scale invariant clustering pattern in the distribution of galaxies. This pattern would violate
homogeneity if this power-law behavior were to extend to arbitrarily large length-scales. Reassuringly, the power law form for $\xi(r)$
does not hold on large scales and it breaks down at $r> 16 h^{-1} {\rm Mpc} $
for SDSS and at $r> 20 h^{-1} {\rm Mpc} $ for 2dFGRS. The fact that the values
of $\xi(r)$ fall off sufficiently fast with increasing $r$ is consistent with the galaxy distribution being homogeneous on
large scales. A point to note is that though the $\xi(r)$ determined from
redshift surveys is consistent with the universe being homogeneous at
large scales it does not actually test this. This is because the way
in which $\xi(r)$ is defined and determined from observations refers
to the mean number density of galaxies  and therefore it presupposes
that the galaxy distribution is homogeneous on large scales.
Further, the mean density which we compute is only that on the scale
of the survey. It will be equal to the mean density in the universe only
if the transition to homogeneity occurs well within the survey region.
To verify
the large scale homogeneity of the galaxy distribution it is necessary
to consider a statistical test which does not presuppose the premise
which is being tested. Here we consider one such test, the
``multi-fractal dimension'' and apply it to the SDSS-DR1.

The fact that the galaxy clustering  is scale-invariant  over a  range
of length-scales led \citet{1987PhyA..144..257P} to propose that the galaxies had a
fractal distribution. The later analysis of \citet{coleman} seemed to
bear out such a proposition whereas \citet{bor} claimed that
the fractal description was valid only on small scales and the
galaxy distribution was consistent with homogeneity on large scales.
A purely fractal distribution
would not be homogeneous on any length-scale and this would violate the
Cosmological Principle. Further,
the mean density would decrease if it were to be
evaluated for progressively  larger volumes and this would manifest
itself as an increase in the correlation length $r_0$ (equation \ref{eq:1})
with the size of the sample. However, this simple prediction of the
fractal interpretation is not
supported by data, instead $r_0$ remains constant for volume limited samples
of CfA2 redshift survey with increasing depth \citep{martinez}.

The analysis of the ESO slice project by \citet{gujo} confirms large scale
 homogeneity whereas  the  analysis  of volume limited samples of
 SSRS2 by \citet{cappi} is consistent with both the   scenarios of
 fractality and homogeneity. A similar analysis \citep{haton} carried
 out on  APM-Stromlo survey exhibits a fractal behavior with a
 fractal dimension of $ 2.1\pm 0.1$ on scales up to $40\, h^{-1}{\rm  Mpc}$. Coming to the fractal analysis of the Las Campanas Redshift Survey (LCRS), \citet{amen} find a fractal behavior on scales less than $\sim$
$30 h^{-1}\,{\rm Mpc}$ but are inconclusive about the transition to
homogeneity. A multi-fractal analysis by \citet{bharad1} shows that
 the   LCRS exhibits homogeneity on the scales $80$ to $200 \,
 h^{-1}\, {\rm Mpc}$. The analysis of \citet{kur}  shows
this to occur at a length-scale of $\sim 30 \, h^{-1}\,{\rm Mpc}$,
whereas \citet{best} fails to find a transition to homogeneity even on
the largest scale analyzed. The fractal analysis of the
PSCz  \citep{pan} shows   a transition to homogeneity on  scales
 of $30\, h^{-1} {\rm Mpc}$.   Recently \citet{baris} have  performed a
 fractal analysis of SDSS EDR and find that a  fractal distribution
 continues to  length-scales of $200\,h^{-1}{\rm Mpc}$ whereas
\citet{hog} analyze the  SDSS LRG  to find
a convergence to homogeneity at a scale of  around $70\, h^{-1} {\rm  Mpc}$.

In this chapter we use the multi-fractal analysis to study the scaling
properties of the galaxy distribution in the  SDSS-DR1 and test if it
is consistent with homogeneity on large scales. The SDSS  is the
largest galaxy survey available at present. For the current analysis
we have used  volume limited subsamples   extracted from the two equatorial
strips of the SDSS-DR1. This reduces the number of galaxies but offers
several advantages. The volume limitedness of the  samples ensures that the variation in the number density in
these samples are independent of the details of the luminosity function and
is caused only by clustering. The larger area and depth of these
samples provide us the scope  to investigate the scale of homogeneity in greater detail .

The \lcdm~  model with $\Omega_{m0}=0.3$, $\Omega_{\Lambda0}=0.7$,
$h=0.7$ and an adiabatic,  scale invariant
primordial power spectrum is currently believed to be the
minimal model which is consistent with most
cosmological data \citep{efst,perci1,teg1}. Estimates of the two point
correlation function $\xi(r)$ \citep{tuck,zevi,haw} and the power spectrum  $P(k)$ \citep{lin,perci,teg2} are all consistent with this  model.  In this chapter we use the particle position inferred from N-body simulations (see subsection
\ref{n-body}) to determine the  length-scale where the transition to homogeneity occurs in the \lcdm~  model and test if the actual data is
consistent with this.

Galaxy surveys provide us information about the visual part of the matter distribution only.
We should also keep in mind the fact that the models of structure formation primarily predict
the clustering of the {\it dark matter} which dominates the dynamics. The process of galaxy
formation and the exact relation between the distribution of the galaxies
and the dark matter is far from well understood. The fact that the
galaxies are a biased tracer of the dark matter distribution is now well accepted
\citep[e.g.][]{kais,mo,dekel,taru,yoshi}. Further, on large scales one expects the fluctuations in the galaxy and the
dark matter distribution to be linearly related through the linear bias parameter $b$.
Determining the bias $b$ is an important issue in cosmology. Not only will it allow the
dark matter distribution to be determined, but it is also expected to throw light on
galaxy formation. There currently exist several ways to determine the bias. Measuring
the redshift space distortion parameter $\beta ={\Omega^{0.6}}_m /b$ \citep[e.g.][]{haw,teg1} in combination with an independent determination of
$\Omega_{m0}$ allows $b$ to be determined. The bispectrum \citep{2002MNRAS.335..432V} provides
a technique to determine the bias from redshift surveys without the need of inputs
from other observations. A combination of weak lensing and the SDSS galaxy survey
has been used by \citet{2005PhRvD..71d3511S} to determine the bias. The multifractal nature of the
galaxy distribution from the N-body simulations is very sensitive to the bias parameter
and holds the possibility of giving accurate estimates for this. We apply this test to the volume limited samples analyzed in this chapter.

There are various other probes which  test the cosmological principle. The fact that  the Cosmic Microwave Background Radiation (CMBR) is  nearly isotropic $(\Delta T/T \sim  10^{-5})$ can be used to infer  that our
 spacetime is locally  very well described by the Friedmann-Robertson-Walker  metric \citep{ehl}. Further, the  CMBR anisotropy  at large angular scales $(\sim 10^{o})$ constrains  the  {\it rms}  density fluctuations to $\delta\rho/{\rho}  \sim 10^{-4} $ on  length-scales of $1000\,h^{-1}{\rm Mpc}$ \citep[e.g.][]{wu}. The analysis of deep radio surveys \citep[e.g. FIRST,][]{bale} suggests the distribution to be  nearly isotropic on large scales. By comparing the predicted
multipoles of the X-ray Background to those observed by HEAO1 \citep{sch} the
fluctuations in amplitude are found to be consistent with the homogeneous universe \citep{lah}. The absence of big  voids in the distribution of Lyman-$\alpha$ absorbers is inconsistent with a purely fractal model \citep{nus}.

A brief outline of this chapter follows. In Section \ref{Data} we describe the
data collected from SDSS as well as N-body simulations. The method of analysis of the acquired data has been discussed in section \ref{method}. We
conclude in  Section \ref{results} with  results and
conclusions.
\section{\bf Galaxy Sample : The Sloan digital Sky Survey } \label{Data}
\begin{figure}
\begin{center}
\rotatebox{0}{\scalebox{1.0}{\includegraphics{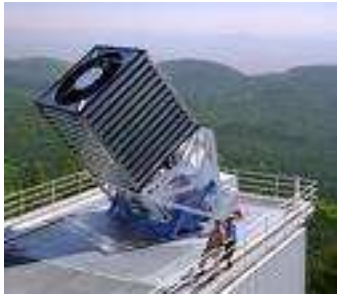}}}
\caption[\sf SDSS Telescope]{\sf The Sloan Digital Sky Survey Telescope}
\label{fig5000}
\end{center}
\end{figure}

The results presented in this chapter are based on a galaxy sample selected from the
largest galaxy survey to date, the Sloan Digital Sky Survey (SDSS). The SDSS is a
wide-field photometric and spectroscopic survey carried out with a dedicated 2.5 meter
telescope (Figure \ref{fig5000}) at Apache Point, New Mexico \citep{2000AJ....120.1579Y}. The SDSS telescope uses the drift scanning technique in  which the telescope is fixed and the earth's rotation is made use of to record small strips of
the sky. The telescope scans continuously the sky on five photometric bandpasses namely $u$, $g$, $r$, $i$ and $z$ . For our analysis, we have used the $r-$band result. The wavelength range for this band is $5500 {\AA} - 7000 {\AA}$,
 which is better for low-redshift galaxy. The telescope scans  to a limiting $r$-band apparent magnitude\footnote{The apparent magnitude (m) of a celestial body is a measure of its brightness as seen by an observer on Earth, normalized to the value it would have in the absence of the atmosphere. The brighter the object appears, the lower the value of its magnitude} of $22.5$ \citep{1996AJ....111.1748F,2002AJ....123.2121S}. All the data obtained from the telescope is processed by dedicated software for astrometry \citep{2003AJ....125.1559P}, identification of sources and candidates \citep{2002SPIE.4836..350L}, candidate selection for the spectroscopy sample \citep{2001AJ....122.2267E,2002AJ....124.1810S}, adaptive tiling \citep{2003AJ....125.2276B} and photometric calibration \citep{2001AJ....122.2129H,2002AJ....123.2121S}. An extensive analysis of possible systematics uncertainties present in the data is described in \citet{2002MNRAS.332..697S}

When finished the SDSS will cover approximately $10^4$ square degrees of the sky. The main part of the survey is located in the Northern Galactic
sky, with an additional small area in the Southern Galactic sky.
In total it will provide $\sim 10^8$ optical images in five bands and $\sim 10^6$ spectra of galaxies
with apparent magnitude $m_r < 17.77$ \citep{1998AJ....116.3040G,2000AJ....120.1579Y}. SDSS is also recording
the redshifts to $\sim 100,000$ quasars (the most distant objects known) giving us unprecedented
knowledge of the distribution of matter to the edge of the visible universe. The spectroscopic targets in the SDSS are divided into three categories.
\bi
\item The main galaxy sample \citep{2002AJ....124.1810S}.
\item The luminous red sample \citep{2001AJ....122.2267E}.
\item The quasar sample \citep{2002AAS...20112505R}.
\ei

The main galaxy sample is complete down to an apparent $r$-band Petrosian magnitude
limit of $m_R < 17.77$. The galaxy sample used in this work was obtained from the sky
server\footnote{http://www.sdss.org/} using the SDSS CasJobs site\footnote{http://casjobs.sdss.org/CasJobs/}. The website
is based on SQL queries which can
perform a large number of pre-processing tasks (see appendix \ref{sql}). All galaxies with $r$ band
apparent magnitude of $r < 20$ were extracted from the spectroscopic main sample.
For each galaxy we obtained the position in the sky (ra, dec) and redshift $z$ as well as
many other properties such as apparent magnitudes in the five bands ($u$, $g$, $r$, $i$ and $z$),
isophotal radius, position angle in the sky, petrosian radius enclosing $90\%$ and $50\%$ of
the total flux, etc. The detailed list of properties queried can be found in appendix \ref{sql}.
\begin{figure}
\begin{center}
\rotatebox{0}{\scalebox{1.0}{\includegraphics{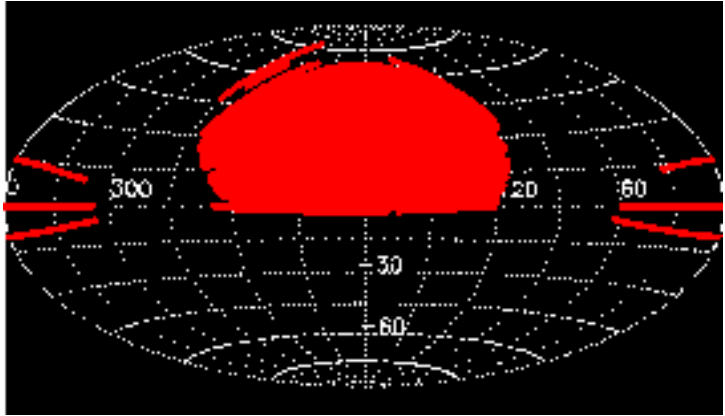}}}
\caption[\sf Projection of SDSS DR6 Imaging sample]{\sf Projection on the sky of the {\bf SDSS-DR6} Imaging (Aitoff projection of Equatorial
coordinates). The relation between Equatorial Coordinates and the survey coordinates is given in appendix \ref{scs}. Figure courtesy the SDSS team}
\label{fig50000}
\end{center}
\end{figure}

\begin{figure}
\begin{center}
\rotatebox{0}{\scalebox{1.0}{\includegraphics{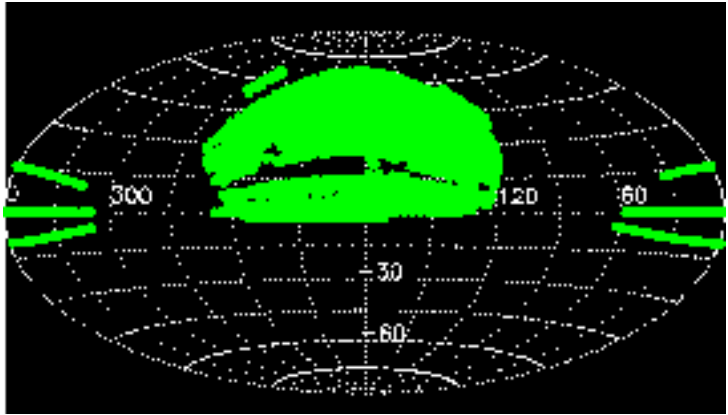}}}
\caption[\sf Projection of SDSS DR6 spectroscopic sample]{\sf Projection on the sky of the {\bf SDSS-DR6} spectroscopic sample. The center of the plot is in the direction $\alpha = 180$, $\delta = 0$. Contour lines delineate the edges of the survey. Courtesy the SDSS team}
\label{fig500}
\end{center}
\end{figure}

Figure \ref{fig50000} and \ref{fig500} show the projection on the sky of the imaging and spectroscopic sample respectively, taken from
the latest data release to date\footnote{by the time of thesis submission}(DR6). There are five large patches. There are also several smaller ones
corresponding to the last observed fields as well as many holes inside the large areas. The holes are the result of bright stars, telescope artefacts
and failures in sky coverage. When finished the SDSS will completely cover the area in the center of the
map in figure \ref{fig500}, covering almost a quarter of the sky.

In order to analyze a galaxy sample from the survey only redshift information is not going to suffice. We have to assume a cosmological model in
order to get physical distance from the observer to the object in the sample.
\subsection{Redshift-Distance Formula}
The expansion of the universe and its (assumed) isotropic nature provides a convenient
way to determine the distance to galaxies by means of their recession velocity:
\be
r_{gal}= \frac{v_{rec} - v_{pec}}{H_0}
\label{rgal}
\ee
where $r_{gal}$ is the distance to the galaxy. $v_{rec}$ is the recession velocity of the unperturbed
Hubble flow and $v_{pec}$ is the line of sight component of the peculiar velocity of the galaxy. The peculiar velocity is the velocity of a galaxy with respect to the Hubble flow. $H_0$
is the Hubble parameter, parameterized by $h$:
\be
H_0 = 100 \,h \left[km s^{-1} Mpc^{-1}\right].
\ee
We can artificially produce a `redshift space' by setting each galaxy at distance $r_{gal}$ obtained by considering
$v_{pec} = 0$ in equation \ref{rgal}.  The `redshift space' is thus distorted representation of the `real space'.

The recession velocity can be inferred from the spectra of the galaxy by measuring the
shift in frequency of emission or absorption lines compared to that in the rest frame. This shift
in the frequency is produced by the expansion of the Universe:
\be
\nu_e = (1+z)\nu_o
\ee
where $\nu_e$ is the frequency of a photon when it was emitted, $\nu_o$ is the observed frequency
and $z$ is the redshift. More generally, we can compute the distance to a galaxy
given its observed redshift according to:
\be
R = a_0 r(z) (1+z)
\ee
where $a_0$ is the present value of scale factor of expansion of the Universe and $r(z)$ is the redshift dependent radial coordinate distance given by \citep{weinberg}:
\be
r(z)=S\left[\frac{1}{a_0 H_0}\int_{1/(1+z)}^{1}\frac{dx}{x^2\sqrt{\Omega_\Lambda +\Omega_m x^{-3} + \Omega_{rad} x^{-4}+(1-\Omega_\Lambda -\Omega_m -\Omega_{rad})x^{-2}}} \right]
\ee
where the function $S[x]$ for different geometries of Universe is given by:
\be
S[x] = \left\{\begin{array}{cl}
            \sin(x)  	& k = +1~~~~(Spherical~~Universe)\\
            x       	&k =~0~~~~(Flat~~Universe) \\
	       \sinh(x) 	&k = -1~~~~(Hyperbolic~~Universe)
    \end{array}\right.
\ee
We have assumed the standard \lcdm ~ model for the value of different cosmological parameters.[See table \ref{tab1} in chapter \ref{chap1}]
\subsection{SDSS-DR1}
Our analysis of multifractal nature of galaxy distribution is based
on the publicly available SDSS-DR1 data \citep{abaz}.
In this work we have analyzed two equatorial strips
which are centred along the celestial equator ($\delta=0^{\circ}$),
one  in the Northern Galactic  Cap (NGP)  spanning  $91^{\circ}$   in
{\it r.a.}(from $145^\circ < \alpha < 236^\circ$) and the other in Southern Galactic Cap (SGP) spanning
$65^{\circ}$ in  {\it r.a.} (from $351^\circ < \alpha < 56^\circ$),  their thickness varying
within $\mid \delta \mid \le 2.5^{\circ}$ in {\it dec.} These regions contains 38,838 galaxies
having redshift in the range $0.02 \leq z \leq 0.2$ with the selection criteria that the extinction corrected Petrosian $r$ band magnitude is $r_p < 17.77$.

For the current purpose we have selected only the galaxies lying within $-1^\circ < \delta < 1^\circ$
as both the equatorial strips have complete coverage in this declination range.
We  have constructed  volume limited subsamples of the data. In this case we have to select the maximum limiting radius $D_{lim}$ of the sample. We also have to find the limiting absolute magnitude $M_{lim}$ of a galaxy that corresponds to the apparent magnitude limit $m_{lim}$ of the survey. The cosmological magnitude-distance relation that connects the three values is given by
\be
m = M + 5 \log_{10}\left(D_L/1 Mpc \right) + 25
\ee
 These volume limited subsamples, used here, extend from
 $z=0.08$ to $0.2$ in redshift ({\it i.e.} $235\,h^{-1}{\rm Mpc} \leq R \leq  571\,h^{-1}{\rm Mpc}$ comoving  in the  radial  direction in the standard \lcdm~ model). They have been formed by restricting  the extinction corrected Petrosian $r$ band apparent magnitude in the range $14.5 \le m_r \le 17.5$ and the absolute
 magnitude   range to  $-22.6\leq M_r  \leq -21.6$. Even though the number of galaxies are reduced this way, there are  several advantages on offer. The radial selection function
is nearly uniform in this case. So the variation in number density of galaxies in the subsample is caused only by
clustering of the galaxies. A volume limited sample defined by an interval in absolute magnitude
translates into an interval in redshift. It has the nice property that in principle each
galaxy could be displaced to any depth within the sample and would still remain
within the apparent magnitude range of the survey. This resulted in $5315$ galaxies distributed
in two wedges, spanning $91^\circ$ (NGP) and $65^\circ$ (SGP) in {\it r. a.}, both with thickness
$2^\circ$ centered along the equatorial plane extending from $235 h^{-1}{\rm Mpc}$ to $571 h^{-1}{\rm Mpc}$
comoving in the radial direction.

\begin{figure}
\rotatebox{-90}{\scalebox{0.6}{\includegraphics{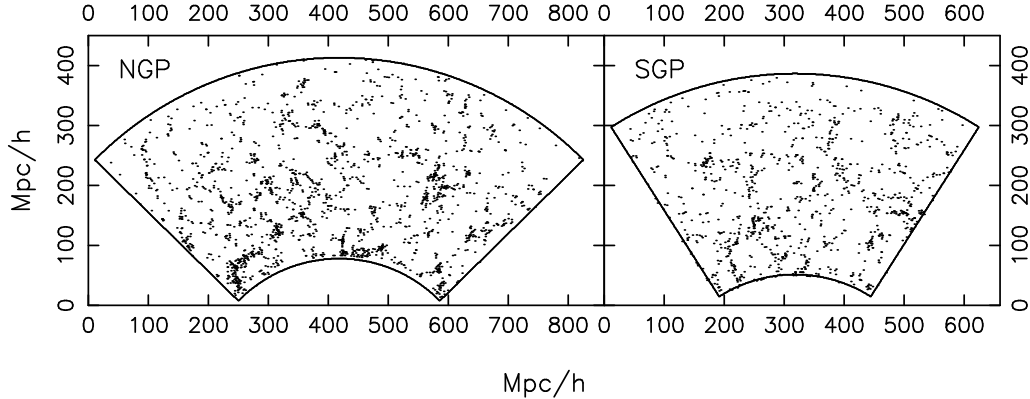}}}
\caption[\sf Two dimensional galaxy distribution of SDSS ]{\sf This shows the two dimensional galaxy distribution in
  the NGP
  and SGP subsamples of SDSS that have been analyzed here.}
\label{fig501}
\end{figure}

The resulting subsamples are two thin wedges of varying thickness (from $8.2 h^{-1}{\rm Mpc}$ to $20 h^{-1}{\rm Mpc}$) aligned with the equatorial plane. Our analysis is restricted to slices of
uniform thickness  $\pm 4.1 \,h^{-1}{\rm Mpc}$ along  the
equatorial plane extracted out of the wedge shaped regions. These
slices are  nearly $2D$ with the radial extent and the
extent along {\it r.a.} being much
larger than the thickness. We have projected the galaxy distribution
on the equatorial plane and analyzed the resulting 2D distribution
(Figure \ref{fig501}). The SDSS-DR1 subsamples that we
analyze here contains a  total of 3032 galaxies.
\subsection{N-Body Data}\label{n-body}
 We have used a Particle-Mesh (PM) N-body code to simulate the dark
 matter distribution at the mean redshift  $z=0.14$ of our
subsample. A comoving volume of $[645  h^{-1} {\rm Mpc}]^3$ is
 simulated using $256^3$ particles on a
$512^3$ mesh with grid spacing $1.26  h^{-1} {\rm Mpc}$. The set of values
$(\Omega_{m0},\Omega_{\Lambda0},h)=(0.3,0.7,0.7)$ were used for the
cosmological parameters, and we used  a \lcdm~ power
spectrum characterized by a spectral index $n_s=1$ at large-scales and
with a value $\Gamma=\Omega_m h = 0.2$  for the shape parameter of the power spectrum.
The power spectrum was normalised to $\sigma_8=0.84$ \citep[WMAP,][]{sperg}
. Theoretical considerations and simulations suggest that galaxies may
 be biased tracer of the underlying dark matter distribution \citep[e.g.,][]{kais,mo,dekel,taru,yoshi}.
A  ``sharp cutoff'' biasing scheme \citep{cole} was used to
generate this kind of biased particle distributions. This is a local biasing scheme where the
probability of a particle being selected as a galaxy is a function of local density only.
In this scheme the final dark-matter distribution generated by the
 N-body simulation was first smoothed with a Gaussian of width $5
 h^{-1} {\rm Mpc}$. Only the particles which lie in regions where the
 density contrast exceeds a critical value were selected as galaxies.
 The values of the critical density contrast were chosen so as to
 produce particle distributions with a low bias $b=1.2$ and a high
 bias $b=1.6$.

\begin{figure}
\begin{center}
\rotatebox{0}{\scalebox{0.6}{\includegraphics{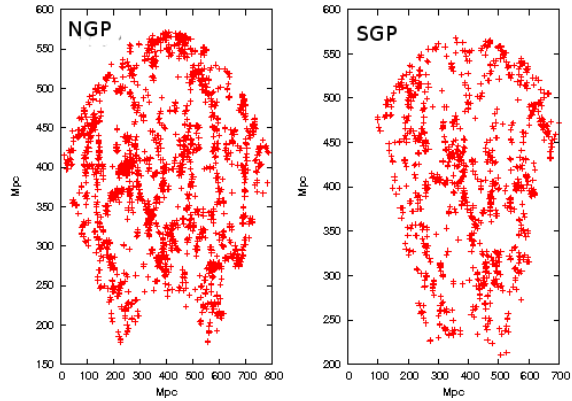}}}
\caption[\sf Two dimensional galaxy distribution of the simulated slices of \lcdm~ model ]{\sf This shows the two dimensional galaxy distribution in
  the NGP and SGP subsamples of simulated slices of \lcdm~ model that have been analyzed here.}
\label{fig5051}
\end{center}
\end{figure}

 An observer is placed at a suitable location inside
 the N-body simulation cube and we use the peculiar velocities to
 determine the particle positions in redshift space.
 Exactly the same number of   particles distributed over  the
 same volume as the actual data was extracted from the simulations
 to produce simulated NGP and SGP slices. The simulated slices (Figure \ref{fig5051}) were
 analyzed in exactly the same way as the actual data.

\section{\bf Method of Analysis}\label{method}
A  fractal point distribution is usually characterized
 in terms of its fractal dimension. As discussed in chapter \ref{chap2} there are different ways to
 calculate this dimension. The correlation dimension happens to be one of the methods
 which is   of  particular relevance to the analysis of galaxy
 distributions.  The formal definition of the correlation dimension
 involves a limit  which is meaningful only when the number of
 particles is infinite and  hence this cannot  be applied in the strict sense to  galaxy
 surveys with a limited number of galaxies. To overcome this we adopt
a  ``working  definition'' which can be applied to a finite  distribution of $N$ galaxies. It should be noted that our galaxy distribution is effectively two dimensional, and we have largely restricted our discussion to this situation.

 Labeling the galaxies from $1$ to $N$,  and using ${\bf x}_i$  to denote the comoving coordinates of the  $i^{th}$   galaxy, the number of
 galaxies within a circle of comoving radius $r$
 centred on the $i$~th galaxy is given by
\begin{equation}
  n_i(r)=\sum_{j=1 \neq i}^{N}\Theta(r-\mid {\bf x}_i-{\bf x}_j \mid)
\label{eq52}
\end{equation}
where  $\Theta(x)$ is the  Heaviside function. Averaging $n_i(r)$
by choosing  $M$ different galaxies as centers and dividing by the
 total number of galaxies gives us
\begin{equation}
  C_2(r)=\frac{1}{MN}\sum_{i=1}^{M}n_i(r)
\end{equation}
which may be interpreted as the probability of finding a galaxy within
a circle of radius $r$ centred on another galaxy.
If  $C_2(r)$ exhibits a power law scaling relation $C_2(r)\propto
r^{D_2}$, the exponent  $D_2$ is defined to be the correlation
dimension. Typically, a power law scaling relation will hold only
over a limited range of length-scales $r_1 \le r \le r_2$, and it may
so happen that the galaxy distribution has different correlation
dimensions over different ranges of length-scales.

It is  clear  that $C_2(r)$ is closely related to the volume
integral of the  two point correlation function  $\xi(r)$. In a situation
where this has  a power law behavior $\xi(r)=(\frac{r}{r_0})^{\gamma}$,
the correlation dimension is $D_2=D-\gamma$ on scales
$r<r_0$. Here $D$ is the dimension of the ambient space in which the galaxies are distributed, which in our analysis happens to be $2$. Further, we expect $D_2=D$ on large scales where the galaxy distribution is expected to be homogeneous and isotropic.

In the usual analysis  the two point correlation does not fully
characterize all the statistical properties of the galaxy
distribution, and it is necessary to also consider the higher order
correlations {\it e. g.} the three point and higher
correlations. Similarly, the full statistical  quantification  of  a
fractal  distribution also requires  a hierarchy of scaling indices.
The multifractal analysis used  here does
 exactly this. It provides a spectrum of
generalized dimension $D_q$, the Minkowski-Bouligand dimension,
which is defined for a range of $q$.

 Closely following the definition of the correlation dimension $D_2$, we can define the generalized dimension  $D_q$ using the $(q-1)^{th}$ moment of the number of neighbors $n_i(r)$. The quantity
 $C_2(r)$ is thus generalized to
\begin{equation}
  C_q(r)=\frac{1}{MN}\sum_{i=1}^{M}[n_i(<r)]^{q-1}
  \label{eq53}
\end{equation}
We would once again like to stress that the $n_i(r)$ is the number of neighbors of the galaxy placed at position $x_i$. Equation \ref{eq53} can now be used to define the generalized Minkowski-Bouligand dimension
\begin{equation}
D_q=\frac{1}{q-1}\frac{d\log{C_q(r)}}{d\log{r}}
\label{eq54}
\end{equation}
Typically $C_q(r)$ will not exhibit the same scaling behavior over
the entire range of length-scales, and it is possible that the
spectrum of  generalized dimension will be different  in different
ranges of length-scales.
As is clear from equation \ref{eq54} the correlation dimension corresponds to the generalized dimension
at $q=2$. The other integer values of $q$ are related to the scaling of higher order
correlation functions.
A mono-fractal is characterized by a single scaling
exponent {\it i.e.} $D_q$ is a constant independent of $q$, whereas the
full spectrum of generalized dimensions is needed to characterize a
multifractal.   It is clear from equation \ref{eq53} that for positive values of $q$ the contribution to $C_q(r)$ will be dominated by the regions for which $n_i(r)$ is higher. This implies that positive value of $q$ gives more weightage to the regions with high number  density. On the other hand, when q is negative the dominant contribution to $C_q(r)$ will come from regions of the survey with lower $n_i(r)$. This is equivalent to saying that the  negative values of $q$ give more weightage to the underdense  regions. Thus we may interpret
$D_q$  for  $q > 0$ as characterizing the scaling behavior of the
galaxy distribution in the high density regions like clusters whereas
$q<0$ characterizes the scaling inside voids. In the situation where
the galaxy distribution, in $2$-Dimensional slices that we have, is homogeneous and isotropic on large scales,
we expect the generalized dimension $D_q$ to take the value $2$ independent of the value of $q$.

There are a variety of different algorithms which can be used  to
calculate the generalized  dimension, the Nearest Neighbour
Interaction \citep{badi} and the Minimal Spanning Tree
\citep{suth} being some of them. We have used the correlation
integral method which we present below.

The two subsamples, NGP and SGP (see figure \ref{fig501}) contain 1936 and 1096 galaxies
respectively and they were analyzed separately. For each galaxy in the
subsample we considered a circle of  radius $r$ centred on the
galaxy and counted the number of other galaxies within the circle to
determine  $n_i(r)$ (equation \ref{eq52}). The radius $r$ was increased
starting from $5\, h^{-1} \, {\rm Mpc}$ to the largest value where the
circle lies entirely within the subsample boundaries.
The values of $n_i(r)$ determined using different galaxies as
centers were then averaged to determine $C_q(r)$ (equation \ref{eq53}). It
should be noted
that the number of centers fall with increasing $r$, and for the NGP
there   are $\sim 800$ centers for $r=80h^{-1} \, {\rm  Mpc}$  with the value
falling  to $\sim 100$ for a radius of $r=150h^{-1} \, {\rm
  Mpc}$. The large scale behavior of $C_q(r)$ was carefully analyzed to
determine the range of length-scales where $C_q(r)$  exhibits a scaling
behavior in order to calculate the scaling exponent $D_q$ as a function
of $q$.

In addition to the actual data,  we have also constructed and analyzed
random distributions of points. The random data  contains
exactly the same number points as there are galaxies in the
actual data distributed over exactly the same region as the actual
NGP and SGP slices. The random data  are homogeneous and
isotropic  by construction, and  the results of the multifractal
analysis of this data gives definite predictions for the
results expected if the galaxy distribution were homogeneous and
isotropic.  The random data and the simulated slices extracted from
the N-body simulations were all analyzed in exactly the same way as
the actual data. We have used $18$ independent realizations of the
random  and simulated slices to estimate the mean and the
$1-\sigma$ error-bars of the spectrum of generalized dimensions
$D_q$. We have also checked that increasing the
number of realizations to $36$ does not significantly change the mean
or the $1 -\sigma$ error bars.
\section{\bf Results and Discussions}\label{results}

\begin{figure}
\rotatebox{-90}{\scalebox{0.75}{\includegraphics{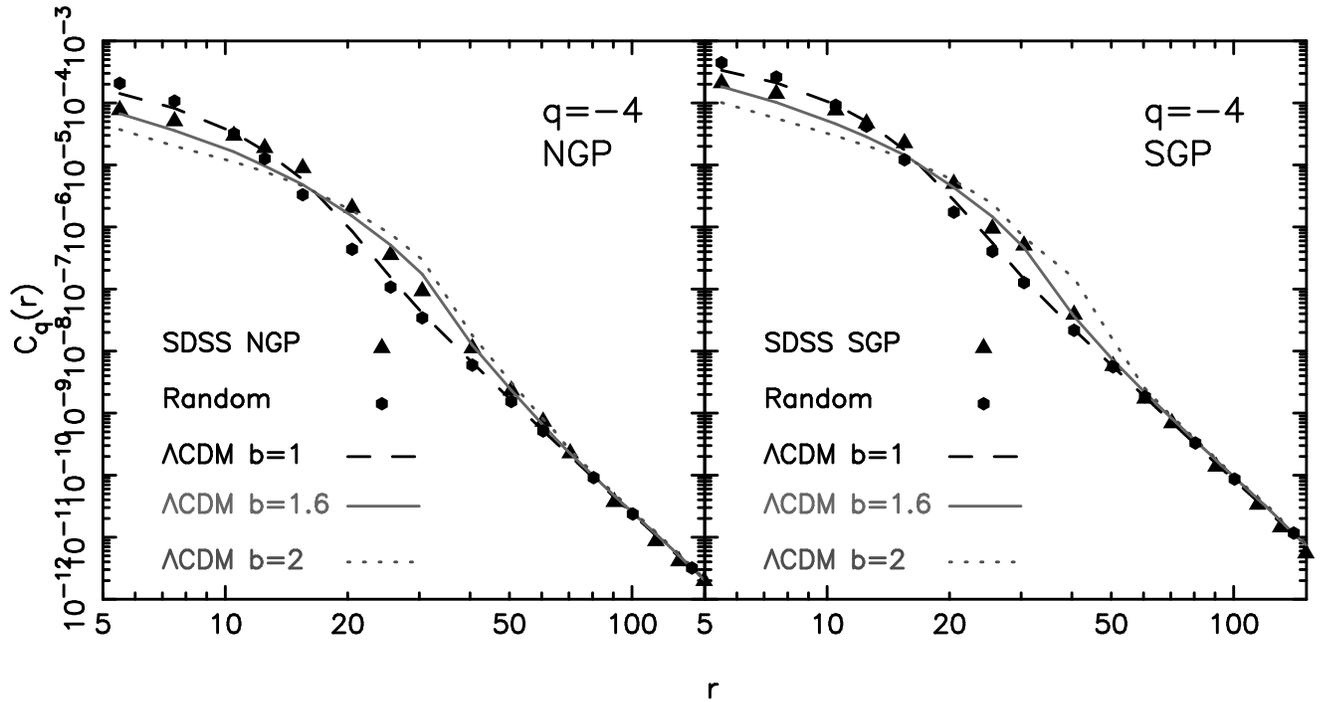}}}
\caption[\sf $C_q(r)$ for $q=-4$]{\sf This shows $C_q(r)$ at $q=-4$ for the actual data, the random
data and the simulated slices. The $1-\sigma$ error bars are not shown in the figure as
they are too small to be seen in the logarithmic scale used here. For the random and unbiased \lcdm ~ model, the error bars are $\sim 20$ \% on small scales
($\le 30 h^{-1} {\rm Mpc}$) and it decreases to $\sim 2$ and $\sim 10$ \% on larger scales for the two models, respectively. For the two biased cases, the error bars are much larger
($\sim 80-100$ \%) on small scales and it decreases to $\sim 20$ per cent at $\sim 70 h^{-1}{\rm Mpc}$ and beyond.}
\label{fig51}
\end{figure}

\begin{figure}
\begin{center}
\rotatebox{-90}{\scalebox{0.75}{\includegraphics{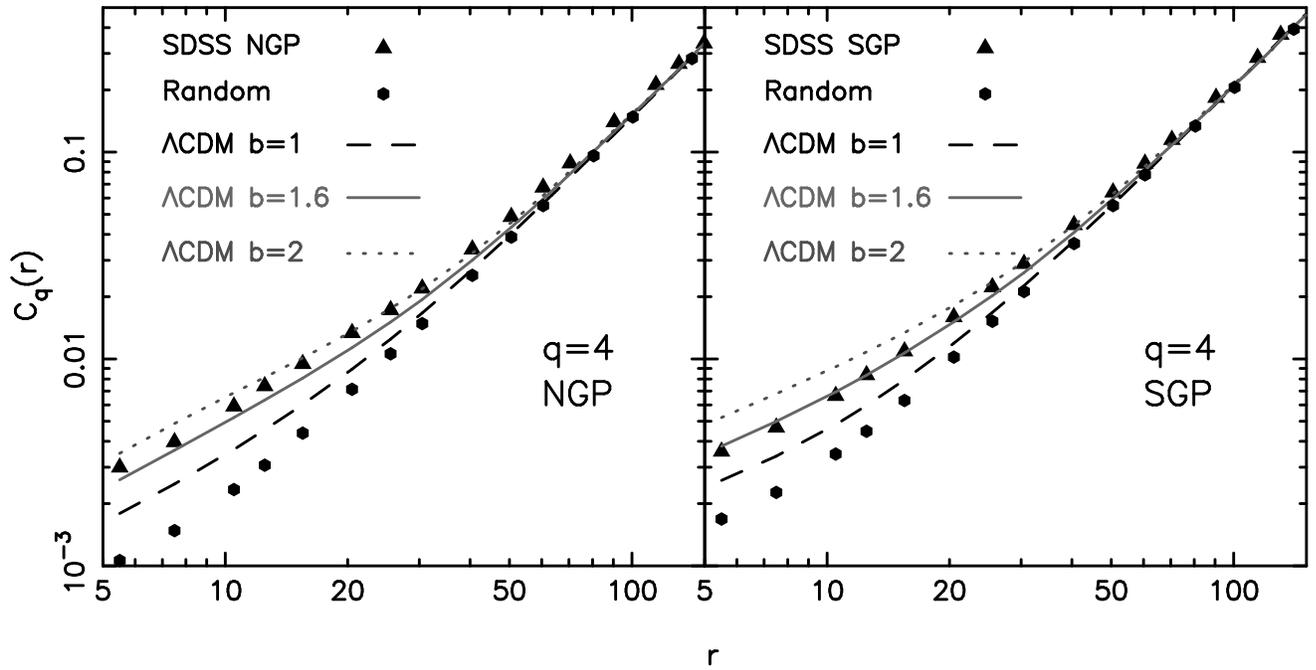}}}
\caption[\sf $C_q(r)$ for $q=4$]{\sf This shows $C_q(r)$ at $q=4$ for the actual data, the random
data and the simulated slices. The $1-\sigma$ error bars are not shown in the figure as they are too
small to be seen in the logarithmic scale used here. The $1-\sigma$ error bars are $\sim 2 - 5$ \% for the entire range of length-scales shown here. It may also be noted
that the error bars decrease monotonically with increasing $q$.}
\label{fig52}
\end{center}
\end{figure}
\begin{figure}
\begin{center}
\rotatebox{-90}{\scalebox{0.75}{\includegraphics{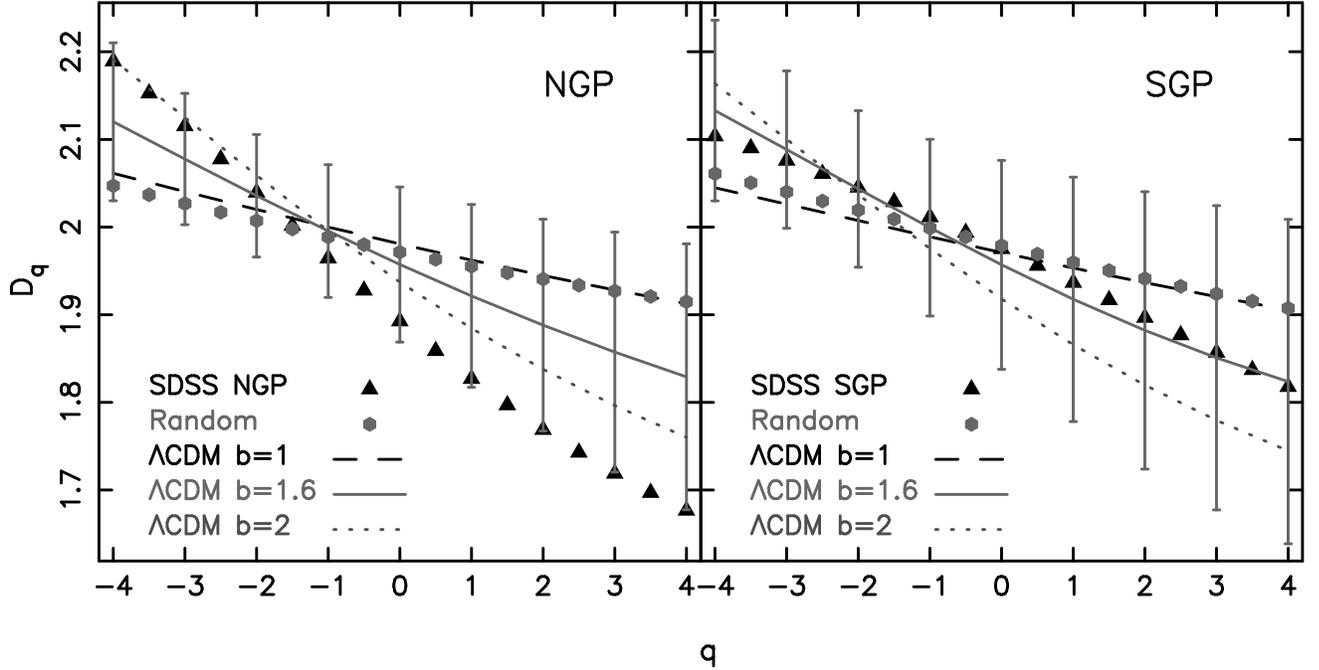}}}
\caption[\sf The spectrum of generalized dimension]{\sf This shows the spectrum of generalized dimensions $D_q$ as a
function of $q$ for the actual data, the random
data and the simulated slices on length scales of from
$60-70 \, h^{-1} \, {\rm Mpc}$ to $150 \, h^{-1} \, {\rm Mpc}$.
The error bars shown are for \lcdm~ model with bias=1.6.}
\label{fig53}
\end{center}
\end{figure}

Figures \ref{fig51} and \ref{fig52} show $C_q(r)$  at $q=-4$
 and $4$, respectively,  for the actual data, for one realization of
 the random
slices and for one realization of the  simulated slices for each value
 of the bias.   The behavior of $C_q(r)$ at other
 values of $q$ is similar to the ones shown here. Our analysis is
 restricted to $-4 \le q \le 4$.
We find that $C_q(r)$ does not exhibit a scaling
behavior at small scales $(5\, h^{-1} \, {\rm Mpc} \le r \le 40 \,
h^{-1} \, {\rm Mpc})$. Further, the small-scale  behavior of $C_q(r)$
in the actual data is different from that of the random slices and is
roughly consistent with the simulated slices for $b=1.6$.  We find
 that $C_q(r)$ shows a scaling behavior on length-scales of
 from somewhere around $60-70 \, h^{-1} \, {\rm Mpc}$ to $150 \,
 h^{-1} \, {\rm Mpc}$.  Although the value of $C_q(r)$ for the actual data, the random and
 simulated slices appear to converge over this range of
 length-scales indicating that they are all roughly consistent with
 homogeneity, there are small differences in the slopes of each line. We have used
 a least-square fit to determine the scaling exponent or  generalized dimension $D_q$ shown in Figure \ref{fig53}.

Ideally we would expect $D_q=2$ for a  two dimensional
homogeneous and isotropic  distribution.
We find that for the actual data $D_q$  varies in the range $1.7$ to
$2.2$ in the NGP and $1.8$ to $2.1$ in the SGP on large-scales in the range $80 \, h^{-1} \, {\rm Mpc}$ to $130 \, h^{-1} \, {\rm Mpc}$. In both the slices the
value of $D_q$  decreases with increasing $q$ as expected. The value of $D_q$  crosses $D_q=2$
somewhere around $q=-1$. The variation of $D_q$ with $q$ shows a
similar behavior in the random slices, but the range of variation is
much smaller $(1.9 \le D_q \le 2.1)$ in the same range of length scales as for the real data. Comparing the actual data with the random data we find that the actual data lies outside the $1-\sigma$ error-bars of the random data (not shown here) for most of the range of $q$ except around $q=-1$ where $D_q=2$ for
both the actual and random data. Accepting this at face value, this would imply that the actual data is not homogeneous at large scales. In order to test if the SDSS subsamples are {\it really} homogeneous and consistent with the \lcdm ~  model we have compared our results with different realizations of the \lcdm ~ model.

Considering the simulated data, we find that the variation in $D_q$
depends on the value of the bias $b$. For the unbiased simulations
$D_q$ shows very small variations $(1.9 \le D_q \le 2.1)$ and the
results are very close to those of the random data. We find that
increasing the bias causes the variations in $D_q$ to increase. In all
cases $D_q$ decreases with increasing $q$ and it crosses $D_q=2$
around $q=-1$. Increasing the bias has another effect in that it
results in larger $1-\sigma$ error-bars.

Comparing the simulated data with the random data and the actual data
we find that the unbiased simulations are consistent with the random
data. This implies that the unbiased \lcdm~
model has a transition to homogeneity at $60-70 \, h^{-1} \, {\rm
  Mpc}$.
  The spectrum of generalized dimensions as determined from
the unbiased simulations on length-scales $60-70 \, h^{-1} \, {\rm  Mpc}$ to $150 \, h^{-1} \, {\rm  Mpc}$ is different from that of the actual data.
The actual data, in fact, lies outside the
$1-\sigma$ error-bars of the unbiased \lcdm~ model.
This indicates that the unbiased \lcdm~ model fails to reproduce the large scale   properties
of the galaxy distribution in our volume limited subsamples of the SDSS-DR1.

 The simulations with bias $b=1.6$ and $b=2$ have larger $1-\sigma$
 error-bars and these are consistent with both the random and the
 actual data.  Interpreting the actual data as being a realization of
 a biased \lcdm~ universe, we conclude that it has a transition to
 homogeneity at $60-70 \, h^{-1} \, {\rm  Mpc}$ and the galaxy
 distribution is homogeneous on scales larger than this.

The galaxy subsample analyzed here contains the most  luminous
galaxies in the SDSS-DR1. Various investigations have shown the bias
to increase with luminosity \citep{nor,zevi} and the
subsample analyzed here is  expected to be
biased with respect to the underlying dark matter distribution. \citet{2005PhRvD..71d3511S}
 have used the halo model in conjunction with weak lensing
to determine the bias for a number of subsamples with different
absolute magnitude ranges. The brightest sample which they have
analyzed has galaxies with absolute magnitudes in the range $-23 \le
M_r \le -22$ for which they find a bias $b=1.94 \pm 0.2$. Our results
are consistent with these findings.

A point to note is that the $1-\sigma$ error-bars of the spectrum of
generalized dimension $D_q$ increases with the bias. This can be
understood in terms of the fact that  $C_q(r)$ is related to  volume
integrals of the correlation functions which receives contribution from
all length-scales. The fluctuations in $C_q(r)$ can also be related to
volume integrals of the correlation functions. Increasing the bias
increases the correlations on small scales $(\le 40-50 \, h^{-1} {\rm
  Mpc})$  which contributes to the fluctuations in $C_q(r)$ at large
scales and causes the fluctuations in $D_q$ to increase.

The galaxies in nearly all redshift surveys appear to be distributed
along filaments. These filaments appear to be interconnected and they
form a complicated network often referred to as the ``cosmic web''.
These filaments are possibly the largest coherent structures in
galaxy redshift surveys. Recent analysis of volume limited subsamples
of the LCRS  by \citet{bharad2} and the same SDSS-DR1 subsamples analyzed
here by \citet{pandey} shows the filaments to be statistically
significant features of the galaxy distribution on  length-scales $\le
70-80 \, h^{-1} {\rm   Mpc}$ and not beyond. Larger filaments present
in the galaxy distribution are not statistically significant and are
the result of chance alignments. Our finding that the galaxy
distribution is homogeneous on scales larger than $60-70 \, h^{-1}
{\rm   Mpc}$ is consistent with the size of the largest statistically
significant coherent structures namely the filaments.

\newpage
\begin{appendices}
\chapter{}
\section{\bf Survey Coordinate System}\label{scs}
The $SDSS$ is mapped in a spherical coordinate system with poles at $\alpha = 95^\circ, \delta = 0^\circ$
and $\alpha = 275^\circ, \delta = 0^\circ ~(J2000)$. The survey equator is a great circle
perpendicular to the $J2000$ celestial equator. The transformations between the equatorial
system $(\alpha, \delta)$ and the survey system $(\lambda, \eta)$ are given by:
\begin{eqnarray}
&\cos(\alpha - 95) \cos(\delta) &= -\sin \lambda \nonumber \\
&\sin(\alpha - 95) \cos (\delta) &= \cos(\lambda) \cos (\eta + 32.5) \\
&\sin(\delta)  &= \cos(\lambda) \sin (\eta + 32.5) \nonumber
\end{eqnarray}
as explained in \citet{2002AJ....123..485S}.
\section{\bf SQL Query to get data from SDSS Data Server}\label{sql}
{\bf SELECT} \\
objID,   
~ra,  
~dec, 
~z,	
~petro,
~$r1.petroMag\_r$  \\
{\bf FROM}  Galaxy \\
{\bf WHERE}  \\
-~-~For Northern Galactic Region\\
ra between 145 and 236 \\
-~-~For Southern Galactic Region\\
-~-~Right Ascention Range\\
and ra between 351 and 56\\
-~-~Declination Range\\
and dec between -1 and +1\\
-~-~Redshift Range\\
and z between 0.002 and 0.2\\
-~-~Extinction corrected Petrosin r band magnitude\\
and $( (petroMag\_r - extinction\_r +
2.5 * LOG10(2 * 3.1415 * petroR50\_r * petroR50\_r) ) < 17.7 ) )$\\
-~-~Apparent magnitude Range\\
and  $r1.petroMag\_r$ BETWEEN 14.5 and 17.5\\
\end{appendices} 

\chapter{Summary and future prospectus}
The present chapter briefly summarizes the main findings of our thesis. In this thesis
we have quantified the multifractal nature of the galaxy distribution
focusing mainly on the SDSS. The main findings are as follows,
\bi
\item
The distribution of galaxies in the SDSS behaves as a multifractal on scales less than $60 \, h^{-1} Mpc$.
It has different scaling index in different range of scales.
\item
The distribution of galaxies in SDSS is homogeneous on scales larger than $60-70  \, h^{-1} Mpc$.
\item
The unbiased $\Lambda CDM$~ model fails to reproduce the large scale properties
of the galaxy distribution in our volume limited subsamples of the SDSS-DR1 whereas a biased $\Lambda CDM$
 model with bias $b=1.6$ is consistent with the nature of galaxy distribution in SDSS.
\item
 The Minkowski- Bouligand Dimension ($D_q(r)$) for a homogeneous distribution of points is given by
\be
D_q(r) = D - \frac{\left( q - 2 \right)}{2} \frac{D }{\bar{N}} - \frac{D
}{\bar{N}}
\label{eqn:homogen1}
\ee
 implying that $D_q(r)$ does not  coincide with the Euclidean dimension ($D$) even if the distribution of points is homogeneous. Thus the benchmark for a uniform sample of points is not ($D$) but $D_q(r)$ as obtained from equation \ref{eqn:homogen1}. Thus  if the Minkowski-Bouligand dimension for a distribution of points coincides with $D_q(r)$ then it may be considered as a homogeneous distribution of points.
\item
   The correction due to a finite size sample always leads to a smaller value for $D_q(r)$
than the $D$.
\item
 The correction is small if ${\bar N} \gg 1$, as expected. The correction arises primarily due
to discreteness. The major advantage of our approach is that we are able to derive an expression for the correction.
\item
For a slightly clustered the Minkowski- Bouligand Dimension ($D_q(r)$) is presented as
\begin{eqnarray}
 D_q(r) &=& D  -
\frac{D \left( q - 2 \right)}{2\bar{N}} -
\frac{Dq}{2} \left(\bar\xi(r) - \xi(r) \right) \nonumber  \\
&=& D - \left(\Delta{D_q}\right)_{\bar{N}} -
  \left(\Delta{D_q}\right)_{clus}
\label{eqn:clus1}
\end{eqnarray}
implying that for hierarchical clustering, the value of  $D_q$ is always smaller than $D$ for positive value of q.
\item
 Unless the correlation function has a feature at some scale, smaller correlation
corresponds to a smaller correction to the Minkowski-Bouligand dimension.
\item
If the correlation function has a feature then it is possible to have a small
correction term $\left(\Delta{D_q}\right)_{clus}$ for a relatively large
$\xi$.
The relation between $\xi$ and $\left(\Delta{D_q}\right)_{clus}$ is not longer
one to one.
\item
For unbiased tracers of mass distribution (e.g.  $L_*$ type of galaxies) in the concordance model of cosmology
$\Delta D_q = D_q - D$ is a very slowly varying function of scale above $100$~h$^{-1}$Mpc. Hence this may be tentatively identified as the scale of homogeneity for this sort of population.
\item
For biased tracers of mass distribution (e.g. Large Redshift Galaxies (LRG)) our model fails to predict homogeneity at $100$~h$^{-1}$Mpc showing that the scale of homogeneity is much above $100$~h$^{-1}$Mpc for this kind of population.
\ei
A more quantitative approach requires us to estimate not only the systematic
offset $\Delta D_q$ but the dispersion in this quantity. The scale of homogeneity can then be identified as the scale where the offset is smaller than the expected dispersion.
In future we plan to undertake estimation of dispersion in $\Delta D_q$ .
We also plan to verify the results of our model using simulated distributions of points. In this thesis we have done a 2-D multifractal analysis of SDSS DR1 due ti its small data size. Now that large data sets is available in the form of SDSS DR6, we plan to do a 3-D multifractal analysis of the distribution.

\end{document}